\documentstyle[preprint,aps]{revtex}
\tightenlines
\begin{document}
\draft

\title{Tetrad Gravity: I) A New Formulation.}

\author{Luca Lusanna}

\address{
Sezione INFN di Firenze\\
L.go E.Fermi 2 (Arcetri)\\
50125 Firenze, Italy\\
E-mail LUSANNA@FI.INFN.IT}

\author{and}

\author{Stefano Russo}

\address
{Condominio dei Pioppi 16\\
6916 Grancia (Lugano)\\
Svizzera}

\maketitle
\begin{abstract}

A new version of tetrad gravity in globally hyperbolic, asymptotically flat 
at spatial infinity spacetimes with Cauchy surfaces diffeomorphic to $R^3$ is
obtained by using a new parametrization of arbitrary cotetrads to define
a set of configurational variables to be used in the ADM metric action. Seven of
the fourteen first class constraints have the form of the vanishing of 
canonical momenta. A comparison is made with other models of tetrad gravity 
and with the ADM canonical formalism for metric gravity. The phase space 
expression of various 4-tensors is explicitly given.

\vskip 1truecm
\noindent \today
\vskip 1truecm
\noindent This work has been partially supported by the network ``Constrained 
Dynamical Systems" of the E.U. Programme ``Human Capital and Mobility".

\end{abstract}

\newpage 

\vfill\eject

\section
{Introduction}

This is the first of a series of papers on the canonical reduction of a new 
formulation of tetrad gravity, motivated by the attempt to arrive at a unified 
description of the four interactions [with the matter being either 
Grassmann-valued Dirac fields or relativistic particles] based on 
Dirac-Bergmann theory of constraints, which is needed for the Hamiltonian
formulation of both gauge theories and general relativity. Therefore, we
shall study general relativity from the canonical point of view
generalizing to it all the results already obtained in the canonical
study of gauge theories in a systematic way, since neither a complete reduction
of gravity with an identification of the physical canonical degrees of
freedom of the gravitational field nor a detailed study of its Hamiltonian
group of gauge transformations (whose infinitesimal generators are the first
class constraints) has ever been pushed till the end in an explicit way.

The research program aiming to express the special relativistic strong, weak and
electromagnetic interactions in terms of Dirac's observables \cite{dirac}
is in an advanced  stage of development\cite{re}. This program is based on
the Shanmugadhasan canonical transformations \cite{sha}: if a system has 1st
class constraints at the Hamiltonian level (so that its dynamics is restricted
to a presymplectic submanifold of phase space), then, at least locally, one can
find a canonical basis with as many new momenta as 1st class constraints 
(Abelianization of 1st class constraints), with their conjugate canonical 
variables as Abelianized gauge variables and with the remaining pairs of 
canonical variables as pairs of canonically conjugate Dirac's observables 
(canonical basis of physical variables adapted to the chosen Abelianization;
they give a trivialization of the BRST construction of observables). Putting 
equal to zero the Abelianized gauge variables one defines a local gauge of the
model. If a system with constraints admits one (or more) global
Shanmugadhasan canonical transformations, one obtains one (or more) privileged 
global gauges in which the physical Dirac observables are globally defined and
globally separated from the gauge degrees of freedom [for
systems with a compact configuration space this is impossible]. These
privileged gauges (when they exist) can be called generalized Coulomb gauges.
Second class constraints, when present, are also taken into account by the
Shanmugadhasan canonical transformation\cite{sha}.

Firstly, inspired by Ref.\cite{dira},
the canonical reduction to noncovariant 
generalized Coulomb gauges, with the determination of the physical Hamiltonian
as a function of a canonical basis of Dirac's observables, has been achieved for
the following isolated systems (for them one can ask that the 10 conserved 
generators of the Poincar\'e algebra are finite so to be able to use group 
theory; theories with external fields can only be recovered as limits in some
parameter of a subsystem of the isolated system): 

a) Yang-Mills theory with Grassmann-valued
fermion fields \cite{lusa} in the case of a trivial principal
bundle over a fixed-$x^o$ $R^3$ slice of Minkowski spacetime with suitable
Hamiltonian-oriented boundary conditions; this excludes monopole solutions and,
since $R^3$ is not compactified, one has only winding number and no instanton
number. After a discussion of the
Hamiltonian formulation of Yang-Mills theory, of its group of gauge
transformations and of the Gribov ambiguity, the theory has been studied in
suitable  weighted Sobolev spaces where the Gribov ambiguity is absent and
the global color charges are well defined.
The global Dirac observables are the transverse quantities ${\vec A}_{a\perp}
(\vec x,x^o)$, ${\vec E}_{a\perp}(\vec x,x^o)$ and fermion fields dressed
with Yang-Mills (gluonic) clouds. The nonlocal and nonpolynomial (due to the
presence of classical Wilson lines along flat 
geodesics) physical Hamiltonian has been obtained: it is nonlocal but without 
any kind of singularities, it has the correct Abelian limit if the structure 
constants are turned off, and it contains the explicit realization of the 
abstract Mitter-Viallet metric.

b) The Abelian and non-Abelian SU(2)
Higgs models with fermion fields\cite{lv1,lv2}, where the
symplectic decoupling is a refinement of the concept of unitary gauge.
There is an ambiguity in the solutions of the Gauss law constraints, which
reflects the existence of disjoint sectors of solutions of the Euler-Lagrange
equations of Higgs models. The physical Hamiltonian and Lagrangian of  the
Higgs phase have been found; the self-energy turns out to be local and
contains a local four-fermion interaction. 

c) The standard SU(3)xSU(2)xU(1) model of elementary particles\cite{lv3}
with Grassmann-valued fermion fields.
The final reduced Hamiltonian contains nonlocal self-energies for the
electromagnetic and color interactions, but ``local ones" for the weak 
interactions implying the nonperturbative emergence of 4-fermions interactions.

The next problem is how to covariantize these results. Again the starting point 
was given by Dirac\cite{dirac} with his reformulation of classical field theory 
on spacelike hypersurfaces foliating Minkowski spacetime $M^4$ [the foliation 
is defined by an embedding $R\times \Sigma \rightarrow M^4$, $(\tau ,\vec 
\sigma ) \mapsto z^{\mu}(\tau ,\vec \sigma )$, with $\Sigma$ an abstract
3-surface diffeomorphic to $R^3$: this is the classical basis of
Tomonaga-Schwinger quantum field theory]. In this way one gets 
parametrized field theory with a covariant 3+1 splitting of flat spacetime and
already in a form suited to the transition to general relativity in its ADM
canonical formulation (see also Ref.\cite{kuchar}
, where a theoretical study of this problem is done in curved spacetimes).
The price is that one has to add as new configuration variables  the points 
$z^{\mu}(\tau ,\vec \sigma )$ of the spacelike hypersurface $\Sigma_{\tau}$ 
[the only ones carrying Lorentz indices; the scalar parameter $\tau$ labels
the leaves of the foliation and $\vec \sigma$ are curvilinear coordinates on
$\Sigma_{\tau}$] and then to define the fields on
$\Sigma_{\tau}$ so that they know  the hypersurface $\Sigma_{\tau}$ of 
$\tau$-simultaneity [for a Klein-Gordon field $\phi (x)$, this new field is
$\tilde \phi (\tau ,\vec \sigma )=\phi (z(\tau ,\vec \sigma ))$: it contains 
the nonlocal information about the embedding]. 
Then one rewrites the Lagrangian 
of the given isolated system in the form required by the coupling to an 
external gravitational field, makes the previous 3+1 splitting of Minkowski 
spacetime and interpretes all the fields of the system as the new fields on 
$\Sigma_{\tau}$ (they are Lorentz scalars, having only surface indices). 
Instead of considering the 4-metric as describing a 
gravitational field (and therefore as an independent field as it is done in 
metric gravity, where one adds the Hilbert action to the action for the matter 
fields), here one replaces the 4-metric with the the induced metric $g_{ AB}[z]
=z^{(\mu )}_{A}\eta_{(\mu )(\nu )}z^{(\nu )}_{B}$ on
$\Sigma_{\tau}$ [a functional of $z^{(\mu )}$;
here we use the notation $\sigma^{A}=(\tau ,\sigma^{r})$; $z^{(\mu )}_{A}=
\partial z^{(\mu )}/\partial \sigma^{A}$ are flat tetrad fields on Minkowski 
spacetime with the $z^{(\mu )}_r$'s tangent to $\Sigma_{\tau}$]
and considers the embedding coordinates $z^{(\mu )}(\tau ,\vec \sigma )$ as
independent fields [this is not possible in metric gravity, because in curved
spacetimes $z^{\mu}_{A}\not= \partial z^{\mu}/\partial \sigma^{A}$ are not
tetrad fields since the holonomic coordinates $z^{\mu}(\tau ,\vec \sigma )$
do not exist]. From this Lagrangian,
besides a Lorentz-scalar form of the constraints of the given system, 
we get four extra primary first class constraints
${\cal H}_{\mu}(\tau ,\vec \sigma )
\approx 0$ implying the independence of the description from the choice of the
foliation with spacelike hypersufaces. In special relativity, it is 
convenient to restrict ourselves to arbitrary spacelike hyperplanes $z^{\mu}
(\tau ,\vec \sigma )=x^{\mu}_s(\tau )+b^{\mu}_{\check r}(\tau ) \sigma^{\check 
r}$. Since they are described by only 10 variables [an origin $x^{\mu}_s(\tau 
)$ and, on it, three orthogonal spacelike unit vectors generating the fixed 
constant timelike unit normal to the hyperplane], we remain only with 10 first 
class constraints determining the 10 variables conjugate to the hyperplane 
[they are a 4-momentum $p^{\mu}_s$ and the six independent degrees of freedom 
hidden in a spin tensor $S^{\mu\nu}_s$] in terms of the variables of the system.

If we now consider only the set of configurations of the isolated system with
 timelike ($p^2_s > 0$) 4-momenta, we can 
restrict the description to the so-called Wigner hyperplanes orthogonal to 
$p^{\mu}_s$ itself. To get this result, we must boost at rest all the 
variables with Lorentz indices by using the standard Wigner boost $L^{\mu}{}
_{\nu}(p_s,{\buildrel \circ \over p}_s)$ for timelike Poincar\'e orbits, and
then add the gauge-fixings $b^{\mu}_{\check r}(\tau )-L^{\mu}{}_{\check r}(p_s,
{\buildrel \circ \over p}_s)\approx 0$. Since these gauge-fixings depend on 
$p^{\mu}_s$, the final canonical variables, apart $p^{\mu}_s$ itself, are of 3
types: i) there is a non-covariant ``external"
center-of-mass variable ${\tilde x}^{\mu}
(\tau )$ [it is only covariant under the little group of timelike Poincar\'e
orbits like the Newton-Wigner position operator]; ii) all
the 3-vector variables become Wigner spin 1 3-vectors [boosts in $M^4$ induce
Wigner rotations on them]; iii) all the other variables are Lorentz scalars. 
Only  four 1st class constraints  are left. One obtains 
in this way a new kind of instant form of the dynamics (see Ref.\cite{dira2}), 
the  ``Wigner-covariant 1-time rest-frame instant form"\cite{lus1} with a 
universal breaking of Lorentz covariance. 
It is the special relativistic generalization of
the nonrelativistic separation of the center of mass from the relative motions
[$H={{ {\vec P}^2}\over {2M}}+H_{rel}$]. The role of the ``external" center of 
mass is taken by the point ${\tilde x}^{\mu}(\tau )$ in the Wigner hyperplane
and by its normal $p^{\mu}_s$. 
The four 1st class constraints can be put in the 
following form: i) the vanishing of the total (Wigner spin 1) 3-momentum of the
system ${\vec p}_{sys}\approx 0$ , saying that 
the Wigner hyperplane $\Sigma_{W\, \tau}$ is the intrinsic rest frame
[instead, ${\vec p}_s$ is left arbitrary, since $p^{\mu}_s$ depends
upon the orientation of
the Wigner hyperplane with respect to arbitrary reference frames in $M^4$]; 
ii) $\pm \sqrt{p^2_s}-M_{sys}\approx 0$, saying that the
invariant mass $M_{sys}$ of the system replaces the nonrelativistic  Hamiltonian
$H_{rel}$ for the relative degrees of freedom, after the addition of the
gauge-fixing $T_s-\tau \approx 0$ [identifying the time parameter $\tau$,
labelling the leaves of the foliation,  with the Lorentz scalar time of the 
center of mass in the rest frame, $T_s=p_s\cdot {\tilde x}_s/M_{sys}$; 
$M_{sys}$ generates the evolution in this time].

Now 3 degrees of freedom of the isolated system [an ``internal" 
center-of-mass 3-variable ${\vec \sigma}_{sys}$ defined inside the Wigner
hyperplane and conjugate to ${\vec p}_{sys}$] become gauge variables [the
natural gauge fixing is ${\vec \sigma}_{sys}\approx 0$, so that it coincides 
with the origin $x^{(\mu )}_s(\tau )=z^{(\mu )}(\tau ,\vec \sigma =0)$ of the 
Wigner hyperplane], while the ${\tilde x}^{(\mu )}$ 
is playing the role of a kinematical external
center of mass for the isolated system and may be interpreted as a decoupled 
observer with his parametrized clock (point particle clock).
All the fields living on the Wigner hyperplane are now either Lorentz scalar 
or with their 3-indices transformaing under Wigner rotations (induced by Lorentz
transformations in Minkowski spacetime) as any Wigner spin 1 index.
The determination of ${\vec \sigma}_{sys}$ may be done with the group 
theoretical methods of Ref.\cite{pauris}: given a realization on the phase space
of a given system of the ten Poincar\'e generators one can build three 
3-position variables only in terms of them, which in our case of a system
on the Wigner hyperplane with ${\vec p}_{sys}\approx 0$ are: i) a canonical 
center of mass (the ``internal" center of mass ${\vec \sigma}_{sys}$); ii)
a noncanonical M\o ller center of energy ${\vec \sigma}^{(E)}_{sys}$; iii)
a noncanonical Fokker-Pryce center of inertia ${\vec \sigma}^{(FP)}_{sys}$. Due 
to ${\vec p}_{sys}\approx 0$, we have ${\vec \sigma}_{sys} \approx
{\vec \sigma}^{(E)}_{sys} \approx {\vec \sigma}^{(FP)}_{sys}$. By adding the
gauge fixings ${\vec \sigma}_{sys}\approx 0$ one can show that the origin
$x_s^{(\mu )}(\tau )$ becomes  simultaneously the Dixon center of mass of
an extended object and both the Pirani and Tulczyjew centroids (see Ref.
\cite{alp,mate} for the application of these methods to find the center of mass
of a configuration of the Klein-Gordon field after the preliminary work of
Ref.\cite{lon}). With similar methods one can construct three ``external"
collective positions (all located on the Wigner hyperplane): i) the ``external"
canonical noncovariant center of mass ${\tilde x}_s^{(\mu )}$; ii) the
``external" noncanonical and noncovariant M\o ller center of energy 
$R^{(\mu )}_s$; iii) the ``external" covariant noncanonical Fokker-Pryce center 
of inertia $Y^{(\mu )}_s$ (when
there are the gauge fixings ${\vec \sigma}_{sys}\approx 0$ it also coincides
with the origin $x^{(\mu )}_s$). It turns out that the Wigner hyperplane is
the natural setting for the study of the Dixon multipoles of extended 
relativistic systems\cite{dixon} and for defining the canonical relative
variables with respect to the center of mass. After having put control on the
relativistic definitions of center of mass of an extended system, the lacking
kinematics of relativistic rotations in now under investigation.
The Wigner hyperplane with its 
natural Euclidean metric structure offers a natural solution to the problem of
boost for lattice gauge theories and realizes explicitly the machian aspect of
dynamics that only relative motions are relevant.

The isolated systems till now analyzed to get their rest-frame 
Wigner-covariant generalized
Coulomb gauges [i.e. the subset of global Shanmugadhasan canonical bases, 
which, for each Poincar\'e stratum, are also adapted to the geometry of the
corresponding Poincar\'e orbits with their little groups; these special bases
can be named Poincar\'e-Shanmugadhasan bases for the given Poincar\'e stratum
of the presymplectic constraint manifold (every stratum requires an independent
canonical reduction); till now only the main stratum with
$P^2$ timelike and $W^2\not= 0$ has been investigated] are:

a) The system of N scalar particles with Grassmann electric charges
plus the electromagnetic field \cite{lus1}. The starting configuration 
variables are a 3-vector ${\vec 
\eta}_i(\tau )$ for each particle [$x^{\mu}_i(\tau )=z^{\mu}(\tau ,{\vec \eta}
_i(\tau ))$] and the electromagnetic gauge potentials 
$A_{\check A}(\tau ,\vec \sigma )={{\partial z^{\mu}(\tau ,\vec \sigma )}\over
{\partial \sigma^{\check A}}} A_{\mu}(z(\tau ,\vec \sigma ))$, 
which know  the embedding of
$\Sigma_{\tau}$ into $M^4$. One has to choose the sign of the energy of each
particle, because there are not mass-shell constraints (like $p_i^2-m^2_i\approx
0$) among the constraints of this formulation, due to the fact that one has only
three degrees of freedom for particle, determining the intersection of a 
timelike trajectory and of the spacelike hypersurface $\Sigma_{\tau}$. For
each choice of the sign of the energy of the N particles, one describes only one
of the branches of the mass spectrum of the manifestly covariant approach based
on the coordinates $x^{\mu}_i(\tau )$, $p^{\mu}_i(\tau )$, i=1,..,N, and on
the constraints $p^2_i-m^2_i\approx 0$ (in the free case). In this way, one 
gets a description of relativistic particles with a given sign of the energy
with consistent couplings to fields and valid independently from the quantum
effect of pair production [in the manifestly covariant approach, containing
all possible branches of the particle mass spectrum, the classical counterpart 
of pair production is the intersection of different branches deformed by the
presence of fields]. The final Dirac's 
observables are: i) the transverse radiation field variables; ii) the particle
canonical variables ${\vec \eta}_i(\tau )$, ${\check {\vec \kappa}}_i(\tau )$,
dressed with a Coulomb cloud. The physical Hamiltonian contains the 
mutual instantaneous Coulomb 
potentials extracted from field theory and there is a regularization of the
Coulomb self-energies due to the Grassmann character of the electric charges
$Q_i$ [$Q^2_i=0$]. In Ref.\cite{lus2} there is the study of the 
Lienard-Wiechert potentials and of Abraham-Lorentz-Dirac equations in this
rest-frame Coulomb gauge and also scalar electrodynamics is reformulated in it.
Also the rest-frame 1-time relativistic statistical mechanics has been developed
\cite{lus1}.

b) The system of N scalar particles with Grassmann-valued color charges plus 
the color SU(3) Yang-Mills field\cite{lus3}: 
it gives the pseudoclassical description of the
relativistic scalar-quark model, deduced from the classical QCD Lagrangian and 
with the color field present. The physical invariant mass of the system is
given in terms of the Dirac observables. From the reduced Hamilton equations  
the second order equations of motion both for the reduced transverse color 
field and the particles are extracted. Then, one studies  the N=2 
(meson) case. A special form of the requirement of having only color singlets, 
suited for a field-independent quark model, produces a ``pseudoclassical 
asymptotic freedom" and a regularization of the quark self-energy. With these
results one can covariantize the bosonic part of the standard model given in
Ref.\cite{lv3}.
 
c) The system of N spinning particles of definite energy [$({1\over 2},0)$ or
$(0,{1\over 2})$ representation of SL(2,C)] with Grassmann electric charges 
plus the electromagnetic field\cite{biga} and that of a Grassmann-valued
Dirac field plus the electromagnetic field (the pseudoclassical basis of QED) 
\cite{bigaz}. In both cases there are geometrical complications connected with 
the spacetime description of the path of electric currents and not only of their
spin structure, suggesting a reinterpretation of the supersymmetric scalar 
multiplet as a spin fibration; a new canonical decomposition of the 
Klein-Gordon field into collective and relative variables \cite{lon,mate} 
will be helpful to clarify these problems. After their solution and after having
obtained the description of Grassmann-valued chiral fields [this will require
the transcription of the front form of the dynamics in the instant one for the
Poincar\'e strata with $P^2=0$] the rest-frame form of the full standard 
$SU(3)\times SU(2)\times U(1)$ model can be achieved.

Finally,  to eliminate the three 1st class constraints $\vec p[system]
\approx 0$ by finding their natural gauge-fixings, when fields are present,
one needs to find a rest-frame canonical basis of center-of-mass and relative
variables for fields (in analogy to particles). A basis with a
``center of phase" has already been found for a real Klein-Gordon field
both in the covariant approach\cite{lon} and on spacelike hypersurfaces 
\cite{mate}. In this case also the ``internal" center of mass has been
found, but not yet a canonical basis containing it. There is the hope that all
these new pieces of information  will allow, after quantization of this new
consistent relativistic mechanics without the classical problems connected
with pair production, to find the  asymptotic states of the covariant
Tomonaga-Schwinger formulation of quantum field theory on spacelike
hypersurfaces: these states are needed for the theory of quantum bound states
[since Fock states do not constitute a Cauchy problem for the field equations,
because an in (or out) particle can be in the absolute future of another one due
to the tensor product nature of these asymptotic states, bound state equations
like the Bethe-Salpeter one have spurious solutions which are excitations in
relative energies, the variables conjugate to relative times (which are gauge
variables\cite{lus1})]. Moreover, it will be possible to include bound states 
among the asymptotic states.

As said in Ref.\cite{lus2,lus3}, the quantization of these rest-frame
models has to overcome two problems. On the particle
side, the complication is the quantization of the square roots associated
with the relativistic kinetic energy terms: in the free case this has been done
in Ref.\cite{lam} [see Refs.\cite{sqroot} for the complications induced by the
Coulomb potential]. On the field side (all physical
Hamiltonian are nonlocal and, with the exception of the Abelian case,
nonpolynomial, but quadratic in the momenta), the obstacle
is the absence (notwithstanding there is no  no-go theorem) of a complete
regularization and renormalization procedure of electrodynamics (to start with) 
in the Coulomb gauge: see Ref.\cite{cou} (and its bibliography)
for the existing results for QED. 

However, as shown in Refs.\cite{lus1,lusa}
[see their bibliography for the relevant references regarding  all the 
quantities introduced in this Section], the rest-frame instant 
form of dynamics automatically gives a physical ultraviolet cutoff in the 
spirit of Dirac and Yukawa: it is the M$\o$ller radius\cite{mol} 
$\rho =\sqrt{-W^2}/P^2=|\vec S|/\sqrt{P^2}$ ($W^2=-P^2{\vec 
S}^2$ is the Pauli-Lubanski Casimir when $P^2 > 0$), namely the classical 
intrinsic radius of the worldtube, around the covariant noncanonical 
Fokker-Pryce center of inertia $Y^{\mu}$, 
inside which the noncovariance of the canonical center of mass ${\tilde
x}^{\mu}$ is concentrated. At the quantum level $\rho$ becomes the Compton 
wavelength of the isolated system multiplied its spin eigenvalue $\sqrt{s(s+1)}$
, $\rho \mapsto \hat \rho = \sqrt{s(s+1)} \hbar /M=\sqrt{s(s+1)} \lambda_M$ 
with $M=\sqrt{P^2}$ the invariant mass and $\lambda_M=\hbar /M$ its Compton
wavelength. Therefore, the criticism to classical relativistic physics, based
on quantum pair production, concerns the testing of distances where, due to the
Lorentz signature of spacetime, one has intrinsic classical covariance problems:
it is impossible to localize the canonical center of mass ${\tilde x}^{\mu}$
adapted to the first class constraints of the system
(also named Pryce center of mass and having the same covariance of the 
Newton-Wigner position operator) in a frame independent way.

Let us remember \cite{lus1}
that $\rho$ is also a remnant in flat Minkowski spacetime of 
the energy conditions of general relativity: since the M$\o$ller
noncanonical, noncovariant center of energy has its noncovariance localized
inside the same worldtube with radius $\rho$ (it was discovered in this way)
\cite{mol}, it turns out that for an extended relativistic system with the
material radius smaller than its intrinsic radius $\rho$ one has: i) its 
peripheral rotation velocity can exceed the velocity of light; ii) its 
classical energy density cannot be positive definite everywhere in every frame. 

Now, the real relevant point is that this ultraviolet cutoff determined by
$\rho$ exists also in Einstein's
general relativity (which is not power counting renormalizable) in the case of
asymptotically flat spacetimes, taking into account the Poincar\'e Casimirs of
its asymptotic ADM Poincar\'e charges (when supertranslations are eliminated 
with suitable boundary conditions; let us remark that Einstein and Wheeler
use closed universes because they don't want to introduce boundary conditions,
but in this way they loose Poincar\'e charges and the possibility to make 
contact  with particle physics and to define spin). The generalization of the 
worldtube of radius $\rho$ to asymptotically flat general relativity with 
matter could be connected with the unproved cosmic censorship hypothesis.

Moreover, the extended Heisenberg relations  of string theory\cite{ven}, i.e.
$\triangle x ={{\hbar}\over {\triangle p}}+{{\triangle p}\over {T_{cs}}}=
{{\hbar}\over {\triangle p}}+{{\hbar \triangle p}\over {L^2_{cs}}}$ implying the
lower bound $\triangle x > L_{cs}=\sqrt{\hbar /T_{cs}}$ due to the $y+1/y$
structure,
have a counterpart in the quantization of the M$\o$ller radius\cite{lus1}:
if we ask that, also at the quantum level, one cannot test the inside of the 
worldtube, we must ask $\triangle x > \hat \rho$ which is the lower bound
implied by the modified uncertainty relation $\triangle x ={{\hbar}\over 
{\triangle p}}+{{\hbar \triangle p}\over {{\hat \rho}^2}}$. This could imply 
that the center-of-mass canonical noncovariant  3-coordinate 
$\vec z=\sqrt{P^2}({\vec {\tilde x}}-{{\vec P}\over {P^o}}{\tilde x}^o)$ 
\cite{lus1} cannot become a
self-adjoint operator. See Hegerfeldt's theorems (quoted in 
Refs.\cite{lusa,lus1}) and his interpretation 
pointing at the impossibility of a good localization of relativistic particles
(experimentally one determines only a worldtube in spacetime emerging from the 
interaction region). Since the eigenfunctions of the canonical center-of-mass
operator are playing the role of the wave function of the universe, one could 
also say that the center-of-mass variable has not to be quantized, because it
lies on the classical macroscopic side of Copenhagen's interpretation and,
moreover, because, in the spirit of Mach's principle that only relative 
motions can be observed, no one can observe it (it is only used to define a
decoupled ``point particle clock"). On the other hand, if one 
rejects the canonical noncovariant center of mass in favor of the covariant
noncanonical Fokker-Pryce center of inertia $Y^{\mu}$, $\{ Y^{\mu},Y^{\nu} \}
\not= 0$, one could invoke the philosophy of quantum groups to quantize 
$Y^{\mu}$ to get some kind of quantum plane for the center-of-mass 
description. Let us remark that the quantization of the square root Hamiltonian
done in Ref.\cite{lam} is consistent with this problematic.

In conclusion, the best set of canonical coordinates adapted to the constraints
and to the geometry of Poincar\'e orbits and naturally predisposed to the
coupling to canonical tetrad gravity is emerging for the electromagnetic, weak
and strong interactions with matter described either by fermion fields or by
relativistic particles with a definite sign of the energy.
Therefore, one can begin to think how to quantize the standard model in the 
Wigner-covariant Coulomb gauge in the rest-frame instant form (the classical
background fo the Tomonaga-Schwinger approach to quantum field theory) with the
M\"oller radius as a ultraviolet cutoff. 

Since our aim is to arrive at a unified description of the four interactions,
in this paper and in the following ones
we shall explore the canonical reduction to Dirac's observables
of tetrad gravity (more natural than metric gravity for the coupling to fermion 
fields) and we shall begin to explore the connection of Dirac's observables 
with Bergmann's definition of observables and the problem of time in general 
relativity \cite{be,ish,kuchar1}. Moreover, in globally hyperbolic, 
asymptotically flat at spatial infinity, spacetimes, we shall arrive at a 
solution of the deparametrization problem of general relativity (how to recover
the rest-frame instant form when the Newton constant is put equal to zero,
G=0), to a solution, till now at order G, of the superhamiltonian constraint,
with the matter represented (to start with) by N massive scalar particles,
allowing to
visualize the instantaneous part of the interaction (think to the Coulomb
potential in the electromagnetic Coulomb gauge), and to the identification of
the volume expression of the ADM energy as the reduced Hamiltonian of the 
universe, containing all the interactions. Then, the replacement of scalar
particles with spinning ones will allow to test the precessional effects 
(gravitomagnetism) of general relativity.

We shall restrict ourselves to the simplest class of spacetimes to have some 
chance to arrive at the end of the canonical reduction. Refs.
\cite{naka,oneil,blee} are used 
for the background in differential geometry. A spacetime is a time-oriented
pseudo-Riemannian (or Lorentzian) 4-manifold $(M^4,{}^4g)$ with signature
$\epsilon \, (+---)$ ($\epsilon =\pm 1$) and with a choice of time orientation
[i.e. there exists a continuous, nowhere vanishing timelike vector field which
is used to separate the nonspacelike vectors at each point of $M^4$ in either
future- or past-directed vectors]. Our spacetimes are assumed to be:

i) Globally hyperbolic 4-manifolds, i.e. topologically they are $M^4=R\times 
\Sigma$, so to have a well posed Cauchy problem [with $\Sigma$ the abstract
model of Cauchy surface] at least till when no singularity develops in $M^4$
[see the singularity theorems]. Therefore, these spacetimes admit regular 
foliations with orientable, complete, non-intersecting spacelike 3-manifolds:
the leaves of the foliation are the embeddings $i_{\tau}:\Sigma \rightarrow 
\Sigma_{\tau} \subset M^4$, $\vec \sigma \mapsto z^{\mu}(\tau ,\vec \sigma )$,
where $\vec \sigma =\{ \sigma^r \}$, r=1,2,3, are local coordinates in a chart
of the $C^{\infty}$-atlas of the abstract 3-manifold $\Sigma$ and $\tau :M^4 
\rightarrow R$, $z^{\mu} \mapsto \tau (z^{\mu})$, is a global timelike
future-oriented function labelling the leaves (surfaces of simultaneity). In 
this way, one obtains 3+1 splittings of $M^4$ and the possibility of a 
Hamiltonian formulation.

ii) Asymptotically flat at spatial infinity, so to have the possibility to 
define asymptotic Poincar\'e charges \cite{adm,reg,reg1,reg2,reg3,ash}:
they allow the definition of a M$\o$ller radius in general relativity and are a
bridge towards a future soldering with the theory of elementary particles in
Minkowski spacetime defined as irreducible representation of its kinematical,
globally implemented Poincar\'e group according to Wigner. In this paper we
will not compactify space infinity
at a point like in the spi approach of Ref.\cite{ash}.

iii) Since we want to be able to introduce Dirac fermion fields, our
spacetimes $M^4$ must admit a spinor (or spin) structure\cite{wald}. Since we 
consider noncompact space- and time-orientable spacetimes, spinors can be
defined if and only if they are ``parallelizable" \cite{ger}. This means that
we have trivial principal frame bundle $L(M^4)=M^4\times GL(4,R)$ with GL(4,R) 
as structure group and trivial orthonormal frame bundle $F(M^4)=M^4\times
SO(3,1)$; the fibers of $F(M^4)$ are the disjoint union of four components and
$F_o(M^4)=M^4\times L^{\uparrow}_{+}$ [with projection $\pi: F_o(M^4)
\rightarrow M^4$] corresponds to the proper subgroup
$L^{\uparrow}_{+}\subset SO(3,1)$ of the Lorentz group. Therefore, global
frames (tetrads) and coframes (cotetrads) exist. A spin structure for $F_o(M^4)$
is, in this case, the trivial spin principal SL(2,C)-bundle $S(M^4)=M^4\times
SL(2,C)$ [with projection $\pi_s:S(M^4)\rightarrow M^4$] and a map $\lambda :
S(M^4)\rightarrow F_o(M^4)$ such that $\pi (\lambda (p))=\pi_s(p)\in M^4$ for
all $p\in S(M^4)$ and $\lambda (pA)=\lambda (p) \Lambda (A)$ for all $p\in 
S(M^4)$, $A\in SL(2,C)$, with $\Lambda :SL(2,C)\rightarrow L^{\uparrow}_{+}$ the
universal covering homomorphism. Then, Dirac fields are defined as cross 
sections of a bundle associated with $S(M^4)$ \cite{blee}. Since $M^4=R\times 
\Sigma$ is time- and space-oriented, the hypersurfaces $\Sigma_{\tau}$ of 
simultaneity are necessarily space-oriented and are parallelizable (as every 
3-manifold\cite{ger}): therefore, global triads and cotriads exist. $F(\Sigma
_{\tau})=\Sigma_{\tau}\times SO(3)$ is the trivial orthonormal frame
SO(3)-bundle and, since one has $\pi_1(SO(3))=\pi_1(L^{\uparrow}_{+})=Z_2$ for
the first homotopy group, one can 
define SU(2) spinors on $\Sigma_{\tau}$ \cite{spinor,spinor1}.

iv) The noncompact parallelizable simultaneity 3-manifolds (the Cauchy surfaces)
$\Sigma_{\tau}$ are assumed to be topologically trivial, geodesically complete
[so that the Hopf-Rinow theorem\cite{oneil} assures metric completeness of the
Riemannian 3-manifold $(\Sigma_{\tau},{}^3g)$] and, finally, diffeomorphic to
$R^3$. These 3-manifolds have the same manifold structure as Euclidean spaces
\cite{oneil}:
a) the geodesic exponential map $Exp_p:T_p\Sigma_{\tau}\rightarrow \Sigma
_{\tau}$ is a diffeomorphism (Hadamard theorem); b) the sectional curvature is 
less or equal  zero everywhere; c) they have no ``conjugate locus" [i.e.
there are no pairs of conjugate Jacobi points (intersection points of distinct
geodesics through them) on any geodesic] and no ``cut locus" [i.e. no closed
geodesics through any point]. In these manifolds two points determine a line, 
so that the ``static" tidal forces in $\Sigma_{\tau}$ due to the 3-curvature 
tensor are repulsive; instead in $M^4$ the tidal forces due to the 4-curvature
tensor are attractive, since they describe gravitation, which is always
attractive, and this implies that the sectional 4-curvature of timelike 
tangent planes must be negative (this is the source of the singularity theorems)
\cite{oneil}. In 3-manifolds not of this class one has to give a physical
(topological)
interpretation of  ``static" quantities like the two quoted loci. In particular,
these 3-manifolds have global charts inherited by $R^3$ through the
diffeomorphism.
Given a Cauchy surface $\Sigma_{\tau_o}$ of this type and a set of Cauchy data 
for the gravitational field (and for matter, if present), the Hamiltonian
evolution we are going to describe will be valid from $\tau_o$ till $\tau_o+
\triangle \tau$, where the interval $\triangle \tau$ is determined by the
appearance of either conjugate points on $\Sigma_{\tau_o+\triangle \tau}$
or 4-dimensional singularities in $M^4$ on its slice $\Sigma_{\tau_o+\triangle 
\tau}$.

v) Like in Yang-Mills case \cite{lusa}, the 3-spin-connection on the orthogonal
frame SO(3)-bundle (and therefore triads and cotriads) will have to be 
restricted to suited weighted Sobolev spaces to avoid Gribov ambiguities. In
turn, this implies the absence of isometries of the noncompact Riemannian
3-manifold $(\Sigma_{\tau},{}^3g)$ [see for instance the review paper in Ref.
\cite{cho}]. All the problems of the boundary conditions on lapse and shift 
functions and on cotriads will be studied in connection with the Poincar\'e
charges in a future paper.

Diffeomorphisms on $\Sigma_{\tau}$ ($Diff\, \Sigma_{\tau}$) will be interpreted 
in the passive way, following Ref.\cite{be}, in accord with the Hamiltonian
point of view that infinitesimal diffeomorphisms are generated by taking the
Poisson bracket with the 1st class supermomentum constraints [passive
diffeomorphisms are also named `pseudodiffeomorphisms']
. The Lagrangian approach based on the Hilbert action, connects general 
covariance with the invariance of the action under spacetime diffeomorphisms 
($Diff\, M^4$) extended to 4-tensors. 
Therefore, the moduli space (or superspace or space of
4-geometries) is the space $Riem\, M^4/Diff\, M^4$ \cite{whe}, where $Riem\, 
M^4$ is the space of Lorentzian 4-metrics; as shown in Refs.\cite{fis,ing},
superspace, in general, is not a manifold [it is a stratified manifold with
singularities\cite{arms}] due to the existence (in Sobolev spaces) of 4-metrics
and 4-geometries with isometries. See Ref.\cite{giul} for the study of great
diffeomorphisms, which are connected with the existence of disjoint components
of the diffeomorphism group 
[in Ref.\cite{lusa} there is the analogous discussion of the connection of 
winding number with the great gauge transformations]. Instead, in 
the ADM Hamiltonian formulation of metric gravity\cite{adm} space 
diffeomorphisms are replaced by $Diff\, \Sigma_{\tau}$ [or better by their
induced action on 3-tensors generated by the supermomentum constraints], while 
time diffeomorphisms are distorted to the transformations generated by the
superhamiltonian 1st class constraint\cite{wa,ish,beig} and by the
momenta conjugate to the lapse and shift functions. In the
Lichnerowicz-York conformal approach to canonical reduction
\cite{conf,york} [see Refs.\cite{cho,yoyo,ciuf} for reviews], one defines, in
the case of closed 3-manifolds, the conformal superspace as the space of
conformal 3-geometries [namely the space of conformal 3-metrics modulo $Diff\,
\Sigma_{\tau}$ or, equivalently, as $Riem\, \Sigma_{\tau}$ (the space of
Riemannian 3-metrics) modulo $Diff\, \Sigma_{\tau}$ and conformal 
transformations ${}^3g \mapsto \phi^4\, {}^3g$ ($\phi > 0$)], because in this
approach gravitational dynamics is regarded as the time evolution of conformal
3-geometry [the momentum conjugate to the conformal factor $\phi$ is 
replaced by York time \cite{york,qadir}, i.e.  the trace of the
extrinsic curvature of $\Sigma_{\tau}$]. However, the gauge transformations
generated by the superhamiltonian constraint are poorly understood.
Moreover, the Hamiltonian group of gauge transformations of the ADM theory has 
8 (and not 4) generators, because, besides the superhamiltonian and 
supermomentum constraints, there are the four primary first class
constraints giving the vanishing of the canonical momenta conjugate to the
lapse and shift functions [ whose gauge nature is connected with the gauge
nature (conventionality) of simultaneity \cite{simul} and of the standards of 
time and length]. A discussion of these problems and of general covariance 
versus Dirac's observables will be given in Ref.\cite{russo2} [as also recently
noted in Ref.\cite{but} the problem of observables is still open in canonical 
gravity].

Our approach to tetrad gravity [see Refs.
\cite{weyl,dirr,schw,kib,tetr,char,clay,maluf,hen1,hen2,hen3,hen4} 
for the existing
versions of the theory] utilizes the ADM action of metric gravity with the
4-metric expressed in terms of arbitrary cotetrads, which are 
parametrized in a particular way in terms of Lorentz-boost parameters and 
cotetrads adapted to $\Sigma_{\tau}$ [which, in turn, depend on cotriads on
$\Sigma_{\tau}$ and on lapse and shift functions].

At the Hamiltonian level, the Hamiltonian gauge group
contains: i) a $R^3\times SO(3)$ subgroup replacing the usual Lorentz subgroup
due to our parametrization which Abelianizes Lorentz boosts; ii) $Diff\, 
\Sigma_{\tau}$ in the sense of the pseudodiffeomorphisms generated by the
supermomentum constraints; iii) the gauge 
transformations generated by a superhamiltonian 1st class constraint; iv) the
gauge transformations generated by the momenta conjugate to the lapse and shift
functions. In the second paper\cite{russo2}
 we shall extract Dirac's observables starting from the symplectic 
action of infinitesimal diffeomorphisms in $Diff\, \Sigma_{\tau}$, ignoring the
problems on the structure in large of the
component of $Diff\, \Sigma_{\tau}$ connected to the identity when a
differential structure is posed on it. Although such 
global properties can be studied in Yang-Mills theory (since the group of
gauge transformations is a Hilbert-Lie group), as shown in Ref.\cite{lusa}, 
and can be applied to the SO(3) gauge transformations of cotriads (in our 
approach the Lorentz boosts are automatically Abelianized), one has that SO(3)
gauge transformations and $Diff\, \Sigma_{\tau}$ do not commute. Therefore, in
tetrad gravity the group of SO(3) gauge transformations is an invariant
subgroup of a larger group, the group of automorphisms of the SO(3) frame 
bundle, containing also $Diff\, \Sigma_{\tau}$ and again
the global situation in the large is of difficult control [$Diff\, \Sigma
_{\tau}$ is an inductive limit of Hilbert-Lie groups \cite{sch}, but the
global properties of its group manifold are not well understood].

In this first paper, after a review of the formalisms needed in this and in 
the future papers, we shall introduce our parametrization of the cotetrads, we 
shall give the Lagrangian and Hamiltonian formulations of tetrad gravity and 
we shall study the algebra of the resulting fourteen first class constraints.

Section II is devoted to a review of 4-dimensional pseudo-Riemannian and
3-dimensional Riemannian manifolds asymptotically flat at spatial infinity,
of the tetrad formalism and of the Lagrangians used for general relativity.

In Section III, $\Sigma_{\tau}$-adapted tetrads and triads are introduced and 
the new parametrization of cotetrads is defined. 

In Section IV such parametrized cotetrads are inserted in the ADM metric action
and the Hamiltonian formulation is performed with the identification of fourteen
first class constraints. The comparison with other formulations of tetrad 
gravity is done.

In Section V there is a comparison with ADM canonical metric gravity and a 
comment on the Hamiltonian formulation of the harmonic gauge.

In the Conclusions the next step, namely the identification of the Dirac
observables with respect to the gauge transfomations generated by thirteen
constraints (only the superhamitonian constraint is not treated), is delineated.

In Appendix A relevant 4-tensors are described in $\Sigma_{\tau}$-adapted
holonomic coordinates and in Appendix B their Hamiltonian expression is given.

\vfill\eject

\section{Notations.}

In this Section we shall introduce the notations needed to define the ADM
tetrads and triads of the next Section. 

Let $M^4$ be a torsion-free, globally hyperbolic, asymptotically flat
pseudo-Riemannian (or Lorentzian) 4-manifold, whose 
nondegenerate 4-metric tensor ${}^4g_{\mu\nu}(x)$ has Lorentzian signature 
$\epsilon (+,-,-,-)$ with $\epsilon =\pm 1$ according to particle physics and
general relativity conventions respectively; 
the inverse 4-metric is ${}^4g^{\mu\nu}(x)$ with
${}^4g^{\mu\rho}(x){}^4g_{\rho\nu}(x)=\delta^{\mu}_{\nu}$.
We shall denote with Greek letters $\mu ,\nu ,..$
($\mu =0,1,2,3$), the world indices and with Greek letters inside round brackets
$(\alpha ), (\beta ),..,$ flat Minkowski indices [with flat 4-metric tensor
${}^4\eta_{(\alpha )(\beta )}=\epsilon (+,-,-,-)$ in Cartesian
coordinates]; analogously, $a, b,..,$ and
$(a), (b),..,$ [a=1,2,3], will denote world and flat 3-space indices.

We shall follow the conventions of Refs.\cite{mtw,ciuf} for $\epsilon =-1$
and those of Ref.\cite{wei} for $\epsilon =+1$ [i.e. the conventions of standard
textbooks; see also Ref.\cite{wald} for many results (this book is consistent
with Ref.\cite{mtw}, even if its index conventions
are different)].

The coordinates of a chart of the atlas of $M^4$ will be denoted $\lbrace 
x^{\mu}\rbrace$. $M^4$ is assumed to be orientable; its volume element in any
right-handed coordinate basis is $-\eta \sqrt{{}^4g}\, d^4x$ [$\eta$ is a sign 
connected with the choice of the orientation and ${}^4g=|det\, {}^4g_{\mu\nu}|$
; with $\eta =\epsilon$ we get 
the choice of Ref.\cite{mtw} for $\epsilon =-1$ and of Ref.\cite{wei} for 
$\epsilon =+1$]. In the coordinate bases $e_{\mu}=\partial_{\mu}$ and 
$dx^{\mu}$ for vector fields [$TM^4$] and one-forms [or covectors; $T^{*}M^4$] 
respectively, the unique metric-compatible 
Levi-Civita affine connection has the symmetric Christoffel symbols
${}^4\Gamma^{\mu}_{\alpha\beta} = {}^4\Gamma^{\mu}_{\beta\alpha}={1\over 2}
{}^4g^{\mu\nu} (\partial_{\alpha}\, {}^4g_{\beta\nu} + \partial_{\beta}\,
{}^4g_{\alpha\nu} - \partial_{\nu}\, {}^4g_{\alpha\beta})$ 
as connection coefficients [${}^4\Gamma^{\mu}_{\mu\nu}=\partial_{\nu}
\sqrt{{}^4g}$] and the associated covariant derivative 
is denoted ${}^4\nabla_{\mu}$ [or with a semicolon ``;"]: ${}^4V^{\mu}{}_{;\nu}
={}^4\nabla_{\nu}\, {}^4V^{\mu}=\partial_{\nu}\, {}^4V^{\mu}+{}^4\Gamma^{\mu}
_{\nu\alpha}\, {}^4V^{\alpha}$, with the metric
compatibility condition being ${}^4\nabla_{\rho}\, {}^4g^{\mu\nu}=0$.
The Christoffel symbols are not tensors. If, instead of the chart of $M^4$
with coordinates $\lbrace x^{\mu}\rbrace$, we choose another chart of $M^4$,
overlapping with the previous one, with coordinates $\lbrace x^{{'}\mu}=
x^{{'}\mu}(x)\rbrace$ [$x^{{'}\mu}(x)$ smooth functions], in the overlap of the
two charts we have the following transformation properties under general
smooth coordinate transformations or diffeomorphisms of $M^4$ [$Diff\, M^4$,
the gauge group  of Einstein-Hilbert Lagrangian] 
of ${}^4g_{\alpha\beta}(x)$ and of ${}^4\Gamma^{\mu}
_{\alpha\beta}(x)$ respectively

\begin{eqnarray}
&&{}^4g^{'}_{\alpha\beta}(x^{'}(x)) = {{\partial x^{\mu}}\over {\partial
x^{{'}\alpha}}}\, {{\partial x^{\nu}}\over {\partial x^{{'}\beta}}}\,
{}^4g_{\mu\nu}(x),\nonumber \\
&&{}^4\Gamma^{{'}\mu}_{\alpha\beta}(x^{'}(x)) = {{\partial x^{{'}\mu}}\over
{\partial x^{\nu}}}\, {{\partial x^{\gamma}}\over {\partial x^{{'}\alpha}}}\,
{{\partial x^{\delta}}\over {\partial x^{{'}\beta}}}\, {}^4\Gamma^{\nu}
_{\gamma\delta}(x) + {{\partial^2x^{\nu}}\over {\partial x^{{'}\alpha} \partial
x^{{'}\beta} }}\, {{\partial x^{{'}\mu}}\over {\partial x^{\nu}}}.
\label{a2}
\end{eqnarray}

For a tensor density of 
weight W, ${}^4{\cal T}^{\mu ...}{}_{\alpha ...}=({}^4g)^{-W/2}\, {}^4T
^{\mu ...}{}_{\alpha ...}$, we have ${}^4{\cal T}^{\mu ...}{}_{\alpha ..;\rho}=
({}^4g)^{-W/2}[({}^4g)^{W/2}\, {}^4{\cal T}^{\mu ...}{}_{\alpha ...}]_{;\rho}=
({}^4g)^{-W/2}\, {}^4T^{\mu ...}{}_{\alpha ..;\rho}=\partial_{\rho}\,
{}^4{\cal T}^{\mu ...}{}_{\alpha ...}+{}^4\Gamma^{\mu}_{\rho\nu}\, 
{}^4{\cal T}^{\nu ...}{}_{\alpha ...}+\cdots -{}^4\Gamma^{\beta}_{\rho\alpha}
{}^4{\cal T}^{\mu ...}{}_{\beta ...} -\cdots +W\, {}^4\Gamma^{\sigma}
_{\sigma\rho}\, {}^4{\cal T}^{\mu ...}{}_{\alpha ...}$ [$\partial_{\rho}
({}^4g)^{-W/2}+W ({}^4g)^{-W/2}\, {}^4\Gamma^{\mu}_{\mu\rho}=0$]. The covariant
divergence of a vector density of weight -1 is equal to its ordinary divergence:
${}^4\nabla_{\mu}\, {}^4{\cal T}^{\mu}=\partial_{\mu}\, {}^4{\cal T}^{\mu}+
{}^4\Gamma^{\mu}_{\mu\nu}\, {\cal T}^{\nu}-{}^4\Gamma^{\mu}_{\mu\nu}\, {}^4{\cal
T}^{\nu}=\partial_{\mu}\, {}^4{\cal T}^{\mu}$.
For the Lie derivatives we have: i) ${\cal L}_{V^{\alpha}\partial
_{\alpha}}\, {}^4g_{\mu\nu}=V_{\mu ;\nu}+V_{\nu ;\mu}$ ; ii) ${\cal L}_{V
^{\alpha}\partial_{\alpha}} \sqrt{{}^4g}={1\over 2}\sqrt{{}^4g}\, {}^4g^{\mu\nu}
{\cal L}_{V^{\alpha}\partial_{\alpha}}\, {}^4g_{\mu\nu}=\partial_{\mu}
(\sqrt{{}^4g} V^{\mu})$ and ${}^4g^{w/2} {\cal L}_{V^{\alpha}\partial_{\alpha}}
\, {}^4g^{-w/2}=-{}^4g^{-w/2} {\cal L}_{V^{\alpha}\partial_{\alpha}}\,
{}^4g^{w/2}=-{w\over {\sqrt{{}^4g}}} {\cal L}_{V^{\alpha}\partial_{\alpha}}
\sqrt{{}^4g}$; iii) ${\cal L}_{V^{\alpha}\partial_{\alpha}}\, {}^4{\cal T}
^{\mu}=-w \partial_{\alpha}V^{\alpha}\, {}^4{\cal T}^{\mu}+V^{\alpha}\partial
_{\alpha}\, {}^4{\cal T}^{\mu}-\partial_{\alpha}V^{\mu}\, {}^4{\cal T}^{\alpha}$
and ${\cal L}_{V^{\alpha}\partial_{\alpha}}(\sqrt{{}^4g} f)=\partial_{\mu}
(\sqrt{{}^4g} f V^{\mu})$.

The Riemann curvature tensor is [this is the definition of Ref.\cite{mtw} for
$\epsilon =-1$; for $\epsilon =+1$ it coincides with minus the definition of
Ref.\cite{wei}]

\begin{eqnarray}
{}^4R^{\alpha}{}_{\mu\beta\nu}&=&{}^4\Gamma^{\alpha}_{\beta\rho}\, {}^4\Gamma
^{\rho}_{\nu\mu} - {}^4\Gamma^{\alpha}_{\nu\rho}\, {}^4\Gamma^{\rho}
_{\beta\mu} + \partial_{\beta}\, {}^4\Gamma^{\alpha}_{\mu\nu}\, -
\partial_{\nu}\, {}^4\Gamma^{\alpha}_{\beta\mu},\nonumber \\
&&{}\nonumber \\
{}^4R_{\alpha\mu\beta\nu}&=&{}^4g_{\alpha\gamma}\, {}^4R^{\gamma}
{}_{\mu\beta\nu}=\nonumber \\
&=&{1\over 2}(\partial_{\alpha}\partial_{\nu}\, {}^4g_{\mu\beta} +\partial
_{\mu}\partial_{\beta} \,{}^4g_{\alpha\nu}-\partial_{\mu}\partial_{\nu}\, 
{}^4g_{\alpha\beta} -\partial_{\alpha}\partial_{\beta}\, {}^4g_{\mu\nu})+
\nonumber \\
&+&{}^4g_{\rho\sigma} ({}^4\Gamma^{\rho}_{\alpha\nu}\, {}^4\Gamma^{\sigma}
_{\mu\beta} - {}^4\Gamma^{\rho}_{\alpha\beta}\, {}^4\Gamma^{\sigma}
_{\mu\nu})
=-{}^4R_{\alpha\mu\nu\beta} =- {}^4R_{\mu\alpha\beta\nu} = {}^4R_{\beta\nu
\alpha\mu},
\label{a3}
\end{eqnarray}

\noindent while the Ricci tensor and the curvature scalar are defined as

\begin{eqnarray}
{}^4R_{\mu\nu} &=& {}^4R_{\nu\mu} ={}^4R^{\beta}{}_{\mu\beta\nu}=\nonumber \\
&=&\partial_{\rho} \, {}^4\Gamma^{\rho}_{\mu\nu}-\partial_{\nu}\, {}^4\Gamma
^{\rho}_{\rho\mu}+{}^4\Gamma^{\rho}_{\mu\nu}\, {}^4\Gamma^{\beta}_{\beta\rho}-
{}^4\Gamma^{\rho}_{\mu\beta}\, {}^4\Gamma^{\beta}_{\nu\rho}=\nonumber \\
&=&{1\over {\sqrt{{}^4g}}} \partial_{\rho} (\sqrt{{}^4g}\, {}^4\Gamma^{\rho}
_{\mu\nu})-\partial_{\mu}\partial_{\nu}\, ln\, \sqrt{{}^4g} -{}^4\Gamma^{\rho}
_{\mu\beta}\, {}^4\Gamma^{\beta}_{\nu\rho},\nonumber \\
{}&&\nonumber \\
{}^4R &=& {}^4g^{\mu\nu}\, {}^4R_{\mu\nu}= {}^4R^{\mu\nu}{}_{\mu\nu}=
\nonumber \\
&=&{}^4g^{\mu\nu}\, {}^4g^{\alpha\beta} [\partial_{\alpha}\partial_{\nu}\,
{}^4g_{\mu\beta}-\partial_{\mu}\partial_{\nu}\, {}^4g_{\alpha\beta}+{}^4g
_{\rho\sigma}({}^4\Gamma^{\rho}_{\alpha\nu}\, {}^4\Gamma^{\sigma}_{\mu\beta}-
{}^4\Gamma^{\rho}_{\alpha\beta}\, {}^4\Gamma^{\sigma}_{\mu\nu})]=
\nonumber \\
&=&{}^4g^{\mu\nu} ({}^4\Gamma^{\rho}_{\mu\beta}\, {}^4\Gamma^{\beta}_{\nu\rho}-
{}^4\Gamma^{\rho}_{\mu\nu}\, {}^4\Gamma^{\beta}_{\beta\rho})+
{1\over {\sqrt{{}^4g}}} \partial_{\rho} [\sqrt{{}^4g} ({}^4g^{\mu\nu}\,
{}^4\Gamma^{\rho}_{\mu\nu}-{}^4g^{\mu\rho}\, {}^4\Gamma^{\nu}_{\nu\mu})].
\label{a4}
\end{eqnarray}

The first and second Bianchi identities have the following expression 
[${}^4G_{\mu\nu}$ is the Einstein tensor and ${}^4\nabla_{\mu}\, {}^4G^{\mu\nu}
\equiv 0$ are the Bianchi identities]

\begin{eqnarray}
&&{}^4R^{\alpha}{}_{\mu\beta\nu}+{}^4R^{\alpha}{}_{\beta\nu\mu}+{}^4
R^{\alpha}{}_{\nu\mu\beta}\equiv 0,\nonumber \\
&&{}\nonumber \\
&&({}^4\nabla_{\gamma}\, {}^4R)^{\alpha}{}_{\mu\beta\nu}+ ({}^4\nabla_{\beta}\,
{}^4R)^{\alpha}{}_{\mu\nu\gamma}+({}^4\nabla_{\nu}\, {}^4R)^{\alpha}
{}_{\mu\gamma\beta}
\equiv 0,\nonumber \\
&&{}\nonumber \\
&&\Rightarrow \, ({}^4\nabla_{\gamma}\, {}^4R^{(ricci)})_{\mu\nu}+({}^4\nabla
_{\alpha}\, {}^4R)^{\alpha}{}_{\mu\nu\gamma}-({}^4\nabla_{\nu}\, 
{}^4R^{(ricci)})_{\mu\gamma}
\equiv 0, \nonumber \\
&&\Rightarrow \, {}^4\nabla_{\mu}\, {}^4G^{\mu\nu}\equiv 0,\quad\quad
{}^4G_{\mu\nu}={}^4R_{\mu\nu} -{1\over 2} {}^4g_{\mu\nu}\, {}^4R,\quad
{}^4G=-{}^4R.
\label{a5}
\end{eqnarray}

\noindent There are 20 independent components of the Riemann tensor in four
dimensions due to its symmetry properties.

The Weyl or conformal tensor (which vanish if and only if $M^4$ is conformally 
flat) is defined as the completely trace-free part of the Riemann tensor
[in empty spacetime Einstein's equations imply ${}^4C_{\alpha\mu\beta\nu}\,
{\buildrel \circ \over =}\, {}^4R_{\alpha\mu\beta\nu}$, where ${\buildrel \circ
\over =}$ means evaluated on the solution of the equations of motion]

\begin{eqnarray}
{}^4C_{\alpha\mu\beta\nu}&=&{}^4R_{\alpha\mu\beta\nu} +{1\over 2}({}^4R
_{\alpha\beta}\, {}^4g_{\mu\nu} -{}^4R_{\mu\beta}\, {}^4g_{\alpha\nu}+
{}^4R_{\mu\nu}\, {}^4g_{\alpha\beta} - {}^4R_{\alpha\nu}\, {}^4g_{\mu\beta})+
\nonumber \\
&+&{1\over 6}({}^4g_{\alpha\beta}\, {}^4g_{\mu\nu} -{}^4g_{\alpha\nu}\,
{}^4g_{\mu\beta})\, {}^4R
\label{a6}
\end{eqnarray}

Let our globally hyperbolic spacetime $M^4$ be foliated with spacelike Cauchy
hypersurfaces $\Sigma_{\tau}$, obtained with the embeddings $i_{\tau}:\Sigma
\rightarrow \Sigma_{\tau} \subset M^4$, $\vec \sigma \mapsto 
x^{\mu}=z^{\mu}(\tau ,\vec
\sigma )$, of a 3-manifold $\Sigma$ in $M^4$ [$\tau :M^4\rightarrow R$ is a
global, timelike, future-oriented function labelling the leaves of the
foliation; $x^{\mu}$ are local coordinates in a chart of $M^4$;
$\vec \sigma =\{ \sigma^r \}$, r=1,2,3, are local coordinates in a 
chart of $\Sigma$, which is diffeomorphic to $R^3$; we shall use the notation
$\sigma^A=(\sigma^{\tau}=\tau ;\vec \sigma )$, $A=\tau ,r$, and $z^{\mu}(\sigma 
)=z^{\mu}(\tau ,\vec \sigma )$]. Let $n^{\mu}(\sigma )$ and $l^{\mu}(\sigma )=
N(\sigma ) n^{\mu}(\sigma )$ be the controvariant timelike normal and unit 
normal [${}^4g_{\mu\nu}(z(\sigma ))l^{\mu}(\sigma ) l^{\nu}(\sigma )=\epsilon$]
to $\Sigma_{\tau}$ at the point $z(\sigma )\in \Sigma_{\tau}$. 
The positive function $N(\sigma ) > 0$ is 
the lapse function: $N(\sigma )d\tau$ measures the proper time interval at 
$z(\sigma )\in \Sigma_{\tau}$ between $\Sigma_{\tau}$ and $\Sigma_{\tau +d\tau}$
. The shift functions $N^r(\sigma )$ are defined so that $N^r(\sigma )d\tau$ 
describes the horizontal shift on $\Sigma_{\tau}$ such that, if $z^{\mu}(\tau 
+d\tau ,\vec \sigma +d\vec \sigma )\in \Sigma_{\tau +d\tau}$, then $z^{\mu}(\tau
+d\tau ,\vec \sigma +d\vec \sigma )\approx z^{\mu}(\tau ,\vec \sigma )+N(\tau ,
\vec \sigma )d\tau l^{\mu}(\tau ,\vec \sigma )+[d\sigma^r+N^r(\tau ,\vec 
\sigma )d\tau ]{{\partial z^{\mu}(\tau ,\vec \sigma )}\over {\partial \sigma^r}
}$; therefore, we have ${{\partial z^{\mu}(\sigma )}\over {\partial \tau}}=
N(\sigma ) l^{\mu}(\sigma )+N^r(\sigma ) {{\partial z^{\mu}(\tau ,\vec 
\sigma )}\over {\partial \sigma^r}}$ for the so called evolution vector. For the
covariant unit normal to $\Sigma_{\tau}$ we have $l_{\mu}(\sigma )={}^4g
_{\mu\nu}(z(\sigma )) l^{\nu}(\sigma )=N(\sigma ) \partial_{\mu}\tau{|}
_{x=z(\sigma )}$.

Instead of local coordinates $x^{\mu}$ for $M^4$,
we use local coordinates $\sigma^A$ on 
$R\times \Sigma \approx M^4$ [$x^{\mu}=z^{\mu}(\sigma )$ with inverse $\sigma^A=
\sigma^A(x)$], i.e. a $\Sigma_{\tau}$-adapted holonomic coordinate basis for 
vector fields $\partial_A={{\partial}\over {\partial \sigma^A}}\in 
T(R\times \Sigma ) \mapsto b^{\mu}_A(\sigma ) \partial_{\mu} ={{\partial z
^{\mu}(\sigma )}\over {\partial \sigma^A}} \partial_{\mu} \in TM^4$, and for 
differential one-forms $dx^{\mu}\in T^{*}M^4 \mapsto d\sigma^A=b^A
_{\mu}(\sigma )dx^{\mu}={{\partial \sigma^A(z)}\over {\partial z^{\mu}}} dx
^{\mu} \in T^{*}(R\times \Sigma )$. Let us note that in the flat Minkowski 
spacetime the transformation coefficients $b^A_{\mu}(\sigma )$ and $b^{\mu}
_A(\sigma )$ become the flat orthonormal cotetrads $z^A_{\mu}(\sigma )=
{{\partial \sigma^A(x)}\over {\partial x^{\mu}}}{|}_{x=z(\sigma )}$ and 
tetrads $z^{\mu}_A(\sigma )={{\partial z^{\mu}(\sigma )}\over {\partial \sigma
^A}}$ of Ref.\cite{lus1}. The induced 4-metric and inverse 4-metric 
become in the new basis

\begin{eqnarray}
{}^4g(x) &=&  {}^4g_{\mu\nu}(x) dx^{\mu} \otimes dx^{\nu} =
{}^4g_{AB}(z(\sigma ))d\sigma^A \otimes d\sigma^B,\nonumber \\
&&{}\nonumber \\
{}^4g_{\mu\nu}&=&b^A_{\mu}\, {}^4g_{AB} b^B_{\nu} =\nonumber \\
&=&\epsilon \, (N^2-{}^3g_{rs}N^rN^s)\partial_{\mu}\tau \partial_{\nu}\tau -
\epsilon \, {}^3g_{rs}N^s(\partial_{\mu}\tau \partial_{\nu}\sigma^r+
\partial_{\nu}\tau \partial_{\mu}\sigma^r)-\epsilon \, {}^3g_{rs}
\partial_{\mu}\sigma^r \partial_{\nu}\sigma^s=\nonumber \\
&=& \epsilon \, l_{\mu} l_{\nu} -\epsilon \, {}^3g_{rs} (\partial_{\mu}
\sigma^r +N^r\, \partial_{\mu}\tau ) (\partial_{\nu}\sigma^s+N^s\,
\partial_{\nu}\tau ),\nonumber \\
&&{}\nonumber \\
&\Rightarrow&{}^4g_{AB}=\lbrace {}^4g_{\tau\tau}=
\epsilon (N^2-{}^3g_{rs}N^rN^s); {}^4g_{\tau r}=-
\epsilon \, {}^3g_{rs}N^s; {}^4g_{rs}=-\epsilon \, {}^3g_{rs}\rbrace =
\nonumber \\
&=&\epsilon [ l_Al_B-{}^3g_{rs}(\delta^r_A+N^r\delta^{\tau}_A)(\delta^s_B+
N^s\delta^{\tau}_B)], \nonumber \\
&&{}\nonumber \\
{}^4g^{\mu\nu}&=& b^{\mu}_A {}^4g^{AB} b^{\nu}_B=\nonumber \\
&=&{{\epsilon}\over {N^2}} \partial_{\tau}z^{\mu}\partial_{\tau}z^{\nu}-
{{\epsilon \, N^r}\over {N^2}} (\partial_{\tau}z^{\mu}\partial_rz^{\nu}+
\partial_{\tau}z^{\nu}\partial_rz^{\mu}) -\epsilon ({}^3g^{rs}-{{N^rN^s}\over 
{N^2}})\partial_rz^{\mu}\partial_sz^{\nu}=\nonumber \\
&=& \epsilon [\, l^{\mu} l^{\nu} - \, {}^3g^{rs} \partial_rz^{\mu} 
\partial_sz^{\nu}],\nonumber \\
&&{}\nonumber \\
&\Rightarrow&{}^4g^{AB}=\lbrace {}^4g^{\tau\tau}=
{{\epsilon}\over {N^2}}; {}^4g^{\tau r}=-{{\epsilon \, N^r}
\over {N^2}}; {}^4g^{rs}=-\epsilon ({}^3g^{rs} - {{N^rN^s}\over {N^2}})
\rbrace =\nonumber \\
&=&\epsilon [l^Al^B -{}^3g^{rs}\delta^A_r\delta^B_s],\nonumber \\
&&{}\nonumber \\
l^A&=&l^{\mu} b^A_{\mu} = N \, {}^4g^{A\tau}={{\epsilon}\over N} (1; -N^r),
\nonumber \\
l_A&=&l_{\mu} b_A^{\mu} = N \partial_A \tau =N \delta^{\tau}_A = (N; \vec 0).
\label{I1}
\end{eqnarray}

Here, we introduced the 3-metric  of $\Sigma_{\tau}$:
$\, {}^3g_{rs}=-\epsilon \, {}^4g_{rs}$ with signature (+++). 
If ${}^4\gamma^{rs}$ is the inverse
of the spatial part of the 4-metric [${}^4\gamma^{ru}\, {}^4g_{us}=\delta^r_s$],
the inverse of the 3-metric is ${}^3g^{rs}=-\epsilon \, {}^4\gamma^{rs}$
[${}^3g^{ru}\, {}^3g_{us}=\delta^r_s$]. ${}^3g_{rs}(\tau ,\vec \sigma )$ 
are the components of
the ``first fundamental form" of the Riemann 3-manifold $(\Sigma_{\tau},{}^3g)$
and we have\hfill\break
\hfill\break
 $ds^2={}^4g_{\mu\nu} dx^{\mu} dx^{\nu}=
\epsilon (N^2-{}^3g_{rs}N^rN^s) (d\tau )^2-2\epsilon
\, {}^3g_{rs}N^s d\tau d\sigma^r -\epsilon \, {}^3g_{rs} d\sigma^rd\sigma^s=
\epsilon \Big[ N^2(d\tau )^2 -{}^3g_{rs}(d\sigma^r+N^rd\tau )(d\sigma^s+
N^sd\tau )\Big]$\hfill\break
\hfill\break
for the line element in $M^4$. We must have $\epsilon \, {}^4g_{oo} >0$,
$\epsilon \, {}^4g_{ij} < 0$, $\left| \begin{array}{cc} {}^4g_{ii}& {}^4g_{ij}
\\ {}^4g_{ji}& {}^4g_{jj} \end{array} \right| > 0$, $\epsilon \, 
det\, {}^4g_{ij} > 0$.

If we define $g={}^4g=|\, det\, ({}^4g_{\mu\nu})\, |$ and $\gamma ={}^3g =|\, 
det\, ({}^3g_{rs})\, |$, we also have

\begin{eqnarray}
&&N=\sqrt{ {{}^4g\over {{}^3g}} }={1\over { \sqrt{{}^4g^{\tau\tau}} }} =
\sqrt{{g\over {\gamma}}}=
\sqrt{{}^4g_{\tau\tau}-\epsilon \, {}^3g^{rs}\, {}^4g_{\tau r}
{}^4g_{\tau s} },\nonumber \\
&&N^r=-\epsilon \, {}^3g^{rs}\, {}^4g_{\tau s} =
-{{{}^4g^{\tau r}}\over {{}^4g^{\tau\tau}}}
,\quad N_r={}^3g_{rs}N^s=-\epsilon \,\, {}^4g_{rs}N^s=-\epsilon {}^4g_{\tau r}.
\label{I2}
\end{eqnarray}

Let us remark [see Ref.\cite{mol}] that in the study of space and time
measurements the equation $ds^2=0$ [use of light signals for the synchronization
of clocks] and the definition $d\bar \tau =\sqrt{\epsilon \, {}^4g_{oo}} dx^o$
of proper time [$\sqrt{\epsilon \, {}^4g_{oo}}$ determines the ratio between
the rates of a standard clock at rest and a coordinate clock at the same point]
imply the use in $M^4$ of a 3-metric ${}^3{\tilde \gamma}_{rs}={}^4g_{rs}-
{{{}^4g_{or}\, {}^4g_{os}}\over {{}^4g_{oo}}}=-\epsilon ({}^3g_{rs}+{{N_rN_s}
\over {\epsilon \, {}^4g_{oo}}} )$ with the covariant shift functions $N_r=
{}^3g_{rs}N^s=-\epsilon \, {}^4g_{or}$, which are connected with the
conventionality of simultaneity \cite{simul} and with the direction dependence 
of the velocity of light [$c(\vec n)=\sqrt{\epsilon \, {}^4g_{oo}} / 
(1+N_rn^r)$ in direction $\vec n$].

See Refs.\cite{in,mtw,ish} for the 3+1 decomposition of 4-tensors on $M^4$. The
horizontal projector ${}^3h^{\nu}_{\mu}=\delta^{\nu}_{\mu}-\epsilon \, l_{\mu}
l^{\nu}$ on $\Sigma_{\tau}$ defines the 3-tensor fields on $\Sigma_{\tau}$
starting from the 4-tensor fields on $M^4$. 

In the standard (not Hamiltonian) description of the 3+1 decomposition we
utilize a 
$\Sigma_{\tau}$-adapted nonholonomic noncoordinate basis [$\bar A=(l;r)$]

\begin{eqnarray}
{\hat b}^{\mu}_{\bar A}(\sigma ) &=&\lbrace {\hat b}^{\mu}_l(\sigma )= 
\epsilon l^{\mu}
(\sigma ) =N^{-1}(\sigma ) [b^{\mu}_{\tau}(\sigma )- N^r(\sigma )b^{\mu}_r
(\sigma )];\nonumber \\
&&{\hat b}^{\mu}_r(\sigma ) = b^{\mu}_r(\sigma ) \rbrace ,\nonumber \\
{\hat b}^{\bar A}_{\mu}(\sigma ) &=& \lbrace {\hat b}^l_{\mu}(\sigma ) =
l_{\mu}(\sigma )= N(\sigma )b^{\tau}_{\mu}(\sigma )=N(\sigma )\partial_{\mu}
\tau (z(\sigma ));\nonumber \\
&&{\hat b}^r_{\mu}(\sigma ) = b^r_{\mu}(\sigma )+ N^r(\sigma ) 
b^{\tau}_{\mu}(\sigma ) \rbrace ,\nonumber \\
&&{}\nonumber \\
&&{\hat b}_{\mu}^{\bar A}(\sigma ) {\hat b}^{\nu}_{\bar A}(\sigma )=
\delta^{\nu}_{\mu},\quad {\hat b}^{\bar A}_{\mu}(\sigma ) {\hat b}^{\mu}
_{\bar B}(\sigma )=\delta^{\bar A}_{\bar B}, \nonumber \\
{}^4{\bar g}_{\bar A\bar B}(z(\sigma ))&=&{\hat b}^{\mu}_{\bar A}(\sigma )
{}^4g_{\mu\nu}(z(\sigma )) {\hat b}^{\nu}_{\bar B}(\sigma )=\nonumber \\
&&=\lbrace {}^4{\bar g}_{ll}(\sigma )=\epsilon ; {}^4{\bar g}_{lr}(\sigma )=0; 
{}^4{\bar g}_{rs}(\sigma )=
{}^4g_{rs}(\sigma )=-\epsilon \, {}^3g_{rs}\rbrace ,\nonumber \\
{}^4{\bar g}^{\bar A\bar B}&=&\lbrace {}^4{\bar g}^{ll}=\epsilon ; {}^4{\bar g}
^{lr}=0; {}^4{\bar g}^{rs}={}^4\gamma^{rs}=-\epsilon {}^3g^{rs}\rbrace ,
\nonumber \\
&&{}\nonumber \\
&&X_{\bar A}={\hat b}^{\mu}_{\bar A}\partial_{\mu}=\{ X_l={1\over
N}(\partial_{\tau}- N^r\partial_r);\partial_r\},\nonumber \\
&&\theta^{\bar A}={\hat b}^{\bar A}_{\mu}dx^{\mu}=\{ \theta^l=Nd\tau ;
\theta^r=d\sigma^r+N^rd\tau \} ,\nonumber \\
&&{}\nonumber \\
&&\Rightarrow l_{\mu}(\sigma )b^{\mu}_r(\sigma )=0,\quad l^{\mu}(\sigma )b^r
_{\mu}(\sigma )=-N^r(\sigma )/N(\sigma ),\nonumber \\
&&{}\nonumber \\
l^{\bar A}&=& l^{\mu} {\hat b}_{\mu}^{\bar A} = (\epsilon ; l^r+N^rl^{\tau})=
(\epsilon ; \vec 0),\nonumber \\
l_{\bar A}&=& l_{\mu} {\hat b}^{\mu}_{\bar A} = (1; l_r) = (1; \vec 0).
\label{I3}
\end{eqnarray}

\noindent We have ${}^3h_{\mu\nu}={}^4g_{\mu\nu}-\epsilon l_{\mu}l_{\nu}=-
\epsilon \, {}^3g_{rs}(b^r_{\mu}+N^rb^{\tau}_{\mu})(b^s_{\mu}+N^sb^{\tau}
_{\mu})=-\epsilon \, {}^3g_{rs}{\hat b}^r_{\mu}{\hat b}^s_{\nu}$ and
for a 4-vector ${}^4V^{\mu}={}^4V^{\bar A}{\hat b}^{\mu}_{\bar A}=
{}^4V^l l^{\mu}+{}^4V^r{\hat b}^{\mu}_r$ we have ${}^3V^{\mu}={}^3V^r{\hat b}
^{\mu}_r={}^3h^{\mu}_{\nu}\, {}^4V^{\nu}$, ${}^3V^r={\hat b}^r_{\mu}\, {}^3V
^{\mu}$.

The nonholonomic basis in $\Sigma_{\tau}$-adapted coordinates is\hfill\break
\hfill\break
${\hat b}_A^{\bar A}={\hat b}^{\bar A}_{\mu}b^{\mu}_A= \{ {\hat b}^l_A=l_A;\,
{\hat b}^r_A=\delta^r_A+N^r \delta^{\tau}_A \}$\hfill\break
${\hat b}^A_{\bar A}={\hat b}^{\mu}_{\bar A}b^A_{\mu}= \{ {\hat b}^A_l=\epsilon
l^A;\, {\hat b}^A_r=\delta^A_r \}$.\hfill\break
\hfill\break
One can show the following results concerning the Lie derivative along the unit
normal: i) ${\cal L}
_l\, {\hat b}^{\mu}_r=-{\cal L}_l\, l^{\mu}=N^{-1} (\partial_rN l^{\mu}+
\partial_rN^s {\hat b}^{\mu}_s)= l^{\nu} {\hat b}^{\mu}_{r;\nu}-{\hat b}^{\nu}_r
l^{\mu}{}_{;\nu}$; ii) ${\cal L}_l {\hat b}^r_{\mu}=-n^{-1} \partial_sN^r 
{\hat b}^s_{\mu}$.

The 3-dimensional covariant derivative [denoted ${}^3\nabla$ or with the
subscript ``$|$"] of a 3-dimensional tensor ${}^3T^{\mu_1..
\mu_p}{}_{\nu_1..\nu_q}$ of rank (p,q) is the  3-dimensional tensor
of rank (p,q+1)
${}^3\nabla_{\rho}\, {}^3T^{\mu_1..\mu_p}{}_{\nu_1..\nu_q}={}^3T^{\mu_1..\mu_p}
{}_{\nu_1..\nu_q | \rho}=
{}^3h^{\mu_1}_{\alpha_1}\cdots {}^3h^{\mu_p}_{\alpha_p}\, {}^3h^{\beta_1}
_{\nu_1}\cdots {}^3h^{\beta_q}_{\nu_q}\, {}^3h^{\sigma}_{\rho}\, {}^4\nabla
_{\sigma}\, {}^3T^{\alpha_1..\alpha_p}{}_{\beta_1..\beta_q}$. For (1,0) and 
(0,1) tensors we have:
${}^3\nabla_{\rho} \, {}^3V^{\mu}={}^3V^{\mu}{}_{| \rho}= {}^3V^r
{}_{| s}\, {\hat b}^{\mu}_r {\hat b}^s_{\rho}$ ,
${}^3\nabla_s\, {}^3V^r={}^3V^r{}_{| s} =\partial_s\, {}^3V^r +
{}^3\Gamma^r_{su}\, {}^3V^u$ and
${}^3\nabla_{\rho}\, {}^3\omega_{\mu}={}^3\omega_{\mu | \rho}={}^3\omega
_{r | s}\, {\hat b}^r_{\mu} {\hat b}^s_{\rho}$,
${}^3\nabla_s\, {}^3\omega_r={}^3\omega_{r | s}=\partial_s\, {}^3\omega_r -
{}^3\Gamma^u_{rs} {}^3\omega_u$ respectively.

The 3-dimensional Christoffel symbols are
${}^3\Gamma^u_{rs}={\hat b}^u_{\mu}\, [{}^3
\nabla_{\rho}\, {\hat b}^{\mu}_r] {\hat b}^{\rho}_s=
{\hat b}^u_{\mu} {\hat b}^{\mu}_{r | \rho} {\hat b}^{\rho}
_s={1\over 2} {}^3g^{uv} (\partial_s\, {}^3g_{vr} +
\partial_r\, {}^3g_{vs} -\partial_v\, {}^3g_{rs})$ and
the metric compatibility [Levi-Civita connection on the
Riemann 3-manifold $(\Sigma_{\tau},{}^3g)$] is
${}^3\nabla_{\rho}\, {}^3g_{\mu\nu} ={}^3g_{\mu\nu | \rho}=0$
[${}^3g_{\mu\nu}=-\epsilon 
\, {}^3h_{\mu\nu}= {}^3g_{rs}{\hat b}^r_{\mu}{\hat b}^s_{\nu}$, 
so that ${}^3{\bar g}_{\bar A\bar B}=\{ {}^3{\bar g}_{ll}
=0; {}^3{\bar g}_{lr}=0; {}^3{\bar g}_{rs}=-\epsilon \, {}^3g_{rs} \}$].
It is then possible to 
define parallel transport on $\Sigma_{\tau}$. The 3-dimensional curvature
Riemann tensor is

\begin{eqnarray}
&&{}^3R^{\mu}{}_{\alpha\nu\beta}\, {}^3V^{\alpha}= {}^3V^{\alpha}{}_{| \beta
| \nu} - {}^3V^{\alpha}{}_{| \nu | \beta},\nonumber \\
&&\Rightarrow {}^3R^r{}_{suv}=\partial_u
\, {}^3\Gamma^r_{sv} -\partial_v\, {}^3\Gamma
^r_{su} + {}^3\Gamma^r_{uw}\,
{}^3\Gamma^w_{sv} - {}^3\Gamma^r_{vw}\, {}^3\Gamma^w_{su}.
\label{I5}
\end{eqnarray}

For 3-manifolds, the Riemann tensor has only 6 independent components
since the Weyl tensor vanishes: this gives the relation
${}^3R_{\alpha\mu\beta\nu}={1\over 2}({}^3R_{\mu\beta}\, {}^3g_{\alpha\nu}+
{}^3R_{\alpha\nu}\, {}^3g_{\mu\beta}-{}^3R_{\alpha\beta}\, {}^3g_{\mu\nu}-
{}^3R_{\mu\nu}\, {}^3g_{\alpha\beta})
-{1\over 6} ({}^3g_{\alpha\beta}\, {}^3g_{\mu\nu}-
{}^3g_{\alpha\nu}\, {}^3g_{\beta\mu})\, {}^3R$, which expresses the
Riemann tensor in terms of  the Ricci tensor. A 3-manifold $M^3$ is
conformally flat if and only if its Weyl-Schouten tensor \hfill\break
\hfill\break
${}^3C_{\lambda\mu\nu}={}^3\nabla_{\nu}\, {}^3R_{\lambda\mu} - {}^3\nabla_{\mu}
\, {}^3R_{\lambda\nu} - {1\over 4} ({}^3g_{\lambda\mu}\, \partial_{\nu}\,
{}^3R -{}^3g_{\lambda\nu}\, \partial_{\mu}\, {}^3R)$\hfill\break
\hfill\break
vanishes\cite{naka}. Equivalently one uses the
Cotton-York tensor \hfill\break
\hfill\break
${}^3{\cal H}_{\mu\nu}={1\over 2}\gamma^{1/3}\,\,
{}^3\epsilon_{\alpha\beta\mu} 
{}^3\nabla^{\alpha}\, {}^3R^{\beta}{}_{\nu}+{}^3\epsilon_{\alpha\beta\nu}
{}^3\nabla^{\alpha}\, {}^3R^{\beta}{}_{\mu}$,\hfill\break
\hfill\break
 which satisfies ${}^3g^{\mu\nu}\,
{}^3{\cal H}_{\mu\nu}= {}^3\nabla^{\mu}\, {}^3{\cal H}_{\mu\nu}=0$ 
\cite{mtw}.

The components of the
``second fundamental form" of $(\Sigma_{\tau},{}^3g)$ is the extrinsic 
curvature \hfill\break
\hfill\break
${}^3K_{\mu\nu}={}^3K_{\nu\mu}=-{1\over 2}{\cal L}_l\, {}^3g_{\mu\nu};$
\hfill\break
\hfill\break
one has ${}^4\nabla_{\rho} \, l^{\mu}=\epsilon \, {}^3a^{\mu} l_{\rho}-
{}^3K_{\rho}{}^{\mu}$, with the acceleration ${}^3a^{\mu}={}^3a^r {\hat b}
^{\mu}_r$ of the observers travelling along the congruence of timelike curves 
with tangent vector $l^{\mu}$ given by ${}^3a_r=\partial_r\, ln\, N$. On 
$\Sigma_{\tau}$ we have\hfill\break
\hfill\break
${}^3K_{rs}={}^3K_{sr}={1\over {2N}}(N_{r|s}+N_{s|r}-{{\partial \, {}^3g_{rs}}
\over {\partial \tau}})$.\hfill\break
\hfill\break
 Moreover, one has: \hfill\break
i) ${\hat b}^{\mu}_{r;\nu}=\epsilon \, {}^3a_rl^{\mu}l_{\nu}- {}^3K_{rs} 
l^{\mu}{\hat b}^s_{\nu}+\epsilon ({}^3K_r{}^s-N^{-1}\partial_rN^s)+ {}^3\Gamma
^u_{rs}{\hat b}^{\mu}_u{\hat b}^s_{\nu}$;\hfill\break
ii) ${\hat b}^r_{\mu ;\nu}=-\, {}^3a^r l_{\mu}l_{\nu}+\epsilon \, {}^3K_s{}^r 
{\hat b}^s_{\nu} l_{\mu}-\epsilon ({}^3K_s{}^r -N^{-1} \partial_sN^r) {\hat b}
^s_{\mu} l_{\nu}-\, {}^3\Gamma^r_{su} {\hat b}^s_{\mu}{\hat b}^u_{\nu}$;
\hfill\break
iii) ${}^3a_{\mu |\nu}={}^3a_{\nu |\mu}={}^3a_{r|s} {\hat b}^r_{\mu}{\hat b}^s
_{\nu} = [\partial_r\partial_sln\, N-\, {}^3\Gamma^u_{rs}\partial_uln\, N] {\hat
b}^r_{\mu}{\hat b}^s_{\nu}$;\hfill\break
iv) ${}^3a^{\mu}{}_{;\mu}={}^3a^{\mu}{}_{|\mu}+{}^3a^{\mu}\, {}^3a_{\mu}$;
\hfill\break
v) $l^{\mu}\, {}^3K_{;\mu}=-(l^{\mu}\, {}^3K)_{;\mu}+{}^3K^2$;\hfill\break
vi) ${\cal L}_l\, {}^3g^{\mu\nu}=l^{\mu}\, {}^3a^{\nu}+l^{\nu}\, {}^3a^{\mu}+2
\, {}^3K^{\mu\nu}$.\hfill\break

The information contained in the 20 independent components ${}^4R^{\mu}
{}_{\nu\alpha\beta}$ of the curvature Riemann tensor of $M^4$ is given by the 
following three projections [see Ref.\cite{carter} for the geometry of
embeddings; one has ${}^4{\bar R}^r{}_{suv}={}^3{\bar R}^r{}_{suv}$]

\begin{eqnarray}
{}^3h^{\mu}_{\rho}\, &{}^3&h^{\sigma}_{\nu}\, {}^3h^{\gamma}_{\alpha}\, {}^3h
^{\delta}_{\beta}\, {}^4R^{\rho}{}_{\sigma\gamma\delta}=
{}^4{\bar R}^r{}_{suv}{\hat b}
^{\mu}_r{\hat b}^s_{\nu}{\hat b}^u_{\alpha}{\hat b}^v_{\beta}=
{}^3R^{\mu}{}_{\nu
\alpha\beta}+{}^3K_{\alpha}{}^{\mu}\, {}^3K_{\beta\nu}-{}^3K_{\beta}{}^{\mu}
\, {}^3K_{\alpha\nu},\nonumber \\
&&GAUSS\, EQUATION,\nonumber \\
\epsilon &l_{\rho}&\, {}^3h^{\sigma}_{\nu}\, {}^3h^{\gamma}_{\alpha}\, {}^3h
^{\delta}_{\beta}\, {}^4R^{\rho}{}_{\sigma\gamma\delta}={}^4{\bar R}^l{}_{suv}
{\hat b}^s_{\nu}{\hat b}^u_{\alpha}{\hat b}^v_{\beta}=
{}^3K_{\alpha\nu | 
\beta} - {}^3K_{\beta\nu | \alpha},\nonumber \\
&&CODAZZI-MAINARDI\, EQUATION,\nonumber \\
{}^4&R&_{\mu\sigma\gamma\delta}\, l^{\sigma}\, l^{\gamma}\, {}^3h^{\delta}
_{\nu}={}^4{\bar R}_{\mu llu}{\hat b}^u_{\nu}=
\epsilon ({\cal L}_l\, {}^3K_{\mu\nu}+{}^3K_{\mu}{}^{\rho}\, {}^3K
_{\rho\nu}+{}^3a_{\mu | \nu}+{}^3a_{\mu}\, {}^3a_{\nu}),\nonumber \\
&&RICCI\, EQUATION,\nonumber \\
&&{}\nonumber \\
&&with\quad\quad {\cal L}_l\, {}^3K_{\mu\nu}=l^{\alpha}\, {}^3K_{\mu\nu 
;\alpha}-2\, {}^3K_{\mu}{}^{\alpha}\, {}^3K_{\alpha\nu}+2\epsilon \, {}^3a
^{\alpha}\, {}^3K_{\alpha (\nu}\, l_{\mu )} .
\label{I6}
\end{eqnarray}

In the nonholonomic basis we have:\hfill\break
\hfill\break
 ${}^4R^u{}_{rst}={}^3R^u{}_{rst}+{}^3K_{rs}
\, {}^3K_t{}^u-{}^3K_{rt}\, {}^3K_s{}^u$, \hfill\break
${}^4R={}^3R+{}^3K_{rs}\, {}^3K^{rs}-({}^3K)^2$, \hfill\break
${}^4{\bar R}^l{}_{rst}={}^3K_{rt | s}-{}^3K_{rs | t}$, \hfill\break
${}^4{\bar R}^u{}_{lrl}={}^3a^u{}_{|r}-{}^3a^u\, {}^3a_r+
{\cal L}_l\, {}^3K_r{}^u-{}^3K_r{}^s\, {}^3K_s{}^u$, \hfill\break
${}^4{\bar R}^u{}_{lrs}=-{}^3g^{ut}\, {}^4{\bar R}^l_{trs}$, \hfill\break
${}^4{\bar R}^l{}_{rls}={}^3g_{ru}{}^4{\bar R}^u{}_{lsl}$.\hfill\break
\hfill\break
Then, we can express ${}^4R_{\mu\nu}=\epsilon {}^4{\bar R}_{ll}
l_{\mu}l_{\nu}+\epsilon \, {}^4{\bar R}_{lr}(l_{\mu}{\hat b}^r_{\nu}+l_{\nu}
{\hat b}^r_{\mu})+{}^4{\bar R}_{rs}{\hat b}^r_{\mu}{\hat b}^s_{\nu}$, 
${}^4R$ and the Einstein tensor ${}^4G_{\mu\nu}={}^4R_{\mu\nu}-{1\over 2}\, 
{}^4g_{\mu\nu}\, {}^4R=\epsilon {}^4{\bar G}_{ll}
l_{\mu}l_{\nu}+\epsilon \, {}^4{\bar G}_{lr}(l_{\mu}{\hat b}^r_{\nu}+l_{\nu}
{\hat b}^r_{\mu})+{}^4{\bar G}_{rs}{\hat b}^r_{\mu}{\hat b}^s_{\nu}$ in the
nonholonomic basis, with the result:\hfill\break
${}^4{\bar R}_{ll}={}^3K_{\mu\nu}\, {}^3K^{\mu\nu}-{}^3K^2+({}^3a^{\mu}-{}^3K 
l^{\mu})_{;\mu}$,\hfill\break
${}^4{\bar R}_{lr}=\epsilon ({}^3K_r{}^s-\delta_r^s\, {}^3K)_{|s}$,\hfill\break
${}^4{\bar R}_{rs}=-({\cal L}_l\, {}^3K_{rs}-{}^3R_{rs}-{}^3K\, {}^3K_{rs}+2\, 
{}^3K_r{}^u\, {}^3K_{us}+{}^3a_{r|s}+{}^3a_r\, {}^3a_s)$,\hfill\break
${}^4R=-\epsilon ({}^3R +{}^3K_{rs}\, {}^3K^{rs}-{}^3K^2)-2\epsilon ({}^3a
^{\mu}-{}^3K l^{\mu})_{;\mu}$,\hfill\break
${}^4{\bar G}_{ll}={1\over 2}({}^3R+{}^3K^2-{}^3K_{rs}\, {}^3K^{rs})$,
\hfill\break
${}^4{\bar G}_{lr}=\epsilon ({}^3K_r{}^s-\delta^s_r\, {}^3K)_{|s}$,\hfill\break
${}^4{\bar G}_{rs}=-{1\over {\sqrt{\gamma}}} {\cal L}_l[\sqrt{\gamma} ({}^3K
_{rs}-{}^3g_{rs}\, {}^3K)]+{}^3R_{rs}-{1\over 2}\, {}^3g_{rs}\, {}^3R+2({}^3K\,
{}^3K_{rs}-{}^3K_r{}^u\, {}^3K_{us})+{1\over 2}\, {}^3g_{rs}({}^3K^2-{}^3K_{uv}
\, {}^3K^{uv})+N_{|r|s} -{}^3g_{rs} N^{|u}{}_{|u}$.\hfill\break

The Bianchi identities ${}^4G^{\mu\nu}{}_{;\nu}\equiv 0$ imply the following
four contracted Bianchi identities [according to which only two of the six
equations ${}^4{\bar G}_{rs}\, {\buildrel \circ \over =}\, 0$ are independent]:
\hfill\break

${1\over N}\partial_{\tau}\, {}^4{\bar G}_{ll}-{{N^r}\over N}\partial_r\, 
{}^4{\bar G}_{ll}-{}^3K\, {}^4{\bar G}_{ll}+\partial_r\, {}^4{\bar G}_l{}^r+
(2\, {}^3a_r+{}^3\Gamma^s_{sr}){}^4{\bar G}_l{}^r-{}^3K_{rs}\, {}^4{\bar G}^{rs}
\equiv 0,$\hfill\break
${1\over N}\partial_{\tau}\, {}^4{\bar G}_l{}^r-{{N^s}\over N}\partial_s\, 
{}^4{\bar G}_l{}^r+{}^3a^r\, {}^4{\bar G}_{ll}-(2\, {}^3K^r{}_s+\delta^r_s\, 
{}^3K+{{\partial_sN^r}\over N}){}^4{\bar G}_l{}^s+\partial_s\, {}^4{\bar G}
^{rs}+({}^3a_s+{}^3\Gamma^u_{us}){}^4{\bar G}^{rs}\equiv 0.$\hfill\break

The vanishing of ${}^4{\bar G}_{ll}$, ${}^4{\bar G}_{lr}$, 
corresponds to the four secondary constraints (restrictions of Cauchy data) of 
the ADM Hamiltonian formalism (see Section V). The four contracted Bianchi
identities, ${}^4G^{\mu\nu}{}_{;\nu}\equiv 0$, imply \cite{wald} that, if the
restrictions of Cauchy data are satisfied initially and the spatial equations
${}^4G_{ij}\, {\buildrel \circ \over =}\, 0$ are satisfied everywhere, then 
the secondary constraints are satisfied also at later times
[see Ref.\cite{cho,wald} 
 for the initial value problem]. The four contracted Bianchi
identities plus the four secondary constraints imply that only two combinations 
of the Einstein equations contain the accelerations (second time derivatives)
of the two (non tensorial)
independent degrees of freedom of the gravitational field and that these
equations can be put in normal form [this was one of the motivations behind the 
discovery of the Shanmugadhasan canonical transformations \cite{sha}].

The ``intrinsic geometry" of $\Sigma_{\tau}$ is defined by the Riemannian metric
${}^3g_{rs}$ [it allows to evaluate the length of space curves], the Levi-Civita
affine connection, i.e. the Christoffel symbols ${}^3\Gamma^u_{rs}$, [for
the parallel transport of 3-dimensional tensors on $\Sigma_{\tau}$] and the
curvature Riemann tensor ${}^3R^r{}_{stu}$ [for the evaluation of the holonomy
and for the geodesic deviation equation].
The ``extrinsic geometry" of $\Sigma_{\tau}$ is defined by the lapse N and shift
$N^r$ fields [which describe the ``evolution" of $\Sigma_{\tau}$ in $M^4$]
and by the ``extrinsic curvature" ${}^3K_{rs}$ [it is needed to evaluate how
much a 3-dimensional vector goes outside $\Sigma_{\tau}$ under spacetime
parallel transport and to rebuild the spacetime curvature from the 3-dimensional
one].

Besides the local dual coordinate bases ${}^4e_{\mu}=\partial_{\mu}$ and 
$dx^{\mu}$ for $TM^4$ and $T^{*}M^4$ respectively, we can introduce 
special `noncoordinate' bases ${}^4{\hat E}_{(\alpha )}={}^4{\hat E}^{\mu}
_{(\alpha )}(x)\partial_{\mu}$ and its dual ${}^4{\hat \theta}^{(\alpha )}=
{}^4{\hat E}^{(\alpha )}_{\mu}(x)dx^{\mu}$ [$i_{{}^4{\hat E}_{(\alpha )}}\, 
{}^4{\hat \theta}^{(\beta )}={}^4E^{(\alpha )}_{\mu}\, {}^4E^{\mu}_{(\beta )}=
\delta^{(\beta )}_{(\alpha )}$ $\Rightarrow {}^4\eta_{(\alpha )(\beta )}=
{}^4E^{\mu}_{(\alpha )}\, {}^4g_{\mu\nu}\, {}^4E^{\nu}_{(\beta )}$; 
$(\alpha )=(0),(1),(2),(3)$  are numerical indices] with the 
``vierbeins or tetrads or (local) frames"  ${}^4{\hat E}^{\mu}
_{(\alpha )}(x)$, which are, 
for each point $x^{\mu}\in M^4$, the matrix elements
of matrices $\lbrace {}^4{\hat E}^{\mu}_{(\alpha )}\rbrace \in GL(4,R)$; the
set of one-forms ${}^4{\hat \theta}^{(\alpha )}$ (with ${}^4{\hat E}^{(\alpha )}
_{\mu}(x)$ being the dual ``cotetrads") is also called ``canonical" or 
``soldering" one-form or ``coframe". Since a ``frame" ${}^4{\hat E}$ at the
point $x^{\mu}\in M^4$ is a linear isomorphism\cite{blee} ${}^4{\hat E}:
R^4\rightarrow T_xM^4$, $\partial_{\alpha}\mapsto {}^4{\hat E}(\partial
_{\alpha})={}^4{\hat E}_{(\alpha )}$, a frame determines a basis ${}^4{\hat
E}_{(\alpha )}$ of $T_xM^4$ [the coframes ${}^4{\hat \theta}$ determine
a basis ${}^4{\hat \theta}^{(\alpha )}$ of $T^{*}_xM^4$] and we can define a
principal fiber bundle with structure group GL(4,R), $\pi : L(M^4)\rightarrow
M^4$ called the ``frame bundle" of $M^4$ [its fibers are the sets of all the
frames over the points $x^{\mu}\in M^4$; it is an affine bundle, i.e. there
is no (global when it exists) cross section playing the role of the
identity cross section of vector bundles]; if $\Lambda \in GL(4,R)$, then the
free right action of GL(4,R) on $L(M^4)$ is denoted $R_{\Lambda}({}^4{\hat
E})={}^4{\hat E}\circ \Lambda$, ${}^4{\hat E}_{(\alpha )}\mapsto {}^4{\hat E}
_{(\beta )}\, (\Lambda^{-1})^{(\beta )}{}_{(\alpha )}$. When $M^4$ is
``parallelizable" [i.e. $M^4$ admits four vector fields which are independent in
each point, so that the tangent bundle $T(M^4)$ is trivial, $T(M^4)=M^4\times
R^4$; this is not possible (no hair theorem) for any compact manifold
except a torus], as we shall assume, then $L(M^4)=M^4\times GL(4,R)$ is a
trivial principal bundle [i.e. it admits a global cross section $\sigma :
M^4\rightarrow L(M^4)$, $x^{\mu}\mapsto {}^4_{\sigma}{\hat E}_{(\alpha )}(x)$].
See Ref.\cite{blee} for the differential structure on $L(M^4)$. With the assumed
pseudo-Riemannian manifold $(M^4,\, {}^4g)$, 
we can use its metric ${}^4g_{\mu\nu}$ to 
define the ``orthonormal frame bundle" of $M^4$, $F(M^4)=M^4\times SO(3,1)$, 
with structure group SO(3,1), of the orthonormal frames (or noncoordinate 
basis or orthonormal tetrads) 
${}^4E_{(\alpha )}={}^4E^{\mu}_{(\alpha )}\partial_{\mu}$ of $TM^4$. The
orthonormal tetrads and their duals, the orthonormal cotetrads ${}^4E_{\mu}
^{(\alpha )}$ [${}^4\theta^{(\alpha )}={}^4E^{(\alpha )}_{\mu}dx^{\mu}$ are
the orthonormal coframes], satisfy the duality and orthonormality conditions

\begin{eqnarray}
&&{}^4E^{(\alpha )}_{\mu}\, {}^4E^{\mu}_{(\beta )}=\delta^{(\alpha )}_{(\beta )}
,\quad\quad
{}^4E^{(\alpha )}_{\mu}\, {}^4E^{\nu}_{(\alpha )}=\delta^{\nu}_{\mu},
\nonumber \\
&&{}^4E^{\mu}_{(\alpha )}\, {}^4g_{\mu\nu}\, {}^4E^{\nu}_{(\beta )}=
{}^4\eta_{(\alpha )(\beta )},\quad\quad
{}^4E_{\mu}^{(\alpha )}\, {}^4g^{\mu\nu}\, {}^4E^{(\beta )}_{\nu} =
{}^4\eta^{(\alpha )(\beta )}.
\label{a7}
\end{eqnarray}

Under a rotation $\Lambda \in SO(3,1)$ [$\Lambda \, {}^4\eta \, \Lambda^T=
{}^4\eta$] we have ${}^4E^{\mu}_{(\alpha )}\mapsto {}^4E^{\mu}_{(\beta )}
(\Lambda^{-1})^{(\beta )}{}_{(\alpha )}$, ${}^4E^{(\alpha )}_{\mu}\mapsto
\Lambda^{(\alpha )}{}_{(\beta )}\, {}^4E^{(\beta )}_{\mu}$. Therefore, while 
the indices $\alpha , \beta ...$ transform under general coordinate
transformations [the diffeomorphisms of $Diff\, M^4$], the indices $(\alpha ),
(\beta )...$ transform under Lorentz rotations. The 4-metric can be expressed
in terms of orthonormal cotetrads or local coframes in the noncoordinate basis

\begin{eqnarray}
&&{}^4g_{\mu\nu}={}^4E^{(\alpha )}_{\mu}\, {}^4\eta_{(\alpha )(\beta )}\, 
{}^4E^{(\beta )}_{\nu},\quad\quad {}^4g^{\mu\nu}={}^4E^{\mu}_{(\alpha )}\,
{}^4\eta^{(\alpha )(\beta )}\, {}^4E^{\nu}_{(\beta )},\nonumber \\
&&{}^4g={}^4g_{\mu\nu}\, dx^{\mu} \otimes dx^{\nu} = {}^4\eta_{(\alpha )
(\beta )}\, \theta^{(\alpha )} \otimes \theta^{(\beta )}.
\label{a8}
\end{eqnarray}

For each vector ${}^4V^{\mu}$ and covector ${}^4\omega_{\mu}$ we have the
decompositions ${}^4V^{\mu}={}^4V^{(\alpha )}\, {}^4E^{\mu}_{(\alpha )}$
[${}^4V^{(\alpha )}={}^4E^{(\alpha )}_{\mu}\, {}^4V^{\mu}$], ${}^4\omega
_{\mu}={}^4E^{(\alpha )}_{\mu}\, {}^4\omega_{(\alpha )}$ [${}^4\omega
_{(\alpha )}={}^4E^{\mu}_{(\alpha )}\, {}^4\omega_{\mu}$].

In a noncoordinate (nonholonomic) basis we have

\begin{eqnarray}
&&[ {}^4E_{(\alpha )}, {}^4E_{(\beta )} ] = c_{(\alpha )(\beta )}{}^{(\gamma )}
\, {}^4E_{(\gamma )},\nonumber \\
&&c_{(\alpha )(\beta )}{}^{(\gamma )} = {}^4E^{(\gamma )}_{\nu} ({}^4E^{\mu}
_{(\alpha )}\, \partial_{\mu}\, {}^4E^{\nu}_{(\beta )} - {}^4E^{\mu}_{(\beta )}
\partial_{\mu}\, {}^4E^{\nu}_{(\alpha )}).
\label{a9}
\end{eqnarray}

Physically, in a coordinate system (chart) $x^{\mu}$ of $M^4$, a tetrad may be 
considered as a collection of accelerated observers described by a congruence of
timelike curves with 4-velocity ${}^4E^{\mu}_{(o)}$; in each point $p\in 
M^4$ consider a coordinate transformation to local inertial coordinates at p,
i.e. $x^{\mu} \mapsto X^{(\mu)}_p(x)$: then we have, in p, ${}^4E^{\mu}_{(\alpha
)}(p)={{\partial x^{\mu}(X_p(p))}\over {\partial X^{(\alpha )}_p}}$ and
${}^4E^{(\alpha )}_{\mu}(p)={{\partial X^{(\alpha )}_p(p))}\over {\partial
x^{\mu}}}$ and locally we have a freely falling observer.

All the connection one-forms $\omega$ on the orthonormal frame bundle 
$F(M^4)=M^4\times SO(3,1)$ have a torsion 2-form [it is ${\cal T}={\cal D}
^{(\omega )} \theta$, where $\theta$ is the canonical or soldering one-form 
(the coframes or cotetrads) and ${\cal D}^{(\omega )}$ is the $F(M^4)$ 
exterior covariant derivative],
except the Levi-Civita connection $\omega_{\Gamma}$. Therefore, since in
general relativity we consider only Levi-Civita connections associated
with pseudo-Riemannian 4-manifolds $(M^4,{}^4g)$, in $F(M^4)$ we consider
only $\omega_{\Gamma}$-horizontal subspaces $H_{\Gamma}$ [$TF(M^4) = V_{\Gamma}
+H_{\Gamma}$ as a direct sum, with $V_{\Gamma}$ the vertical subspace
isomorphic to the Lie algebra o(3,1) of SO(3,1)]. Given a global cross section
$\sigma : M^4\rightarrow F(M^4)=M^4\times SO(3,1)$, the associated gauge
potentials on $M^4$, ${}^4\omega =\sigma^{*} \omega$, are the connection
coefficients ${}^4\omega^{(T)}=\sigma^{*}\omega$
in the noncoordinate basis ${}^4E_{(\alpha )}$ [the second line
defines them through the covariant derivative in the noncoordinate basis]

\begin{eqnarray}
&&{}^4\omega^{(T)(\gamma )}_{(\alpha )(\beta )} =  {}^4E^{(\gamma )}_{\nu} {}^4E
^{\mu}_{(\alpha )} (\partial_{\mu}\, {}^4E^{\nu}_{(\beta )} +{}^4E^{\lambda}
_{(\beta )}\, {}^4\Gamma^{(T)\nu}_{\mu\lambda}) = {}^4E^{(\gamma )}_{\nu}
{}^4E^{\mu}_{(\alpha )}\, {}^4\nabla_{\mu}\, {}^4E^{\nu}_{(\beta )},
\nonumber \\
&&{}^4{\tilde \nabla}_{{}^4E_{(\alpha )}}\, {}^4E_{(\beta )} = {}^4\nabla
_{{}^4E_{(\alpha )}}\, {}^4E_{(\beta )} - {}^4\omega^{(T)(\gamma )}_{(\alpha )
(\beta )}\, {}^4E_{(\gamma )} =0.
\label{a10}
\end{eqnarray}

The components of the  Riemann tensors in the noncoordinate bases are
${}^4R^{(\alpha )}
{}_{(\beta )(\gamma )(\delta )}= {}^4E_{(\gamma )} ({}^4\omega^{(T)(\alpha )}
_{(\delta )(\beta )}) - {}^4E_{(\delta )} ({}^4\omega^{(T)(\alpha )}
_{(\gamma )(\beta )}) + {}^4\omega^{(T)(\epsilon )}_{(\delta )(\beta )}\, 
{}^4\omega^{(T)(\alpha )}_{(\gamma )(\epsilon )} - {}^4\omega^{(T)(\epsilon )}
_{(\gamma )(\beta )}\, {}^4\omega^{(T)(\alpha )}_{(\delta )(\epsilon )} - 
c_{(\gamma )(\delta )}{}^{(\epsilon )}\, {}^4\omega^{(T)(\alpha )}
_{(\epsilon )(\beta )}$. The connection
(gauge potential) one-form ${}^4\omega^{(T)(\alpha )}{}_{(\beta )}={}^4\omega
^{(T)(\alpha )}_{(\gamma )(\beta )}\, {}^4\theta^{(\gamma )}$ [it is called
improperly ``spin connection", while its components are called Ricci
rotation coefficients] and the curvature (field strength) 
2-form ${}^4\Omega^{(T)(\alpha )}{}_{(\beta )} = {1\over 2} {}^4\Omega
^{(T)(\alpha )}
{}_{(\beta )(\gamma )(\delta )}\, {}^4\theta^{(\gamma )} \wedge {}^4\theta
^{(\delta )}$ satisfy the Cartan's structure equations

\begin{eqnarray}
&&d {}^4\theta^{(\alpha )} + {}^4\omega^{(T)(\alpha )}{}_{(\beta )} \wedge
{}^4\theta^{(\beta )} = {}^4T^{(\alpha )},\nonumber \\
&&d {}^4\omega^{(T)(\alpha )}{}_{(\beta )} + {}^4\omega^{(T)(\alpha )}
{}_{(\gamma )}\wedge {}^4\omega^{(T)(\gamma )}{}_{(\beta )} = 
{}^4\Omega^{(T)(\alpha )}{}_{(\beta )},
\label {a11}
\end{eqnarray}

\noindent whose exterior derivatives $0=d {}^4T^{(\alpha )}+{}^4\omega
^{(T)(\alpha )}{}_{(\beta )} \wedge {}^4T^{(\beta )} ={}^4\Omega^{(T)(\alpha )}
{}_{(\beta )} \wedge {}^4\theta^{(\beta )}$, $d {}^4\Omega^{(T)(\alpha )}
{}_{(\beta )} + {}^4\omega ^{(T)(\alpha )}{}_{(\gamma )} \wedge {}^4\Omega
^{(T)(\gamma )}{}_{(\beta )} - {}^4\Omega^{(T)(\alpha )}{}_{(\gamma )} \wedge 
{}^4\omega^{(T)(\gamma )}{}_{(\beta )}
\equiv 0$ are the two Bianchi identities.

With the Levi-Civita connection [which, as said, has zero torsion 2-form 
${}^4T^{(\alpha )}={1\over 2} T^{(\alpha )}{}_{(\beta )(\gamma )}\, {}^4\theta
^{(\beta )} \wedge {}^4\theta^{(\gamma )}=0$, namely 
${}^4T^{(\alpha )}{}_{(\beta )
(\gamma )}= {}^4\omega^{(T)(\alpha )}
_{(\beta )(\gamma )} - {}^4\omega^{(T)(\alpha )}_{(\gamma )(\beta )} - 
c_{(\beta )(\gamma )}{}^{(\alpha )}=0$  ], in a noncoordinate basis the
spin connection takes the form

\begin{eqnarray}
{}^4\omega^{(\alpha )}{}_{(\beta )}&=&{}^4\omega^{(\alpha )}
_{(\gamma )(\beta )}\, {}^4\theta^{(\gamma )}={}^4\omega^{(\alpha )}_{\mu
(\beta )} dx^{\mu},\nonumber \\
{}^4\omega_{(\alpha )(\gamma )(\beta )}&=&{}^4\eta_{(\alpha )(\delta )}\,
{}^4E^{(\delta )}_{\nu}\, {}^4E^{\mu}_{(\gamma )}\, {}^4\nabla_{\mu}\,
{}^4E^{\nu}_{(\beta )}={}^4\eta_{(\alpha )(\delta )}
{}^4\omega^{(\delta )}_{(\gamma )(\beta )},\nonumber \\
{}^4\omega^{(\alpha )}_{\mu (\beta )}&=&{}^4\omega^{(\alpha )}_{(\gamma )
(\beta )}\, {}^4E^{(\gamma )}_{\mu}={}^4E^{(\alpha )}_{\nu}\, {}^4\nabla_{\mu}
\, {}^4E^{\nu}_{(\beta )}={}^4E^{(\alpha )}_{\nu}[\partial_{\mu}\, {}^4E^{\nu}
_{(\beta )}+{}^4\Gamma^{\nu}_{\mu\rho}\, {}^4E^{\rho}_{(\beta )}],
\nonumber \\
\Rightarrow {}^4&\Gamma^{\mu}_{\rho\sigma}&={1\over 2}[{}^4E^{(\beta )}
_{\sigma}({}^4E^{\mu}_{(\alpha )}\, {}^4E^{(\gamma )}_{\rho}\, {}^4\omega
^{(\alpha )}_{(\gamma )(\beta )}-\partial_{\rho}\, {}^4E^{\mu}_{(\beta )})+
\nonumber \\
&+&{}^4E^{(\beta )}_{\rho}({}^4E^{\mu}_{(\alpha )}\, 
{}^4E^{(\gamma )}_{\sigma}\,
{}^4\omega^{(\alpha )}_{(\gamma )(\beta )}-\partial_{\sigma}\, {}^4E^{\mu}
_{(\beta )})],
\label{a12}
\end{eqnarray}

\noindent and
the metric compatibility ${}^4\nabla_{\rho}\, {}^4g_{\mu\nu}=0$ becomes the
following condition

\begin{equation}
{}^4\omega_{(\alpha )(\beta )}= {}^4\eta_{(\alpha )(\delta )}\, {}^4\omega
^{(\delta )}{}_{(\beta )} = {}^4\eta_{(\alpha )(\delta )} {}^4\omega^{(\delta )}
_{(\gamma )(\beta )}\, {}^4\theta^{(\gamma )} = {}^4\omega_{(\alpha )(\gamma )
(\beta )}\, {}^4\theta^{(\gamma )} = - {}^4\omega_{(\beta )(\alpha )}
\label{a13}
\end{equation}

\noindent or ${}^4\omega_{(\alpha )(\gamma )(\beta )}=-{}^4\omega_{(\beta )
(\gamma )(\alpha )}$ [${}^4\omega_{(\alpha )(\gamma )(\beta )}$ are called
Ricci rotation coefficients, only 24 of which are independent] 

Given a vector ${}^4V^{\mu}={}^4V^{(\alpha )}\, {}^4E^{\mu}_{(\alpha )}$ and a
covector ${}^4\omega_{\mu}={}^4\omega_{(\alpha )}\, {}^4E^{(\alpha )}_{\mu}$, 
we define the covariant derivative of the components ${}^4V^{(\alpha )}$ and 
${}^4\omega_{(\alpha )}$ as ${}^4\nabla_{\nu}\, {}^4V^{\mu}={}^4V^{\mu}{}_{; 
\nu}\equiv [{}^4\nabla_{\nu}\, {}^4V^{(\alpha )}]\, {}^4E^{\mu}_{(\alpha )}=
{}^4V^{(\alpha )}{}_{; \nu}\, {}^4E^{\mu}_{(\alpha )}$ and ${}^4\nabla_{\nu}\,
{}^4\omega_{\mu}={}^4\omega_{\mu ; \nu}\equiv [{}^4\nabla_{\nu}\, {}^4\omega
_{(\alpha )}]\, {}^4E^{(\alpha )}_{\mu}={}^4\omega_{(\alpha ) ; \nu}\, {}^4E
^{(\alpha )}_{\mu}$, so that

\begin{eqnarray}
{}^4V^{\mu}_{; \nu}&=&\partial_{\nu}\, {}^4V^{(\alpha )}\, {}^4E^{\mu}
_{(\alpha )}+{}^4V^{(\alpha )}\, {}^4E^{\mu}_{(\alpha ) ; \nu},\nonumber \\
&&\Rightarrow {}^4V^{(\alpha )}{}_{; \nu}=\partial_{\nu}\, {}^4V^{(\alpha )}+
{}^4\omega^{(\alpha )}_{\nu (\beta )}\, {}^4V^{(\beta )},\nonumber \\
{}^4\omega_{\mu ; \nu}&=&\partial_{\nu}\, {}^4\omega_{(\alpha )}\, {}^4E
^{(\alpha )}_{\mu}+{}^4\omega_{(\alpha )}\, {}^4E^{(\alpha )}_{\mu ; \nu},
\nonumber \\
&&\Rightarrow {}^4\omega_{(\alpha ) ; \nu}=\partial_{\nu}\, {}^4\omega_{(\alpha
)}-{}^4\omega_{(\beta )}\, {}^4\omega^{(\beta )}_{\nu (\alpha )}.
\label{a14}
\end{eqnarray}

Therefore,  for the ``internal tensors" ${}^4T^{(\alpha )...}{}_{(\beta )...}$, 
the spin connection
${}^4\omega^{(\alpha )}_{\mu (\beta )}$ is a gauge potential associated with a
gauge group SO(3,1). For
internal vectors ${}^4V^{(\alpha )}$ at $p\in M^4$ the cotetrads ${}^4E
^{(\alpha )}_{\mu}$ realize a soldering of this internal vector space at p
with the tangent space $T_pM^4$: ${}^4V^{(\alpha )}={}^4E^{(\alpha )}_{\mu}\,
{}^4V^{\mu}$. For tensors with mixed world and internal indices, like
tetrads and cotetrads, we could define a generalized covariant derivative
acting on both types of indices ${}^4{\tilde \nabla}_{\nu}\, {}^4E^{\mu}
_{(\alpha )}=\partial_{\nu}\, {}^4E^{\mu}_{(\alpha )}+{}^4\Gamma^{\mu}_{\nu
\rho}\, {}^4E^{\rho}_{(\alpha )}-{}^4E^{\mu}_{(\beta )}\, {}^4\omega^{(\beta )}
_{\nu (\alpha )}$: then ${}^4\nabla_{\nu}\, {}^4V^{\mu}={}^4\nabla_{\nu}\,
{}^4V^{(\alpha )}\, {}^4E^{\mu}_{(\alpha )}+{}^4V^{(\alpha )}\, {}^4{\tilde 
\nabla}_{\nu}\, {}^4E^{\mu}_{(\alpha )}\equiv {}^4\nabla_{\nu}\, {}^4V
^{(\alpha )}\, {}^4E^{\mu}_{(\alpha )}$ implies ${}^4{\tilde \nabla}_{\nu}\,
{}^4E^{\mu}_{(\alpha )}=0$ [or ${}^4\nabla_{\nu}\, {}^3E^{\mu}_{(\alpha )}={}^4E
^{\mu}_{(\beta )}\, {}^4\omega^{(\beta )}_{\nu (\alpha )}$]
which is nothing else that the definition 
(\ref{a12}) of the spin connection ${}^4\omega^{(\alpha )}_{\mu (\beta )}$.

We have

\begin{eqnarray}
[ {}^4E_{(\alpha )}, {}^4E_{(\beta )}] &=& c_{(\alpha )(\beta )}{}^{(\gamma )}
{}^4E_{(\gamma )} = {}^4\nabla_{{}^4E_{(\alpha )}}\, {}^4E_{(\beta )} -
{}^4\nabla_{{}^4E_{(\beta )}}\, {}^4E_{(\alpha )} =\nonumber \\
&=& ({}^4\omega^{(\gamma )}_{(\alpha )(\beta )} - {}^4\omega^{(\gamma )}
_{(\beta )(\alpha )})\, {}^4E_{(\gamma )},
\label{a15}
\end{eqnarray}

\begin{eqnarray}
{}^4\Omega^{(\alpha )}{}_{(\beta )(\gamma )(\delta )}&=& {}^4E_{(\gamma )}({}^4
\omega^{(\alpha )}_{(\delta )(\beta )}) - {}^4E_{(\delta )} ({}^4\omega
^{(\alpha )}_{(\gamma )(\beta )})+\nonumber \\
&+& {}^4\omega^{(\epsilon )}_{(\delta )(\beta )}\, {}^4\omega^{(\alpha )}
_{(\gamma )(\epsilon )} - {}^4\omega^{(\epsilon )}_{(\gamma )(\beta )}\,
{}^4\omega^{(\alpha )}_{(\delta )(\epsilon )} - ({}^4\omega^{(\epsilon )}
_{(\gamma )(\delta )} - {}^4\omega^{(\epsilon )}_{(\delta )(\gamma )})
{}^4\omega^{(\alpha )}_{(\epsilon )(\beta )}=\nonumber \\
&=&{}^4E^{(\alpha )}_{\mu}\, {}^4R^{\mu}{}_{\rho\nu\sigma}\, {}^4E^{\rho}
_{(\beta )}\, {}^4E^{\nu}_{(\gamma )}\, {}^4E^{\sigma}_{(\delta )},
\nonumber \\
&&{}\nonumber \\
{}^4\Omega_{\mu\nu}{}^{(\alpha )}{}_{(\beta )}&=&
{}^4E^{(\gamma )}_{\mu}\, {}^4E^{(\delta )}_{\nu}\, {}^4\Omega
^{(\alpha )}{}_{(\beta )(\gamma )(\delta )}={}^4R^{\rho}{}_{\sigma\mu\nu}\, 
{}^4E^{(\alpha )}_{\rho}\, {}^4E^{\sigma}_{(\beta )}=\nonumber \\
&=&\partial_{\mu}
{}^4\omega^{(\alpha )}_{\nu (\beta )}-\partial_{\nu}\, {}^4\omega^{(\alpha )}
_{\mu (\beta )} +{}^4\omega^{(\alpha )}_{\mu (\gamma )}\, {}^4\omega^{(\gamma )}
_{\nu (\beta )} - {}^4\omega^{(\alpha )}_{\nu (\gamma )}\, {}^4\omega
^{(\gamma )}_{\mu (\beta )},\nonumber \\
{}^4\Omega_{\mu\nu (\alpha )(\beta )}&=&
{}^4\eta_{(\alpha )(\gamma )}\, {}^4\Omega_{\mu\nu}{}^{(\gamma )}{}_{(\beta )}=
-{}^4\Omega_{\nu\mu (\alpha )(\beta )}=-{}^4\Omega_{\mu\nu (\beta )(\alpha )},
\nonumber \\
&&{}\nonumber \\
{}^4R^{\alpha}{}_{\beta\mu\nu}&=&{}^4E^{\alpha}_{(\gamma )}\, {}^4E^{(\delta )}
_{\beta}\, {}^4\Omega_{\mu\nu}{}^{(\gamma )}{}_{(\delta )},\nonumber \\
{}^4R_{\mu\nu}&=&{}^4E^{\alpha}_{(\gamma )}\, {}^4E^{(\delta )}_{\nu}\,
{}^4\Omega_{\alpha\mu}{}^{(\gamma )}{}_{(\delta )},\nonumber \\
{}^4R&=&{}^4E^{\mu}_{(\gamma )}\, {}^4E^{(\delta )}_{\rho}\, {}^4g^{\rho\nu}
{}^4\Omega_{\mu\nu}{}^{(\gamma )}{}_{(\delta )},
\label{a16}
\end{eqnarray}

\begin{eqnarray}
&&d {}^4\theta^{(\alpha )} + {}^4\omega^{(\alpha )}{}_{(\beta )} \wedge
{}^4\theta^{(\beta )} =0,\nonumber \\
&&d {}^4\omega^{(\alpha )}{}_{(\beta )} + {}^4\omega^{(\alpha )}{}_{(\gamma )}
\wedge {}^4\omega^{(\gamma )}{}_{(\beta )} = 
{}^4\Omega^{(\alpha )}{}_{(\beta )},
\label{a17}
\end{eqnarray}

\noindent with the Bianchi identities ${}^4\Omega^{(\alpha )}{}_{(\beta )} 
\wedge {}^4\theta^{(\beta )}\equiv 0$, $d {}^4\Omega^{(\alpha )}{}_{(\beta )} 
+ {}^4\omega^{(\alpha )}{}_{(\gamma )} \wedge {}^4\Omega^{(\gamma )}
{}_{(\beta )} -{}^4\Omega^{(\alpha )}{}_{(\gamma )} \wedge 
{}^4\omega^{(\gamma )}{}_{(\beta )}
\equiv 0$.

Let us remark that Eqs.(\ref{a10}) and (\ref{a12}) imply ${}^4\Gamma^{\rho}
_{\mu\nu}={}^4\triangle^{\rho}_{\mu\nu}+{}^4\omega^{\rho}_{\mu\nu}$ with
${}^4\omega^{\rho}_{\mu\nu}={}^4E^{\rho}_{(\alpha )}\, {}^4E^{(\beta )}_{\nu}
\, {}^4\omega^{(\alpha )}_{\mu (\beta )}$ and ${}^4\triangle^{\rho}
_{\mu\nu}={}^4E^{\rho}_{(\alpha )}\, \partial_{\mu}\, {}^4E^{(\alpha )}_{\nu}$;
the Levi-Civita connection (i.e. the Christoffel symbols) turn out to be
decomposed in a flat connection ${}^4\triangle^{\rho}_{\mu\nu}$ (it produces
zero Riemann tensor as was already known to Einstein\cite{dav}) and in a
tensor, like in the Yang-Mills case\cite{lusa}.

Let us finish this Section with a review of some action principles used for
general relativity. In metric gravity, one 
uses the generally covariant Hilbert action  depending on the
4-metric and its first and second derivatives [G is Newton gravitational 
constant; $U\subset M^4$ is a subset of spacetime; we use units with $x^o=ct$]

\begin{equation}
S_H={{c^3}\over {16\pi G}}\, \int_U\, d^4x\, \sqrt{{}^4g}\, {}^4R= \int_U
d^4x\, {\cal L}_H.
\label{I7}
\end{equation}

\noindent The variation of $S_H$ is [$d^3\Sigma_{\gamma}=d^3\Sigma l_{\gamma}$]

\begin{eqnarray}
&&\delta S_H = \delta S_E +\Sigma_H = -{{c^3}\over {16\pi G}} \int_U
d^4x\, \sqrt{{}^4g}\, {}^4G^{\mu\nu} \delta {}^4g_{\mu\nu} +\Sigma_H,
\nonumber \\
&&{}\nonumber \\
&&\Sigma_H ={{c^3}\over {16\pi G}} \int_{\partial U} {d^3\Sigma}_{\gamma}
\sqrt{{}^4g}\, ({}^4g^{\mu\nu} \delta^{\gamma}_{\delta} - {}^4g^{\mu\gamma}
\delta^{\nu}_{\delta}) \delta \, {}^4\Gamma^{\delta}_{\mu\nu}=\nonumber \\
&&={{c^3}\over {8\pi G}} \int_{\partial U} d^3\Sigma \, \sqrt{{}^3\gamma}\,\,
\delta \, {}^3K,\nonumber \\
&&{}\nonumber \\
&&\delta {}^4\Gamma^{\delta}_{\mu\nu} ={1\over 2} {}^4g^{\delta\beta}
[{}^4\nabla_{\mu} \delta \, {}^4g_{\beta\nu} +{}^4\nabla_{\nu} \delta \,
{}^4g_{\beta\mu} -{}^4\nabla_{\beta} \delta \, {}^4g_{\mu\nu}].
\label{I8}
\end{eqnarray}

\noindent where ${}^3\gamma_{\mu\nu}$ is the metric induced on $\partial U$ and
$l_{\mu}$ is the outer unit covariant normal to $\partial U$. The trace of the
extrinsic curvature ${}^3K_{\mu\nu}$ of $\partial U$ is ${}^3K=-l^{\mu}
{}_{;\mu}$.
The surface term $\Sigma_H$ takes care of the second derivatives of the 4-metric
and to get Einstein equations ${}^4G_{\mu\nu}={}^4R_{\mu\nu}-
{1\over 2}{}^4g_{\mu\nu}\, {}^4R \, {\buildrel \circ \over =}\, 0$ 
one must take constant certain normal 
derivatives of the 4-metric on the boundary of $U$ [${\cal L}_l\, ({}^4g
_{\mu\nu}-l_{\mu}l_{\nu})=0$] to have $\delta S_H=0$ \cite{yo}.

The term $\delta S_E$ in Eq.(\ref{I8}) means the variation of the action $S_E$,
which is the (not generally covariant) Einstein action
depending only on the 4-metric and its first derivatives [$\delta
S_E=0$ gives ${}^4G_{\mu\nu}\, {\buildrel \circ \over =}\, 0$ if ${}^4g
_{\mu\nu}$ is held fixed on $\partial U$]

\begin{eqnarray}
S_E&=& \int_U d^4x\, {\cal L}_E=
{{c^3}\over {16\pi G}}\, \int_U\, d^4x\, \sqrt{{}^4g}\, 
{}^4g^{\mu\nu}({}^4\Gamma^{\rho}_{\nu\lambda}\, {}^4\Gamma^{\lambda}_{\rho\mu} 
- {}^4\Gamma^{\lambda}_{\lambda\rho}\, {}^4\Gamma^{\rho}_{\mu\nu})=\nonumber \\
&=&S_H-{{c^3}\over {16\pi G}} \int_Ud^4x\, \partial_{\lambda} [\sqrt{{}^4g} 
({}^4g^{\mu\nu}\, {}^4\Gamma^{\lambda}_{\mu\nu}-{}^4g^{\lambda\mu}\, {}^4\Gamma
^{\rho}_{\rho\mu})],\nonumber \\
{}&&\nonumber \\
\delta S_E&=&{{c^3}\over {16\pi G}} \int_U d^4x\, ({{\partial {\cal L}_E}\over
{\partial \, {}^4g^{\mu\nu}}}-\partial_{\rho}\, {{\partial {\cal L}_E}\over
{\partial \partial_{\rho}\, {}^4g^{\mu\nu}}})\, \delta \, {}^4g^{\mu\nu}=
-{{c^3}\over {16\pi G}} \int_U d^4x\, \sqrt{{}^4g}\, {}^4G_{\mu\nu} \delta
\, {}^4g^{\mu\nu}.
\label{I9}
\end{eqnarray}

We shall not consider the first-order Palatini action; see for instance
Ref.\cite{ro}, where there is also  a review of the variational principles of
the connection-dependent formulations of general relativity.

In Ref.\cite{yo} (see also Ref.\cite{in}), it is shown that the DeWitt-ADM 
action\cite{witt,adm} for a 3+1
decomposition of $M^4$ can be obtained from $S_H$ in the following way
[$\sqrt{{}^4g} {}^4R =-\epsilon \sqrt{{}^4g}({}^3R+{}^3K_{\mu\nu}{}^3K^{\mu\nu}-
({}^3K)^2)-2\epsilon \,
\partial_{\lambda}(\sqrt{{}^4g} ({}^3K l^{\lambda}+a^{\lambda}))$,
with $a^{\lambda}$ the 4-acceleration ($l^{\mu}a_{\mu}=0$); the 4-volume U is
$[\tau_f,\tau_i]\times S$]

\begin{eqnarray}
&&S_H = S_{ADM}+\Sigma_{ADM}, \nonumber \\
&&S_{ADM} = -\epsilon
{{c^3}\over {16\pi G}} \int_{U} d^4x\, \sqrt{{}^4g} 
[{}^3R+{}^3K_{\mu\nu}\, {}^3K^{\mu\nu} -({}^3K)^2],\nonumber \\
&&\Sigma_{ADM}=-\epsilon {{c^3}\over {8\pi G}} \int d^4x \partial_{\alpha}
[\sqrt{{}^4g} ({}^3K l^{\alpha} +l^{\beta} l^{\alpha}{}_{;\beta})]=\nonumber \\
&&=-\epsilon {{c^3}\over {8\pi G}} \Big[
\int_S d^3\sigma \, [\sqrt{\gamma}\,\, {}^3K](\tau ,\vec \sigma ){|}^{\tau_f}
_{\tau_i}+\nonumber \\
&&+\int^{\tau_f}_{\tau_i}d\tau \int_{\partial S}d^2\Sigma^r [{}^3\nabla_r
(\sqrt{\gamma}N)- {}^3K N_r](\tau ,\vec \sigma )\Big] ,\nonumber \\
&&{}\nonumber \\
&&\delta S_{ADM}=-\epsilon {{c^3}\over {16\pi G}} \int d\tau d^3\sigma 
\sqrt{\gamma}\Big[ 2\, {}^4{\bar G}_{ll} \delta N+{}^4{\bar G}_l{}^r\delta N_r-
{}^4{\bar G}^{rs} \delta \, {}^3g_{rs}\Big] (\tau ,\vec \sigma )+\nonumber \\
&&+\delta S_{ADM} {|}_{{}^4G_{\mu\nu}=0}-\epsilon \int^{\tau_f}_{\tau_i}d\tau
\int_{\partial U}d^3\Sigma_r [N_{|s}\delta \, {}^3g^{rs}-N\delta \, {}^3g
^{rs}{}_{|s}](\tau ,\vec \sigma ),\nonumber \\
&&\delta S_{ADM} {|}_{{}^4G_{\mu\nu}=0} =-\epsilon
 {{c^3}\over {16\pi G}} \int_{\partial
U} d^3\sigma \, {}^3{\tilde \Pi}^{\mu\nu} \delta {}^3\gamma_{\mu\nu},\nonumber\\
&& {}^3{\tilde \Pi}^{\mu\nu}=\sqrt{\gamma} 
({}^3K^{\mu\nu}-{}^3g^{\mu\nu}\, {}^3K)=
{{16\pi G}\over {c^3}} \epsilon {\hat b}^{\mu}_r{\hat b}^{\nu}_s\, 
{}^3{\tilde \Pi}^{rs},
\label{I10}
\end{eqnarray}

\noindent so that $\delta S_{ADM}=0$ gives ${}^4G_{\mu\nu}\, {\buildrel \circ 
\over =}\, 0$ if one holds 
fixed the intrinsic 3-metric ${}^3\gamma_{\mu\nu}$ on the boundary 
[${}^3{\tilde  \Pi}^{\mu\nu}$ is the ADM momentum with world indices, whose 
form in a 3+1 splitting is given in Section V]. This action is not generally 
covariant, but it is quasi-invariant under the 8 types of gauge transformations
generated by the ADM first class constraints, as it will be shown in the
third paper of the series. As shown in
Refs.\cite{dew,reg,yo,hh} in this way one obtains a well defined gravitational
energy. However, in so doing one still neglects some boundary terms.
Following Ref.\cite{hh}, let us assume that, given a subset $U\subset M^4$ of 
spacetime, $\partial U$ consists of two
slices, $\Sigma_{\tau_i}$ (the initial one) and $\Sigma_{\tau_f}$ (the final
one) with outer normals $-l^{\mu}(\tau_i,\vec \sigma )$ and $l^{\mu}(\tau_f,
\vec \sigma )$ respectively, and of a surface $S_{\infty}$ near space infinity
with outer unit (spacelike) normal $n^{\mu}(\tau ,\vec \sigma )$ tangent
to the slices [so that the normal $l^{\mu}(\tau ,\vec \sigma )$ to every slice
is asymptotically tangent to $S_{\infty}$]. The
3-surface $S_{\infty}$ is foliated by a family of 2-surfaces $S^2_{\tau , 
\infty}$ coming from its intersection with the slices $\Sigma_{\tau}$ 
[therefore, asymptotically $l^{\mu}(\tau ,\vec \sigma )$ is normal to the
corresponding $S^2_{\tau ,\infty}$]. The vector $b^{\mu}_{\tau}=z^{\mu}
_{\tau}=   N l^{\mu}+N^rb^{\mu}_r$ is not in general tangent to $S
_{\infty}$. It is assumed that there are no inner boundaries (see Ref.
\cite{hh} for their treatment), so that the slices $\Sigma_{\tau}$ do not
intersect and are complete. This does not rule out the existence of horizons,
but it implies that, if horizons form, one continues to evolve the
spacetime inside the horizon as well as outside. 
Then, in Ref.\cite{hh} it is shown that one gets [${}^2K$ the trace of the
2-dimensional extrinsic curvature of the 2-surface $S^2_{\tau ,\infty}=
S_{\infty}\cap \Sigma_{\tau}$; to get this result
one assumes that the lapse function
$N(\tau ,\vec \sigma )$ on $\Sigma_{\tau}$ tends asymptotically to a function
$N_{(as)}(\tau )\,\,$ and that the term on $\partial S$ vanishes due to the 
boundary conditions]

\begin{eqnarray}
\Sigma_{ADM}&=&-\epsilon {{c^3}\over {8\pi G}} [\int_{\Sigma_{\tau_f}} d^3\Sigma
-\int_{\Sigma_{\tau_i}} d^3\Sigma ]\, N\, \sqrt{\gamma} \, {}^3K=\nonumber \\
&=&-\epsilon
{{c^3}\over {8\pi G}} \int_{\tau_i}^{\tau_f} d\tau \, N_{(as)}(\tau ) \int
_{S^2_{\tau ,\infty}} d^2\Sigma \, \sqrt{\gamma}\, {}^2K.
\label{I11}
\end{eqnarray}

Instead, in tetrad
gravity\cite{weyl,dirr,schw,kib,tetr,char,maluf,hen1,hen2,hen3,hen4}, 
in which ${}^4g_{\mu\nu}$ is no more the independent variable, 
the new independent 16 variables are a set of cotetrads ${}^4E^{(\alpha )}
_{\mu}$ so that ${}^4g_{\mu\nu}={}^4E^{(\alpha )}_{\mu}\, {}^4\eta_{(\alpha )
(\beta )}\, {}^4E^{(\beta )}_{\nu}$. Tetrad gravity 
has not only the invariance under
$Diff\, M^4$ but also under local Lorentz transformations on $TM^4$ [acting
on the flat indices $(\alpha )$]. An action principle with these local
invariances is obtained by replacing the 4-metric in the Hilbert action
$S_H$ with its expression in terms of the cotetrads. The action acquires the 
form

\begin{equation}
S_{HT}={{c^3}\over {16\pi G}} \int_{U} d^4x\, {}^4\tilde E\, {}^4E^{\mu}
_{(\alpha )}\, {}^4E^{\nu}_{(\beta )}\, {}^4\Omega_{\mu\nu}{}^{(\alpha )
(\beta )},
\label{I13}
\end{equation}

\noindent where ${}^4\tilde E=det\, ({}^4E_{\mu}^{(\alpha )})=\sqrt{{}^4g}$
and ${}^4\Omega_{\mu\nu}{}^{(\alpha )(\beta )}$ is the spin 4-field strength
. One has

\begin{eqnarray}
\delta S_{HT}&=& {{c^3}\over {16\pi G}} \int_U d^4x\, {}^4\tilde E\, 
{}^4G_{\mu\nu}\, {}^4E^{\mu}_{(\alpha )}\, {}^4\eta^{(\alpha )(\beta )}\, 
\delta \, {}^4E^{\nu}_{(\beta )}+\nonumber \\
&+&{{c^3}\over {8\pi G}} \int_U d^4x\, \partial_{\mu} [{}^4\tilde E\, 
({}^4E^{(\rho )}_{\nu} \delta ({}^4g^{\mu\lambda}\, {}^4\nabla_{\lambda}
{}^4E^{\nu}_{(\rho )})-{}^4\eta^{(\rho )(\sigma )}\, {}^4E^{\nu}_{(\rho )}
\delta ({}^4\nabla_{\nu}\, {}^4E^{\mu}_{(\sigma )}))].
\label{I14}
\end{eqnarray}

Again $\delta S_{HT}=0$ produces Einstein equations if complicated derivatives
of the tetrads vanish at the boundary. In Ref.\cite{char}, 
by using ${}^4\tilde E
{}^4E^{\mu}_{(\alpha )}\, {}^4E^{\nu}_{(\beta )}\, {}^4\Omega_{\mu\nu}
{}^{(\alpha )(\beta )}=2\, {}^4\tilde E\, {}^4E^{\mu}_{(\alpha )}\, 
{}^4E^{\nu}_{(\beta )} [{}^4\omega_{\mu}\, {}^4\omega_{\nu} - {}^4\omega_{\nu}
{}^4\omega_{\mu}]^{(\alpha )(\beta )}+2\, \partial_{\mu}({}^4\tilde E\,
{}^4E^{\mu}_{(\alpha )}\, {}^4E^{\nu}_{(\beta )}\, {}^4\omega_{\nu}{}^{(\alpha
)(\beta )})$, the analogue
of $S_E$, i.e. the (not locally Lorentz invariant, therefore not expressible
only in terms of the 4-metric) Charap action, is defined as

\begin{equation}
S_C=-{{c^3}\over {8\pi G}}\int_U d^4x\, {}^4\tilde E\, {}^4E^{\mu}
_{(\alpha )}\, {}^4E^{\nu}_{(\beta )} ({}^4\omega_{\mu}\, {}^4\omega_{\nu}-
{}^4\omega_{\nu}\, {}^4\omega_{\mu})^{(\alpha )(\beta )}.
\label{I15}
\end{equation}

\noindent Its variation $\delta S_C$ vanishes if $\delta \, {}^4E^{\mu}
_{(\alpha )}$ vanish at the boundary and the Einstein equations hold. However 
its Hamiltonian formulation gives too complicated first class constraints
to be solved.

In Einstein metric gravity the gravitational field, described by the 4-metric
${}^4g_{\mu\nu}$ depends on 2, and not 10,
physical degrees of freedom in each point; this is not explicitly evident if one
starts with the Hilbert action, which is invariant under $Diff\, M^4$, a group
with only four generators. Instead in ADM canonical gravity (see 
Section V) there are in each point 20 canonical variables and 8 first class
constraints, implying the determination of 8 canonical variables and the
arbitrariness of the 8 conjugate ones. At the Lagrangian level, only 6 of the
ten Einstein equations are independent, due to the contracted Bianchi
identities, so that four components of the metric tensor ${}^4g_{\mu\nu}$
(the lapse and shift functions) are arbitrary not being determined by the
equation of motion. Moreover, the four combinations ${}^4{\bar G}_{ll}\,
{\buildrel \circ \over =}\, 0$, ${}^4{\bar G}_{lr}\, {\buildrel \circ \over
=}\, 0$, of the Einstein equations do not depend on the second time derivatives
or accelerations (they
are restrictions on the Cauchy data and become the secondary first class
constraints of the ADM canonical theory): the general theory \cite{sha} implies 
that four generalized velocities (and therefore other four components of the 
metric) inherit the arbitrariness of the lapse and shift functions. Only two
combinations of the Einstein equations depend on the accelerations 
(second time derivatives) of the
two (non tensorial) independent degrees of freedom of the gravitational field
and are
genuine equations of motion. Therefore, the ten components of every 4-metric 
${}^4g_{\mu\nu}$, compatible with the Cauchy data, depend on 8 arbitrary 
functions not determined by the Einstein equations.

Tetrad gravity with action $S_{HT}$, in which
the elementary natural Lagrangian object is the soldering or canonical
one-form (or orthogonal coframe) $\theta^{(\alpha )}={}^4E^{(\alpha )}_{\mu} 
dx^{\mu}$, is gauge invariant simultaneously under diffeomorphisms [$Diff\, 
M^4$] and Lorentz transformations [SO(3,1)]. Instead in phase space
(see Section IV)  only two
of the 16 components of the cotetrad ${}^4E^{(\alpha )}_{\mu}(x)$ are
physical degrees of freedom in each point, since  the 32
canonical variables present in each point are restricted by 14 first class
constraints, so that the 16 components of a cotetrad compatible with the Cauchy 
data depend on 14 arbitrary functions not determined by the equation of motion.

The gauge transformations of tetrad gravity with action $S_{HT}$ are 
[$x^{\mu}\mapsto x^{{'}\mu}(x)$, 
$\Lambda (x)\in SO(3,1)$ for each $x^{\mu}\in M^4$]

\begin{eqnarray}
{}^4E^{(\alpha )}_{\mu}(x)&\mapsto& {}^4E^{{'}(\alpha )}_{\mu}(x^{'}(x))=
{{\partial x^{\nu}}\over {\partial x^{{'}\mu}}} \Lambda^{(\alpha )}{}_{(\beta )}
(x)\, {}^4E^{(\beta )}_{\nu}(x),\nonumber \\
{}^4g_{\mu\nu}(x)&\mapsto& {}^4g^{'}_{\mu\nu}(x^{'}(x))={{\partial x^{\alpha}}
\over {\partial x^{{'}\mu}}}\, {{\partial x^{\beta}}\over {\partial x^{{'}\nu}
}}\, {}^4g_{\alpha\beta}(x),\nonumber \\ 
{}^4\Gamma^{\mu}_{\alpha\beta}(x)&\mapsto& {}^4\Gamma^{{'}\mu}_{\alpha\beta}
(x^{'}(x))={{\partial x^{{'}\mu}}\over {\partial x^{\nu}}}\, {{\partial 
x^{\gamma}}\over {\partial x^{{'}\alpha}}}\, {{\partial x^{\delta}}\over 
{\partial x^{{'}\beta}}}\, {}^4\Gamma^{\nu}_{\gamma\delta}(x)+{{\partial^2
x^{\nu}}\over {\partial x^{{'}\alpha}\partial x^{{'}\beta}}} {{\partial
x^{{'}\mu}}\over {\partial x^{\nu}}},\nonumber \\
{}^4R^{\mu}{}_{\alpha\nu\beta}(x)&\mapsto& {}^4R^{{'}\mu}{}_{\alpha\nu\beta}
(x^{'}(x))={{\partial x^{{'}\mu}}\over {\partial x^{\rho}}}\, {{\partial 
x^{\gamma}}\over {\partial x^{{'}\alpha}}}\, {{\partial x^{\sigma}}\over
{\partial x^{{'}\nu}}}\, {{\partial x^{\delta}}\over {\partial x^{{'}\beta}}}
\, {}^4R^{\rho}{}_{\gamma\sigma\delta}(x),\nonumber \\
{}^4\omega^{(\alpha )}_{(\gamma )(\beta )}(x)&\mapsto& {}^4\omega^{{'}(\alpha )}
_{(\gamma )(\beta )}(x^{'}(x))=\Lambda^{(\alpha )}{}_{(\mu )}(x) (\Lambda^{-1})
^{(\rho )}{}_{(\gamma )}(x) (\Lambda^{-1})^{(\sigma )}{}_{(\beta )}(x)\,
{}^4\omega^{(\mu )}_{(\rho )(\sigma )}(x)+\nonumber \\
&+&(\Lambda^{-1})^{(\rho )}
{}_{(\gamma )}(x)\, {}^4E^{\nu}_{(\rho )}(x)\, \partial_{\nu}\, \Lambda
^{(\alpha )}{}_{(\mu )}(x)\, (\Lambda^{-1})^{(\mu )}{}_{(\beta )}(x),
\nonumber \\
{}^4\Omega^{(\alpha )}{}_{(\beta )(\gamma )(\delta )}(x)&\mapsto& {}^4\Omega
^{{'}(\alpha )}{}_{(\beta )(\gamma )(\delta )}(x^{'}(x))=\nonumber \\
&=&\Lambda^{(\alpha )}
{}_{(\mu )}(x)\, (\Lambda^{-1})^{(\rho )}{}_{(\beta )}(x)\, (\Lambda^{-1})
^{(\nu )}{}_{(\gamma )}(x)\, (\Lambda^{-1})^{(\sigma )}{}_{(\delta )}(x)\,
{}^4\Omega^{(\mu )}{}_{(\rho )(\nu )(\sigma )}(x).\nonumber \\
&&{}
\label{a18}
\end{eqnarray}

With the Lie derivative one can characterize the action of infinitesimal
diffeomorphisms

\begin{eqnarray}
x^{{'}\mu}(x)&=& x^{\mu} + \xi^{\mu}(x)=x^{\mu}+\delta_ox^{\mu},\quad\quad
\Rightarrow x^{\mu}(x^{'})\approx x^{{'}\mu} - \xi^{\mu}(x^{'}),\nonumber \\
&&{}\nonumber \\
\delta {}^4E^{\mu}_{(\alpha )}(x)&=& {}^4E^{{'}\mu}_{(\alpha )}(x^{'}(x))-
{}^4E^{\mu}_{(\alpha )}(x)= \delta_o{}^4E^{\mu}_{(\alpha )}(x) + \xi^{\nu}(x)
\partial_{\nu} {}^4E^{\mu}_{(\alpha )}(x)=\nonumber \\
&=&{{\partial x^{{'}\mu}}\over {\partial x^{\nu}}} {}^4E^{\nu}_{(\alpha )}(x)-
{}^4E^{\mu}_{(\alpha )}(x)= \partial_{\nu}\xi^{\mu}(x) {}^4E^{\nu}_{(\alpha )}
(x),\nonumber \\
\delta_o {}^4E^{\mu}_{(\alpha )}(x)&=&{}^4E^{{'}\mu}_{(\alpha )}(x)-{}^4E^{\mu}
_{(\alpha )}(x) = [\partial_{\nu}\xi^{\mu}(x)-\delta^{\mu}_{\nu} \xi^{\rho}(x)
\partial_{\rho}] {}^4E^{\nu}_{(\alpha )}(x)= [{\cal L}_{-\xi^{\rho}\partial
_{\rho}} {}^4E^{\nu}_{(\alpha )}(x)\partial_{\nu}]^{\mu},\nonumber \\
&&{}\nonumber \\
\delta{}^4E^{(\alpha )}_{\mu}(x)&=&{}^4E^{{'}(\alpha )}_{\mu}(x^{'}(x))-
{}^4E^{(\alpha )}_{\mu}(x)=\delta_o {}^4E^{(\alpha )}_{\mu}+\xi^{\nu}(x)
\partial_{\nu} {}^4E^{(\alpha )}_{\mu}(x)=\nonumber \\
&=&{{\partial x^{\nu}}\over {\partial x^{{'}\mu}}} {}^4E^{(\alpha )}_{\nu}(x)-
{}^4E^{(\alpha )}_{\mu}(x) = -\partial_{\mu}\xi^{\nu}(x) {}^4E^{(\alpha )}
_{\nu}(x),\nonumber \\
\delta_o {}^4E^{(\alpha )}_{\mu}(x)&=&{}^4E^{{'}(\alpha )}_{\mu}(x)-{}^4E
^{(\alpha )}_{\mu}(x)=-[\partial_{\mu}\xi^{\nu}(x)+\delta^{\nu}_{\mu}\xi^{\rho}
(x)\partial_{\rho}] {}^4E^{(\alpha )}_{\nu}(x)= \nonumber \\
&=&[{\cal L}_{-\xi^{\rho}\partial
_{\rho}} {}^4E^{(\alpha )}_{\nu}(x)dx^{\nu}]_{\mu},\nonumber \\
&&{}\nonumber \\
\delta {}^4g_{\mu\nu}(x)&=& {}^4g^{'}_{\mu\nu}(x^{'}(x))-{}^4g_{\mu\nu}(x)=
\delta_o {}^4g_{\mu\nu}(x)+\xi^{\rho}(x)\partial_{\rho} {}^4g_{\mu\nu}(x)=
\nonumber \\
&=&{{\partial x^{\alpha}}\over {\partial x^{{'}\mu}}}\, {{\partial x^{\beta}}
\over {\partial x^{{'}\nu}}}\, {}^4g_{\alpha\beta}(x)- {}^4g_{\mu\nu}(x)=-
[\delta^{\alpha}_{\mu}\partial_{\nu}\xi^{\beta}(x) + \delta^{\beta}_{\nu}
\partial_{\mu}\xi^{\alpha}(x)] {}^4g_{\alpha\beta}(x),\nonumber \\
\delta_o {}^4g_{\mu\nu}(x)&=& {}^4g^{'}_{\mu\nu}(x)-{}^4g_{\mu\nu}(x)=-
[\delta^{\alpha}_{\mu}\partial_{\nu}\xi^{\beta}(x)+\delta^{\beta}_{\nu}
\partial_{\mu}\xi^{\alpha}(x)+\delta^{\alpha}_{\mu}\delta^{\beta}_{\nu}
\xi^{\rho}(x)\partial_{\rho}] {}^4g_{\alpha\beta}(x)=\nonumber \\
&=& -[{}^4\nabla_{\mu} \xi_{\nu}(x)+{}^4\nabla_{\nu} \xi_{\mu}(x)]
=[{\cal L}_{-\xi^{\rho}
\partial_{\rho}} {}^4g_{\alpha\beta} dx^{\alpha} \otimes dx^{\beta}]_{\mu\nu}.
\label{a20}
\end{eqnarray}

With the spin connection coefficients ${}^4\omega^{(\alpha )}
_{\mu (\beta )}={}^4\omega^{(\alpha )}_{(\gamma )(\beta )}\, {}^4E^{(\gamma )}
_{\mu}$, and the field strengths ${}^4\Omega_{\mu\nu}{}^{(\alpha )}
{}_{(\beta )}$, we get the transformation properties of the
gauge potentials and field strengths of a SO(3,1) connection on the
orthonormal frame bundle $F(M^4)$

\begin{eqnarray}
{}^4\omega^{(\alpha )}_{\mu (\beta )}(x)&\mapsto& {}^4\omega^{{'}(\alpha )}
_{\mu (\beta )}(x^{'}(x))={{\partial x^{\nu}}\over {\partial x^{{'}\mu}}}
[\Lambda (x)\, {}^4\omega_{\nu}(x) \Lambda^{-1}(x)+\partial_{\nu}\, \Lambda (x)
\Lambda^{-1}(x)]^{(\alpha )}{}_{(\beta )},\nonumber \\
{}^4\Omega_{\mu\nu}{}^{(\alpha )}{}_{(\beta )}(x)&\mapsto& {}^4\Omega^{'}
_{\mu\nu}{}^{(\alpha )}{}_{(\beta )}(x^{'}(x))={{\partial x^{\rho}}\over
{\partial x^{{'}\mu}}}\, {{\partial x^{\sigma}}\over {\partial x^{{'}\nu}}}
\Lambda^{(\alpha )}{}_{(\gamma )}(x)\, {}^4\Omega_{\rho\sigma}{}^{(\gamma )}
{}_{(\delta )}(x)\, (\Lambda^{-1})^{(\delta )}{}_{(\beta )}(x).
\label{a19}
\end{eqnarray}

Instead in Refs.\cite{hen1,hen2,hen3,hen4} it was implicitly used the
metric ADM action $S_{ADM}[{}^4g_{\mu\nu}]$ with the metric expressed in
terms of cotetrads in the Schwinger time gauge\cite{schw}
as independent Lagrangian variables $S_{ADMT}
[{}^4E^{(\alpha )}_{\mu}]$. This is the action we shall study in this paper 
after having expressed arbitrary cotetrads in terms of $\Sigma_{\tau}$-adapted 
ones in the next Section.

Like $S_{ADM}$ is not manifestly invariant under $Diff\, M^4$, also $S_{ADMT}$ 
is not manifestly invariant under the transformations of Eqs.(\ref{a18}). 
However both theories are quasi-invariant under the gauge transformations 
generated by their first class constraints. This aspect of the theory, till 
now poorly explored due to the prevalence of the idea of general covariance, is
the fundamental one in the presymplectic approach on which our discussion is
based. A completely open point is the physical relevance of the special 
canonical Shanmugadhasan coordinate systems adapted to the constraints in
generally covariant theories, since only in these coordinate systems there is a
manifest (even if not tensorial)
identification of which are the degrees of freedom, underlying general
covariance, which are left undetermined by Einstein equations. The physical 
meaning of the so called gauge variables (conjugate to the Abelianized first
class constraints) and of the resulting Dirac's observables is an open problem 
in theories with general covariance (it points at the existence of privileged 
structures natural from the presymplectic point of view, at least for noncompact
spacetimes) on which we shall return in the next paper (no such problem exists 
with `internal' gauge invariances like in Yang-Mills theory).

In what follows we shall use the notation
$k={{c^3}\over {16\pi G}}$.

\vfill\eject

\section{$\Sigma_{\tau}$-Adapted  Tetrads and Triads.}

On $\Sigma_{\tau}$ with local coordinate system $\{ \sigma^r \}$ and Riemannian 
metric ${}^3g_{rs}$ of signature (+++) we can introduce
orthonormal frames (triads) ${}^3e_{(a)}={}^3e_{(a)}^r{{\partial}\over 
{\partial \sigma^r}}$, a=1,2,3, and coframes (cotriads) ${}^3\theta^{(a)}=
{}^3e^{(a)}_r d\sigma^r$ satisfying

\begin{eqnarray}
&&{}^3e^r_{(a)}\, {}^3g_{rs}\, {}^3e^s_{(b)}=\delta_{(a)(b)},\quad\quad
{}^3e^{(a)}_r\, {}^3g^{rs}\, {}^3e^{(b)}_s=\delta^{(a)(b)},\nonumber \\
&&{}^3e^r_{(a)}\, \delta^{(a)(b)}\, {}^3e^s_{(b)}={}^3g^{rs},\quad\quad
{}^3e_r^{(a)}\, \delta_{(a)(b)}\, {}^3e_s^{(b)}={}^3g_{rs}.
\label{II1}
\end{eqnarray}

\noindent and consider the orthonormal frame bundle $F(\Sigma_{\tau})$
over $\Sigma_{\tau}$ with structure group SO(3). See Ref.\cite{gr13} for
geometrical properties of triads.

The 3-dimensional spin connection 1-form ${}^3\omega^{(a)}_{r(b)}d\sigma^r$ is

\begin{eqnarray}
{}^3\omega^{(a)}_{r(b)}&=&{}^3\omega^{(a)}_{(c)(b)}\, {}^3e^{(c)}_r={}^3e_s
^{(a)}\, {}^3\nabla_r\, {}^3e^s_{(b)}=\nonumber \\
&=&{}^3e^{(a)}_s\, {}^3e^s_{(b) | r}={}^3e^{(a)}_s [\partial_r\, {}^3e^s_{(b)}+
{}^3\Gamma^s_{ru}\, {}^3e^u_{(b)}],\nonumber \\
&&{}\nonumber \\
{}^3\omega_{(a)(b)}&=&\delta_{(a)(c)}\, {}^3\omega^{(c)}_{r(b)} d\sigma^r=-
{}^3\omega_{(b)(a)},\quad\quad
{}^3\omega_{r(a)}={1\over 2}\epsilon_{(a)(b)(c)}\, {}^3\omega_{r(b)(c)},
\nonumber \\
{}^3\omega_{r(a)(b)}&=&\epsilon_{(a)(b)(c)}\, {}^3\omega_{r(c)}=[{\hat R}^{(c)}
{}^3\omega_{r(c)}]_{(a)(b)}=[{}^3\omega_r]_{(a)(b)},\nonumber \\
&&{}\nonumber \\
&&[{}^3e_{(a)},
{}^3e_{(b)}]=({}^3\omega^{(c)}_{(a)(b)}-{}^3\omega^{(c)}_{(b)(a)}){}^3e_{(c)},
\label{II2}
\end{eqnarray}

\noindent where $\epsilon_{(a)(b)(c)}$ is the standard Euclidean antisymmetric 
tensor and $({\hat R}^{(c)})_{(a)(b)}=\epsilon_{(a)(b)(c)}$ is the adjoint
representation of SO(3) generators.

Given vectors and covectors ${}^3V^r={}^3V^{(a)}\, {}^3e^r_{(a)}$, ${}^3V_r=
{}^3V_{(a)}\, {}^3e^{(a)}_r$, we have [remember that ${}^3\nabla_s\, {}^3e^r
_{(a)}={}^3e^r_{(b)}\, {}^3\omega^{(b)}_{s (a)}$]

\begin{eqnarray}
{}^3\nabla_s\, &{}^3V^r&={}^3V^r{}_{| s}\equiv {}^3V^{(a)}_{| s}\, 
{}^3e^r_{(a)},\nonumber \\
&&\Rightarrow {}^3V^{(a)}{}_{| s}=\partial_s\, {}^3V^{(a)}+
{}^3\omega^{(a)}_{s(b)}
\, {}^3V^{(b)}=\partial_s\, {}^3V^{(a)}+\delta^{(a)(c)}\epsilon_{(c)(b)(d)}{}^3
\omega_{s(d)}{}^3V^{(b)},\nonumber \\
{}^3\nabla_s\, &{}^3V_r&={}^3V_{r | s}={}^3V_{(a) | s} {}^3e^{(a)}_r,
\nonumber \\
&&\Rightarrow {}^3V_{(a) | s}=\partial_s\, {}^3V_{(a)}-{}^3V_{(b)}\, {}^3\omega
^{(b)}_{s(a)}=\partial_s\, {}^3V_{(a)}-{}^3V_{(b)}\delta^{(b)(c)}\, 
\epsilon_{(c)(a)(d)}{}^3\omega_{s(d)}.
\label{II3}
\end{eqnarray}

For the field strength and the curvature tensors we have

\begin{eqnarray}
{}^3\Omega^{(a)}{}_{(b)(c)(d)}&=&{}^3e_{(c)}({}^3\omega^{(a)}_{(d)(b)})-{}^3e
_{(d)}({}^3\omega^{(a)}_{(c)(b)})+\nonumber \\
&+&{}^3\omega^{(n)}_{(d)(b)}\, {}^3\omega^{(a)}_{(c)(n)}-{}^3\omega^{(n)}
_{(c)(b)}\, {}^3\omega^{(a)}_{(d)(n)}-({}^3\omega^{(n)}_{(c)(d)}-{}^3\omega
^{(n)}_{(d)(c)}){}^3\omega^{(a)}_{(a)(b)}=\nonumber \\
&=&{}^3e^{(a)}_r\, {}^3R^r{}_{stw}\, {}^3e^s_{(b)}\, {}^3e^t_{(c)}\, {}^3e^w
_{(d)},\nonumber \\
&&{}\nonumber \\
{}^3\Omega_{rs}{}^{(a)}{}_{(b)}&=&{}^3e^{(c)}_r\, {}^3e^{(d)}_s\, {}^3\Omega
^{(a)}{}_{(b)(c)(d)}={}^3R^t{}_{wrs}\, {}^3e_t^{(a)}\, {}^3e^w_{(b)}=
\nonumber \\
&=&\partial_r\, {}^3\omega^{(a)}_{s(b)} -\partial_s\, {}^3\omega^{(a)}_{r(b)} 
+{}^3\omega^{(a)}_{r(c)}\, {}^3\omega^{(c)}_{s(b)} -{}^3\omega^{(a)}_{s(c)}\,
{}^3\omega^{(c)}_{r(b)}=\nonumber \\
&=&\delta^{(a)(c)}\, {}^3\Omega_{rs(c)(b)}=\delta^{(a)(c)}\, \epsilon_{(c)(b)
(d)}\, {}^3\Omega_{rs(d)},\nonumber \\
&&{}\nonumber \\
{}^3\Omega_{rs(a)}&=&{1\over 2}\epsilon_{(a)(b)(c)}\, {}^3\Omega_{rs(b)(c)}=
\partial_r\, {}^3\omega_{s(a)}-\partial_s\, {}^3\omega_{r(a)} -\epsilon
_{(a)(b)(c)}\, {}^3\omega_{r(b)}\, {}^3\omega_{s(c)},\nonumber \\
&&{}\nonumber \\
{}^3R^r{}_{stw}&=& 
\epsilon_{(a)(b)(c)}\, {}^3e^r_{(a)}\, \delta_{(b)(n)}\, 
{}^3e^{(n)}_s\, {}^3\Omega_{tw(c)},\nonumber \\
{}^3R_{rs}&=&
\epsilon_{(a)(b)(c)}\, {}^3e^u_{(a)}\, \delta_{(b)(n)}\, {}^3e
^{(n)}_r\, {}^3\Omega_{us(c)},\nonumber \\
{}^3R&=&
\epsilon_{(a)(b)(c)}\, {}^3e^r_{(a)}\, {}^3e^s_{(b)}\, {}^3\Omega
_{rs(c)}.
\label{II4}
\end{eqnarray}

The first Bianchi identity (\ref{a5}) ${}^3R^t{}_{rsu}+{}^3R^t{}_{sur}+{}^3R^t{}
_{urs}\equiv 0$ implies the cyclic identity ${}^3\Omega_{rs(a)}\, {}^3e^s_{(a)}
\equiv 0$.

Under local SO(3) rotations R [$R^{-1}=R^t$] we have 

\begin{eqnarray}
{}^3\omega^{(a)}_{r(b)} &\mapsto& [R\, {}^3\omega_r\, R^T-R \partial_r\, R^T]
^{(a)}{}_{(b)},\nonumber \\
{}^3\Omega_{rs}{}^{(a)}{}_{(b)} &\mapsto& [R\, {}^3\Omega_{rs}\, R^T]^{(a)}
{}_{(b)}.
\label{II5}
\end{eqnarray}

Since the flat metric $\delta_{(a)(b)}$ has signature (+++), we have
${}^3V^{(a)}=\delta^{(a)(b)}\, {}^3V_{(b)}={}^3V_{(a)}$ and one can simplify 
the notations by using only lower (a) indices [${}^3e^{(a)}_r={}^3e_{(a)r}$].
For instance, we have

\begin{eqnarray}
{}^3\Gamma^u_{rs}&=&{}^3\Gamma^u_{sr}=
{1\over 2}\, {}^3e^u_{(a)} \Big[ \partial_r\, {}^3e_{(a)s}+
\partial_s\, {}^3e_{(a)r}+\nonumber \\
&+&{}^3e^v_{(a)} \Big( {}^3e_{(b)r}(\partial_s\, {}^3e_{(b)v}-\partial_v\,
{}^3e_{(b)s})+{}^3e_{(b)s}(\partial_r\, {}^3e_{(b)v}-\partial_v\, {}^3e_{(b)r})
\Big) \Big] =\nonumber \\
&=&{1\over 2}\epsilon_{(a)(b)(c)}\, {}^3e^u_{(a)}({}^3e_{(b)r}\, {}^3\omega
_{s(c)}+{}^3e_{(b)s}\, {}^3\omega_{r(c)})-{1\over 2}({}^3e_{(a)r}\partial_s\,
{}^3e^u_{(a)}+{}^3e_{(a)s}\partial_r\, {}^3e^u_{(a)}),\nonumber \\
{}^3\omega_{r(a)(b)}&=&-{}^3\omega_{r(b)(a)}=
{1\over 2}\Big[ {}^3e^s_{(a)}(\partial_r\, {}^3e_{(b)s}-
\partial_s\, {}^3e_{(b)r})+\nonumber \\
&+&{}^3e^s_{(b)}(\partial_s\, {}^3e_{(a)r}-\partial_r\, {}^3e_{(a)s})+{}^3e^u
_{(a)}\, {}^3e^v_{(b)}\, {}^3e_{(c)r}(\partial_v\, {}^3e_{(c)u}-\partial_u\,
{}^3e_{(c)v}) \Big] =\nonumber \\
&=&{1\over 2} \Big[ {}^3e_{(a)u} \partial_r\, {}^3e^u_{(b)}-{}^3e_{(b)u} 
\partial_r\, {}^3e^u_{(a)}+{}^3\Gamma^u_{rs} ({}^3e_{(a)u}\, {}^3e^s_{(b)}-
{}^3e_{(b)u}\, {}^3e^s_{(a)})\Big] ,\nonumber \\
{}^3\omega_{r(a)}&=&{1\over 2} \epsilon_{(a)(b)(c)} \Big[ {}^3e^u_{(b)}
(\partial_r\, {}^3e_{(c)u}-\partial_u\, {}^3e_{(c)r})+\nonumber \\
&+&{1\over 2}\, {}^3e^u_{(b)}\, {}^3e^v_{(c)}\, {}^3e_{(d)r}(\partial_v\, 
{}^3e_{(d)u}-\partial_u\, {}^3e_{(d)v})\Big] ,\nonumber \\
{}^3\Omega_{rs(a)}&=&{1\over 2} \epsilon_{(a)(b)(c)} \Big[ \partial_r\, 
{}^3e^u_{(b)}\partial_s\, {}^3e_{(c)u}-\partial_s\, {}^3e^u_{(b)}\partial_r\,
{}^3e_{(c)u}+\nonumber \\
&+&{}^3e^u_{(b)}(\partial_u\partial_s\, {}^3e_{(c)r}-\partial_u\partial_r\,
{}^3e_{(c)s})+\nonumber \\
&+&{1\over 2}\Big( {}^3e^u_{(b)}\, {}^3e^v_{(c)}(\partial_r\, {}^3e_{(d)s}-
\partial_s\, {}^3e_{(d)r})(\partial_v\, {}^3e_{(d)u}-\partial_u\, {}^3e_{(d)v})
+\nonumber \\
&+&({}^3e_{(d)s}\partial_r-{}^3e_{(d)r}\partial_s)[{}^3e^u_{(b)}\, {}^3e^v
_{(c)}(\partial_v\, {}^3e_{(d)u}-\partial_u\, {}^3e_{(d)v})]\Big) \Big] -
\nonumber \\
&-&{1\over 8}[\delta_{(a)(b_1)}\epsilon_{(c_1)(c_2)(b_2)}+\delta_{(a)(b_2)}
\epsilon_{(c_1)(c_2)(b_1)}+\delta_{(a)(c_1)}\epsilon_{(b_1)(b_2)(c_2)}+\delta
_{(a)(c_2)}\epsilon_{(b_1)(b_2)(c_1)}]\times \nonumber \\
&&{}^3e^{u_1}_{(b_1)}\, {}^3e^{u_2}_{(b_2)}\Big[ (\partial_r\, {}^3e_{(c_1)u_1}
-\partial_{u_1}\, {}^3e_{(c_1)r})(\partial_s\, {}^3e_{(c_2)u_2}-\partial_{u_2}\,
{}^3e_{(c_2)s})+\nonumber \\
&+&{1\over 2}\Big( {}^3e^{v_2}_{(c_2)}\, {}^3e_{(d)s}(\partial_r\, {}^3e
_{(c_1)u_1}-\partial_{u_1}\, {}^3e_{(c_1)r})(\partial_{v_2}\, {}^3e_{(d)u_2}-
\partial_{u_2}\, {}^3e_{(d)v_2})+\nonumber \\
&+&{}^3e^{v_1}_{(c_1)}\, {}^3e_{(d)r}(\partial_s\, {}^3e_{(c_2)u_2}-\partial
_{u_2}\, {}^3e_{(c_2)s})(\partial_{v_1}\, {}^3e_{(d)u_1}-\partial_{u_1}\,
{}^3e_{(d)v_1})\Big) +\nonumber \\
&+&{1\over 4}\, {}^3e^{v_1}_{(c_1)}\, {}^3e^{v_2}_{(c_2)}\, {}^3e_{(d_1)r}\, 
{}^3e_{(d_2)s}(\partial_{v_1}\, {}^3e_{(d_1)u_1}-\partial_{u_1}\, {}^3e
_{(d_1)v_1})(\partial_{v_2}\, {}^3e_{(d_2)u_2}-\partial_{u_2}\, {}^3e
_{(d_2)v_2}) \Big] ,\nonumber \\
{}^3\Omega_{rs(a)(b)}&=&\epsilon_{(a)(b)(c)}\, {}^3\Omega_{rs(c)},\nonumber \\
{}^3R_{rsuv}&=&\epsilon_{(a)(b)(c)}\, {}^3e_{(a)r}\, {}^3e_{(b)s}\, {}^3\Omega
_{uv(c)},\nonumber \\
{}^3R_{rs}&=&{1\over 2} \epsilon_{(a)(b)(c)}\, {}^3e^u_{(a)} \Big[ {}^3e_{(b)r}
\, {}^3\Omega_{us(c)}+{}^3e_{(b)s}\, {}^3\Omega_{ur(c)}\Big] ,\nonumber \\
{}^3R&=&\epsilon_{(a)(b)(c)}\, {}^3e^r_{(a)}\, {}^3e^s_{(b)}\, {}^3\Omega
_{rs(c)}.
\label{II5a}
\end{eqnarray}

In the family of $\Sigma_{\tau}$-adapted frames and coframes on $M^4$, we can
select special tetrads and cotetrads ${}^4_{(\Sigma )}{\check E}_{(\alpha )}$
and ${}^4_{(\Sigma )}{\check \theta}^{(\alpha )}$ also adapted to a given
set of triads and cotriads on $\Sigma_{\tau}$

\begin{eqnarray}
{}^4_{(\Sigma )}{\check E}^{\mu}_{(\alpha )}&=&\lbrace {}^4_{(\Sigma )}{\check 
E}^{\mu}_{(o)}=l^{\mu}= {\hat b}^{\mu}_l={1\over N}(b^{\mu}
_{\tau}-N^rb^{\mu}_r);
\,\, {}^4_{(\Sigma )}{\check E}^{\mu}_{(a)}={}^3e^s_{(a)}
b^{\mu}_s \rbrace ,\nonumber \\
{}^4_{(\Sigma )}{\check E}_{\mu}^{(\alpha )}&=&\lbrace {}^4_{(\Sigma )}{\check
E}_{\mu}^{(o)}=\epsilon l_{\mu}= {\hat b}^l_{\mu}= N b^{\tau}
_{\mu};\,\, {}^4_{(\Sigma )}{\check E}_{\mu}^{(a)}={}^3e_s^{(a)}
{\hat b}^s_{\mu}\rbrace ,\nonumber \\
&&{}\nonumber \\
{}^4_{(\Sigma )}{\check E}^{\mu}_{(\alpha )}&& {}^4g_{\mu\nu}\,\,\, 
{}^4_{(\Sigma )}{\check E}^{\nu}_{(\beta )} = {}^4\eta_{(\alpha )(\beta )},
\label{II6}
\end{eqnarray}

\noindent where $b^{\mu}_r$ and $b_{\mu}^r$ are defined in Eqs.(\ref{I1}).
The components of these tetrads and cotetrads in the holonomic bases are
(\cite{dirr,hen1}; ${}^4_{(\Sigma )}{\check {\tilde E}}^{(o)}_r=0$ is the
Schwinger time gauge condition\cite{schw})

\begin{eqnarray}
{}^4_{(\Sigma )}{\check {\tilde E}}^A_{(\alpha )}&=&{}^4_{(\Sigma )}{\check E}
^{\mu}_{(\alpha )}\, b^A_{\mu},\quad \Rightarrow {}^4_{(\Sigma )}{\check 
{\tilde E}}^A_{(o)}=\epsilon l^A,\nonumber \\
&&{}^4_{(\Sigma )}{\check {\tilde E}}^{\tau}_{(o)}={1\over N},\quad\quad
{}^4_{(\Sigma )}{\check {\tilde E}}^{\tau}_{(a)}=0,\nonumber \\
&&{}^4_{(\Sigma )}{\check {\tilde E}}^r_{(o)}=-{{N^r}\over N},\quad\quad
{}^4_{(\Sigma )}{\check {\tilde E}}^r_{(a)}={}^3e^r_{(a)};\nonumber \\
{}^4_{(\Sigma )}{\check {\tilde E}}_A^{(\alpha )}&=&{}^4_{(\Sigma )}{\check E}
_{\mu}^{(\alpha )}\, b^{\mu}_A,\quad \Rightarrow {}^4_{(\Sigma )}{\check 
{\tilde E}}_A^{(o)} = l_A,\nonumber \\
&&{}^4_{(\Sigma )}{\check {\tilde E}}^{(o)}_{\tau}=N,\quad\quad
{}^4_{(\Sigma )}{\check {\tilde E}}^{(a)}_{\tau}=N^r\, {}^3e^{(a)}_r=N^{(a)},
\nonumber \\
&&{}^4_{(\Sigma )}{\check {\tilde E}}^{(o)}_r=0,\quad\quad
{}^4_{(\Sigma )}{\check {\tilde E}}^{(a)}_r={}^3e^{(a)}_r,\nonumber \\
&&{}\nonumber \\
&&{}^4_{(\Sigma )}{\check E}^A_{(\alpha )}\, {}^4g_{AB}\, {}^4_{(\Sigma 
)}{\check E}^B_{(\beta )}={}^4\eta_{(\alpha )(\beta )}.
\label{II7}
\end{eqnarray}

With the cotetrads ${}^4_{(\Sigma )}{\check E}^{(\alpha )}_{\mu}(z(\sigma ))$
we can build the vector ${\buildrel \circ
\over V}^{(\alpha )}=l^{\mu}(z(\sigma ))\, {}^4_{(\Sigma )}{\check E}^{(\alpha 
)}_{\mu}(z(\sigma ))= (1 ;\vec 0)$: it is the same unit timelike
future-pointing Minkowski 4-vector in the tangent plane of each point
$z^{\mu}(\sigma )=z^{\mu}(\tau ,\vec \sigma )\in \Sigma_{\tau}\subset M^4$
for every $\tau$ and $\vec \sigma$; we have ${\buildrel \circ \over V}
^{(\alpha )}\, {}^4\eta_{(\alpha )(\beta )}\, {\buildrel \circ \over V}
^{(\beta )}=\epsilon$.

Let ${}^4E^{\mu}_{(\alpha )}(z)$ and ${}^4E^{(\alpha )}_{\mu}(z)$ be arbitrary 
tetrads and cotetrads on $M^4$. Let us define the point-dependent Minkowski
4-vector $V^{(\alpha )}(z(\sigma ))=l^{\mu}(z(\sigma ))\, {}^4E^{(\alpha )}
_{\mu}(z(\sigma ))$ (assumed to be future-pointing), which satisfies
$V^{(\alpha )}(z(\sigma ))\, {}^4\eta_{(\alpha )(\beta )}\, V^{(\beta )}
(z(\sigma ))=\epsilon$, so that $V^{(\alpha )}(z(\sigma ))=(
V^{(o)}(z(\sigma ))=+\sqrt{1+\sum_rV^{(r) 2}(z(\sigma ))}; V^{(r)}(z(\sigma ))
{\buildrel {def}\over =}\, \varphi^{(r)}(\sigma 
) )$ : therefore, the point-dependent Minkowski 4-vector $V^{(\alpha )}
(z(\sigma ))$ depends only on the three functions $\varphi^{(r)}(\sigma 
)$ [one has $\varphi^{(r)}(\sigma )=-\epsilon \varphi_{(r)}(\sigma )$ since
${}^4\eta_{rs}=-\epsilon \, \delta_{rs}$; having the Euclidean signature (+++)
for both $\epsilon =\pm 1$, we shall define the Kronecker delta as $\delta
^{(i)(j)}=\delta^{(i)}_{(j)}=\delta_{(i)(j)}$].
If we introduce the point-dependent Lorentz transformation

\begin{eqnarray}
L^{(\alpha )}{}_{(\beta )}(V(z(\sigma ));{\buildrel \circ \over V})&=&\delta
^{(\alpha )}_{(\beta )}+2\epsilon V^{(\alpha )}(z(\sigma )){\buildrel \circ
\over V}_{(\beta )}-\epsilon { {(V^{(\alpha )}(z(\sigma ))+{\buildrel \circ
\over V}^{(\alpha )})(V_{(\beta )}(z(\sigma ))+{\buildrel \circ \over V}
_{(\beta )})}\over {1+V^{(o)}(z(\sigma ))}}=\nonumber \\
&=& \left( \begin{array}{cc} V^{(o)} & -\epsilon V_{(j)} \\
V^{(i)} & \delta^{(i)}_{(j)}-\epsilon {{V^{(i)}V_{(j)}}\over {1+
V^{(o)} }} \end{array} \right) (z(\sigma )),
\label{II8}
\end{eqnarray}

\noindent which is the standard Wigner boost for timelike Poincar\'e orbits
[see Ref.\cite{longhi}], one has by construction

\begin{equation}
V^{(\alpha )}(z(\sigma ))=l^{\mu}(z(\sigma ))\, {}^4E^{(\alpha )}_{\mu}
(z(\sigma ))=L^{(\alpha )}{}_{(\beta )}(V(z(\sigma ));{\buildrel \circ \over
V})\, {\buildrel \circ \over V}^{(\beta )}.
\label{II9}
\end{equation}

Therefore, we shall define an arbitrary cotretad ${}^4E^{(\alpha )}
_{\mu}(z(\sigma ))$ on $M^4$ starting from the special $\Sigma_{\tau}$- and 
cotriad-adapted cotetrad ${}^4_{(\Sigma )}{\check E}^{(\alpha )}_{\mu}
(z(\sigma ))$ by means of the formula

\begin{equation}
{}^4E^{(\alpha )}_{\mu}(z(\sigma ))=L^{(\alpha )}{}_{(\beta )}(V(z(\sigma ));
{\buildrel \circ \over V})\, {}^4_{(\Sigma )}{\check E}^{(\beta )}_{\mu}
(z(\sigma )).
\label{II10}
\end{equation}

\noindent Let us remark that with this definition we are putting equal to zero,
by convention, the angles of an arbitrary 3-rotation of $b^s_{\mu}(z(\sigma ))$
[i.e. of the choice of the three axes tangent to $\Sigma_{\tau}$] inside
${}^4_{(\Sigma )}{\check E}^{(\alpha )}_{\mu}(z(\sigma ))$.

Since $\varphi^{(a)}(\sigma )=V^{(a)}(z(\sigma ))=l^{\mu}(z(\sigma ))\,
{}^4E^{(a)}_{\mu}(z(\sigma ))$ are the three parameters of the Wigner
boost [$\varphi^{(a)}=\bar \gamma\beta^{(a)}$, $\bar \gamma 
=\sqrt{1+\sum_{(c)}\varphi
^{(c) 2}}$, $\beta^{(a)}=\varphi^{(a)}/ \sqrt{1+\sum_{(c)}\varphi^{(c) 2}}$],
the previous equation can be rewritten in the following form [remembering
that $\varphi^{(a)}=-\epsilon \varphi_{(a)}$]
 
\begin{equation}
\left( \begin{array}{l} {}^4E^{(o)}_{\mu}\\ {}^4E^{(a)}_{\mu} \end{array}
\right) (z(\sigma ))=\left( \begin{array}{cc}  \sqrt{1+\sum_{(c)}
\varphi^{(c) 2}} &-\epsilon \varphi_{(b)}\\  \varphi^{(a)} &
\delta^{(a)}_{(b)}-\epsilon {{\varphi^{(a)}\varphi_{(b)} }\over {1+
\sqrt{1+\sum_{(c)}\varphi^{(c) 2}} }} \end{array} \right) (z(\sigma ))
\left( \begin{array}{l} l_{\mu} \\ {}^3e^{(b)}_s\, b^s_{\mu} \end{array}
\right) (\sigma ).
\label{II11}
\end{equation}

If we go to holonomic bases, ${}^4E^{(\alpha )}_A(z(\sigma ))={}^4E^{(\alpha )}
_{\mu}(z(\sigma ))\, b^{\mu}_A(\sigma )$ and ${}^4_{(\Sigma )}{\check {\tilde
E}}^{(\alpha )}_A(z(\sigma ))={}^4_{(\Sigma )}{\check E}^{(\alpha )}_{\mu}
(z(\sigma ))\, b^{\mu}_A(\sigma )$, one has

\begin{eqnarray}
\left( \begin{array}{l} {}^4E^{(o)}_A \\ {}^4E^{(a)}_A \end{array} \right) &=&
\left( \begin{array}{cc}  \sqrt{1+\sum_{(c)}
\varphi^{(c) 2}} &-\epsilon \varphi_{(b)}\\  \varphi^{(a)} &
\delta^{(a)}_{(b)}-\epsilon {{\varphi^{(a)}\varphi_{(b)} }\over {1+
\sqrt{1+\sum_{(c)}\varphi^{(c) 2}} }} \end{array} \right) \times\nonumber \\
&&\left( \begin{array}{l} {}^4_{(\Sigma )}{\check {\tilde E}}^{(o)}_A=(N;\vec 
0)\\ {}^4_{(\Sigma )}{\check {\tilde E}}^{(b)}_A=(N^{(b)}={}^3e^{(b)}_rN^r; 
{}^3e^{(b)}_r) \end{array} \right) ,
\label{II12}
\end{eqnarray}

\noindent so that we get that the cotetrad in holonomic basis can be
expressed in terms of N, $N^{(a)}={}^3e^{(a)}_sN^s=N_{(a)}$, $\varphi^{(a)}$ and
${}^3e^{(a)}_r$ [${}^3g_{rs}=\sum_{(a)} {}^3e_{(a)r}\, {}^3e_{(a)s}$]

\begin{eqnarray}
&&{}^4E^{(o)}_{\tau}(z(\sigma ))= \sqrt{1+\sum_{(c)}\varphi^{(c) 2}
(\sigma )}\, N(\sigma )+\sum_{(a)}\varphi^{(a)}(\sigma ) N^{(a)}(\sigma ),
\nonumber \\
&&{}^4E^{(o)}_r(z(\sigma ))= \sum_{(a)}\varphi^{(a)}(\sigma )\,
{}^3e^{(a)}_r(\sigma ),\nonumber \\
&&{}^4E^{(a)}_{\tau}(z(\sigma ))= \varphi^{(a)}(\sigma )N(\sigma )+
\sum_{(b)}[\delta^{(a)}_{(b)}-\epsilon {{\varphi^{(a)}(\sigma )\varphi_{(b)}
(\sigma )}\over {1+\sqrt{1+\sum_{(c)}\varphi^{(c) 2}(\sigma )} }}]N^{(b)}
(\sigma ),\nonumber \\
&&{}^4E^{(a)}_r(z(\sigma ))=\sum_{(b)}[\delta^{(a)}_{(b)}-\epsilon   
{{\varphi^{(a)}(\sigma )\varphi_{(b)}
(\sigma )}\over {1+\sqrt{1+\sum_{(c)}\varphi^{(c) 2}(\sigma )} }}] {}^3e
^{(b)}_r(\sigma ),\nonumber \\
&&{}\nonumber \\
&&\Rightarrow {}^4g_{AB}={}^4E^{(\alpha )}_A\, {}^4\eta_{(\alpha )(\beta )}\,
{}^4E^{(\beta )}_B=
{}^4_{(\Sigma )}{\check E}^{(\alpha )}_A\, {}^4\eta_{(\alpha )(\beta )}\,
{}^4_{(\Sigma )}{\check E}^{(\beta )}_B=\nonumber \\
&=&\epsilon \left( \begin{array}{cc} (N^2- {}^3g_{rs}N^rN^s) &
-{}^3g_{st}N^t\\ -{}^3g_{rt}N^t & -{}^3g
_{rs} \end{array} \right) ,
\label{II13}
\end{eqnarray}

\noindent with the last line in accord with Eqs.(\ref{I1}); we have used
$L^T {}^4\eta L={}^4\eta$, valid for every Lorentz transformation. We find 
$L^{-1}(V,{\buildrel \circ \over V})={}^4\eta L^T(V,{\buildrel \circ \over V})
{}^4\eta=L(V,{\buildrel \circ \over V}){|}_{\varphi^{(a)}\mapsto -\varphi^{(a)}
}$ and [${}^4E^A_{(\alpha )}={}^4E^{\mu}_{(\alpha )}b^A_{\mu}$,
${}^4_{(\Sigma )}{\check {\tilde E}}^A_{(\alpha )}={}^4_{(\Sigma )}{\check E}
^{\mu}_{(\alpha )}b^A_{\mu}$]

\begin{eqnarray}
&&\left( \begin{array}{l} {}^4E^{\mu}_{(o)}\\ {}^4E^{\mu}_{(a)}\end{array}
\right) =\left( \begin{array}{cc}  \sqrt{1+\sum_{(c)}
\varphi^{(c) 2}} &- \varphi^{(b)}\\ \epsilon \varphi_{(a)} &
\delta_{(a)}^{(b)}-\epsilon {{\varphi_{(a)}\varphi^{(b)} }\over {1+
\sqrt{1+\sum_{(c)}\varphi^{(c) 2}} }} \end{array} \right) \,
\left( \begin{array}{l} l^{\mu} \\ b^{\mu}_s\, {}^3e^s_{(b)} \end{array}
\right) , \nonumber \\
&&\left( \begin{array}{l} {}^4E^A_{(o)} \\ {}^4E^A_{(a)} \end{array} \right)
=\left( \begin{array}{cc}  \sqrt{1+\sum_{(c)}
\varphi^{(c) 2}} &- \varphi^{(b)}\\ \epsilon \varphi^{(a)} &
\delta_{(a)}^{(b)}-\epsilon {{\varphi_{(a)}\varphi^{(b)} }\over {1+
\sqrt{1+\sum_{(c)}\varphi^{(c) 2}} }} \end{array} \right) \,
\left( \begin{array}{l} {}^4_{(\Sigma )}{\check {\tilde E}}^A_{(o)}=(1/N;
-N^r/N) \\ {}^4_{(\Sigma )}{\check {\tilde E}}^A_{(b)}=(0; {}^3e^r_{(b)})
\end{array} \right) ,\nonumber \\
&&{}\nonumber \\
&&{}^4E^{\tau}_{(o)}(z(\sigma ))= \sqrt{1+\sum_{(c)}\varphi^{(c) 2}
(\sigma )} {1\over {N(\sigma )}},\nonumber \\
&&{}^4E^r_{(o)}(z(\sigma )=- \sqrt{1+\sum_{(c)}\varphi^{(c) 2}
(\sigma )} {{N^r(\sigma )}\over {N(\sigma )}}-\varphi^{(b)}(\sigma )
\, {}^3e^r_{(b)}(\sigma ),\nonumber \\
&&{}^4E^{\tau}_{(a)}(z(\sigma ))=\epsilon {{\varphi_{(a)}(\sigma )}\over
{N(\sigma )}},\nonumber \\
&&{}^4E^r_{(a)}(z(\sigma ))=-\epsilon \varphi_{(a)}(\sigma ) {{N^r(\sigma )}
\over {N(\sigma )}}+\sum_{(b)}[\delta_{(a)}^{(b)}-\epsilon
{{\varphi_{(a)}(\sigma )\varphi^{(b)}(\sigma )}\over {1+\sqrt{1+\sum_{(c)}
\varphi^{(c) 2}(\sigma )}}}]\, {}^3e^r_{(b)}(\sigma ),\nonumber \\
&&{}\nonumber \\
&&\Rightarrow {}^4g^{AB}={}^4E^A_{(\alpha )}\, {}^4\eta^{(\alpha )(\beta )}\, 
{}^4E^B_{(\beta )}=
{}^4_{(\Sigma )}{\check E}_{(\alpha )}^A\, {}^4\eta_{(\alpha )(\beta )}\,
{}^4_{(\Sigma )}{\check E}_{(\beta )}^B=\nonumber \\
&=&\epsilon \left( \begin{array}{cc} {1\over {N^2}} & -
{{N^s}\over {N^2}} \\ - {{N^r}\over {N^2}} & 
- ({}^3g^{rs}-{{N^rN^s}\over {N^2}})\end{array} \right) ,
\label{II14}
\end{eqnarray}

\noindent with the last line in accord with Eqs.(\ref{I1}).

From ${}^4_{(\Sigma )}{\check {\tilde E}}^{(\alpha )}_A(z(\sigma ))=(L^{-1})
^{(\alpha )}{}_{(\beta )}(V(z(\sigma ));{\buildrel \circ \over V})\, {}^4E
^{(\beta )}_A(z(\sigma ))$ and ${}^4_{(\Sigma )}{\check {\tilde E}}^A_{(\alpha 
)}(z(\sigma ))={}^4E^A_{(\beta )}\, (L^{-1})^{(\beta )}{}_{(\alpha )}
(V(z(\sigma ));{\buildrel \circ \over V})$ it turns out\cite{longhi} that the
flat indices $(a)$ of the adapted tetrads ${}^4_{(\Sigma )}{\check E}^{\mu}
_{(a)}$ and of the triads ${}^3e^r_{(a)}$ and cotriads ${}^3e_r^{(a)}$ on 
$\Sigma_{\tau}$ transform as Wigner spin 1 indices under point-dependent
SO(3) Wigner rotations $R^{(a)}{}_{(b)}(V(z(\sigma ));\Lambda (z(\sigma )))$
associated with Lorentz transformations $\Lambda^{(\alpha )}{}_{(\beta )}(z)$ 
in the tangent plane to $M^4$ in the same point [$R^{(\alpha )}{}_{(\beta )}
(\Lambda (z(\sigma ));V(z(\sigma )))=[L({\buildrel \circ \over V};V(z(\sigma ))
)\, \Lambda^{-1}(z(\sigma ))\, L(\Lambda (z(\sigma ))V(z(\sigma ));{\buildrel
\circ \over V})]^{(\alpha )}{}_{(\beta )}=\left( \begin{array}{cc} 1&0\\
0& R^{(a)}{}_{(b)}(V(z(\sigma ));\Lambda (z(\sigma )))\end{array} \right)$]. 
Instead the index ${(o)}$ of the adapted tetrads ${}^4_{(\Sigma )}{\check E}
^{\mu}_{(o)}$ is a local Lorentz scalar in each point. Therefore, the adapted
tetrads in the holonomic basis should be denoted as ${}^4_{(\Sigma )}{\check 
{\tilde E}}^A_{(\bar \alpha )}$, with $(\bar o)$ and $A=(\tau ,r)$ Lorentz 
scalar indices and with $(\bar a)$ Wigner spin 1 indices; we shall go on
with the indices $(o),(a)$ without the overbar for the sake of simplicity.
In this way the tangent planes to $\Sigma_{\tau}$ in $M^4$ are described in
a Wigner covariant way, reminiscent of the flat rest-frame covariant instant
form of dynamics introduced in Minkowski spacetime in Ref.\cite{lus1}. 
Similar conclusions are reached independently
in Ref.\cite{ltt} in the framework of nonlinear Poincar\'e gauge theory
[the vector fields $e_{\alpha}$ and the 1-forms $\theta^{\alpha}$ of that
paper correspond to $X_{\tilde A}$ and $\theta^{\tilde A}$ in
Eq.(\ref{I3}) respectively].

Therefore, an arbitrary tetrad field, namely a (in general nongeodesic)
congruence of observers' timelike worldlines with 4-velocity field $u^A(\tau 
,\vec \sigma )={}^4E^A_{(o)}(\tau ,\vec \sigma )$, can be obtained with a
pointwise Wigner boost from the special surface-forming timelike congruence
whose 4-velocity field is the normal to $\Sigma_{\tau}$ $l^A(\tau ,\vec \sigma )
=\epsilon \, {}^4_{(\Sigma )}{\check {\tilde E}}^A_{(o)}(\tau ,\vec \sigma )$
[it is associated with the 3+1 splitting of $M^4$ with leaves $\Sigma_{\tau}$;
see Appendix A].

We can invert Eqs.(\ref{II14}) to get N, $N^r={}^3e^r_{(a)}N^{(a)}$, $\varphi
^{(a)}$ and ${}^3e^r_{(a)}$ in terms of the tetrads ${}^4E^A_{(\alpha )}$

\begin{eqnarray}
N&=&  \frac{1}{ \sqrt{[{}^4E^{\tau}_{(o)}]^2 
          -\sum_{(c)} [{}^4E^{\tau}_{(c)}]^2}}.
 \nonumber \\
N^r&=& - \frac{{}^4E^{\tau}_{(o)}\, {}^4E^{r}_{(0)} -\sum_{(c)}{}^4E^{\tau}
_{(c)}\, {}^4E^r_{(c)}}
{[{}^4E^{\tau}_{(0)}]^2-\sum_{(c)}[{}^4E^{\tau}_{(c)}]^2 }\nonumber \\
\varphi_{(a)}&=&\frac{\epsilon~ {}^4E^{\tau}_{(a)}  }{
          \sqrt{[{}^4E^{\tau}_{(o)}]^2 
          -\sum_{(c)} [{}^4E^{\tau}_{(c)}]^2}}\nonumber\\
{}^3e^r_{(a)}&=&\sum_{(b)} B_{(a)(b)}
    \big( {}^4E^r_{(b)} + N^r~ {}^4E^\tau_{(b)} \big)
\nonumber \\
&&{}\nonumber \\
&& B_{(a)(b)} = \delta_{(a)(b)}
    -\frac{ {}^4E^{\tau}_{(a)} {}^4E^{\tau}_{(b)}  }{
           {}^4E^{\tau}_{(0)} \big[  {}^4E^{\tau}_{(0)} 
       + \sqrt{[{}^4E^{\tau}_{(0)}]^2 
          -\sum_{(c)} [{}^4E^{\tau}_{(c)}]^2}] }.
\label{II15}
\end{eqnarray}

If ${}^3e^{-1}=det\, ({}^3e^r_{(a)})$, then from the orthonormality condition
we get ${}^3e_{(a)r}= {}^3e ({}^3e^s_{(b)}\, {}^3e^t_{(c)}-{}^3e^t_{(b)}\,
{}^3e^s_{(c)})$ [with $(a),(b),(c)$ and $r,s,t$ cyclic] and it allows to 
express the cotriads in terms of the tetrads ${}^4E^A_{(\alpha )}$. Therefore,
given the tetrads ${}^4E^A_{(\alpha )}$ [or equivalently the cotetrads 
${}^4E_A^{(\alpha )}$] on $M^4$, an equivalent set of variables with the
local Lorentz covariance replaced with local Wigner covariance are the lapse
N, the shifts $N^{(a)}=N_{(a)}={}^3e_{(a)r}N^r$, the Wigner-boost parameters 
$\varphi^{(a)}=-\epsilon \varphi_{(a)}$ and either the triads ${}^3e^r_{(a)}$ 
or the cotriads ${}^3e_{(a)r}$.

In Appendix A there is the expression in terms of the variables $N$, $N_{(a)}$,
$\varphi_{(a)}$ and ${}^3e_{(a)r}$ [and/or ${}^3e^r_{(a)}$] of the connection
coefficients ${}^4\Gamma^B_{AC}$, of the spin connection ${}^4\omega
_{A(\alpha )(\beta)}$, of the field strength ${}^4\Omega_{AB(\alpha )(\beta )}$,
of the Riemann tensor ${}^4R^A{}_{BCD}$ and of the Weyl tensor ${}^4C^A{}_{BCD}$
in the $\Sigma_{\tau}$-adapted holonomic coordinate basis, where the 4-metric is
${}^4g_{AB}$. These formulas give the bridge to the reconstruction of the
spacetime $M^4$ starting from the ADM tetrad description and show explicitly
the dependence of 4-tensors on the undetermined lapse and shift functions.

\vfill\eject

\section
{The Lagrangian and the Hamiltonian in the New Variables.}

Let us consider the ADM action (\ref{I10}) $S_{ADM}$;
its independent variables in metric gravity have now the following expression
in terms of N, $N^{(a)}=N_{(a)}={}^3e^r_{(a)}N_r$, $\varphi^{(a)}=-\epsilon 
\varphi_{(a)}$, ${}^3e^{(a)}_r={}^3e_{(a)r}$ [$\gamma =det\, ({}^3g_{rs})= 
({}^3e)^2=(det\, (e_{(a)r}))^2$]

\begin{eqnarray}
N,&&{}\quad\quad
N_r={}^3e_r^{(a)}N_{(a)}={}^3e_{(a)r}N_{(a)},\nonumber \\
{}^3g_{rs}&=&{}^3e^{(a)}_r\, \delta_{(a)(b)}\, {}^3e^{(b)}_s={}^3e_{(a)r}\, 
{}^3e_{(a)s},
\label{III1}
\end{eqnarray}

\noindent so that the line element of $M^4$ becomes\hfill\break
\hfill\break 
$ds^2=\epsilon (N^2-
N_{(a)}N_{(a)})(d\tau )^2-2\epsilon N_{(a)}\, {}^3e_{(a)r} d\tau d\sigma^r-
\epsilon \, {}^3e_{(a)r}\, {}^3e_{(a)s} d\sigma^rd\sigma^s=\epsilon \Big[ N^2
(d\tau )^2-({}^3e_{(a)r}d\sigma^r +N_{(a)}d\tau )({}^3e_{(a)s}d\sigma^s 
+N_{(a)}d\tau )\Big]$.\hfill\break
\hfill\break
The extrinsic curvature takes the form
[$N_{(a)|r}={}^3e^s_{(a)}N_{s|r}=
\partial_r N_{(a)}-\epsilon_{(a)(b)(c)}\, {}^3\omega_{r(b)}N_{(c)}$ from
Eq.(\ref{II3})]

\begin{eqnarray}
{}^3K_{rs}&=& {\hat b}^{\mu}_r{\hat b}^{\nu}_s {}^3K_{\mu\nu}={1\over {2N}}
(N_{r | s}+N_{s | r}-\partial_{\tau} {}^3g_{rs})=\nonumber \\
&=&{1\over {2N}} ({}^3e_{(a)r} \delta^w_s+{}^3e_{(a)s}\delta^w_r)(N_{(a) | w}-
\partial_{\tau}\, {}^3e_{(a)w}),\nonumber \\
{}^3K_{r(a)}&=&{}^3K_{rs}\, {}^3e^s_{(a)}={1\over {2N}}(\delta_{(a)(b)}\delta
^w_r+{}^3e^w_{(a)}\, {}^3e_{(b)r})(N_{(b)|w}-\partial_{\tau}\, {}^3e_{(c)w}),
\nonumber \\
{}^3K&=&{1\over N}\, {}^3e^r_{(a)} (N_{(a)|r}-\partial_{\tau}\, {}^3e_{(a)r}),
\label{III2}
\end{eqnarray}

\noindent so that the ADM action in the new variables is 

\begin{eqnarray}
{\hat S}_{ADMT}&=&\int d\tau {\hat L}_{ADMT}=\nonumber \\
&=&-\epsilon k \int d\tau d^3\sigma \lbrace N\, {}^3e\, \epsilon_{(a)(b)(c)}\,
{}^3e^r_{(a)}\, {}^3e^s_{(b)}\, {}^3\Omega_{rs(c)}+\nonumber \\
&+&{{{}^3e}\over {2N}} ({}^3G_o^{-1})_{(a)(b)(c)(d)} {}^3e^r_{(b)}(N_{(a) | r}-
\partial_{\tau}\, {}^3e_{(a)r})\, {}^3e^s_{(d)}(N_{(c) | s}-\partial_{\tau}
\, {}^3e_{(c) \ s})\rbrace,
\label{III3}
\end{eqnarray}

\noindent where we introduced the flat (with lower indices) inverse 
Wheeler-DeWitt supermetric

\begin{equation}
({}^3G_o^{-1})_{(a)(b)(c)(d)}=\delta_{(a)(c)}\delta_{(b)(d)}+\delta_{(a)(d)}
\delta_{(b)(c)}-2\delta_{(a)(b)}\delta_{(c)(d)}.
\label{III4}
\end{equation}

\noindent The flat supermetric is

\begin{eqnarray}
{}^3G_{o(a)(b)(c)(d)}&=&{}^3G_{o(b)(a)(c)(d)}={}^3G_{o(a)(b)(d)(c)}=
{}^3G_{o(c)(d)(a)(b)}=\nonumber \\
&=&\delta_{(a)(c)}\delta_{(b)(d)}+\delta_{(a)(d)}\delta_{(b)(c)}-\delta
_{(a)(b)}\delta_{(c)(d)},\nonumber \\
&&{}\nonumber \\
&&{1\over 2}\, {}^3G_{o(a)(b)(e)(f)}\, {1\over 2}\, {}^3G^{-1}_{o(e)(f)(c)(d)}
={1\over 2}[\delta_{(a)(c)}\delta_{(b)(d)}+\delta_{(a)(d)}\delta_{(b)(c)}].
\label{III5}
\end{eqnarray}

The new action does not depend on the 3 boost variables $\varphi^{(a)}$
[like the Higgs model Lagrangian in the unitary gauge does not depend on some 
of the Higgs fields\cite{lv1,lv2}],
contains lapse N and modified shifts $N_{(a)}$ as Lagrange multipliers, and is a
functional independent from the second time derivatives of the
fields. The canonical momenta and the Poisson brackets are

\begin{eqnarray}
&&{\tilde \pi}^{\vec \varphi}_{(a)}(\tau ,\vec \sigma )={{\delta {\hat S}
_{ADMT}}\over {\delta \partial_{\tau} \varphi_{(a)}(\tau ,\vec \sigma )}}=0,
\nonumber \\
&&{\tilde \pi}^N(\tau ,\vec \sigma )={{\delta {\hat S}_{ADMT}}\over {\delta
\partial_{\tau} N(\tau ,\vec \sigma )}}=0,\nonumber \\
&&{\tilde \pi}^{\vec N}_{(a)}(\tau ,\vec \sigma )={{\delta {\hat S}_{ADMT}}
\over {\delta \partial_{\tau} N_{(a)}(\tau ,\vec \sigma )}}=0,\nonumber \\
&&{}^3{\tilde \pi}^r_{(a)}(\tau ,\vec \sigma )={{\delta {\hat S}_{ADMT}}
\over {\delta \partial_{\tau} {}^3e_{(a)r}(\tau ,\vec \sigma )}}= 
[{{\epsilon k{}^3e}\over
N} ({}^3G^{-1}_o)_{(a)(b)(c)(d)}\, {}^3e_{(b)}^r\,
{}^3e^s_{(d)}\, (N_{(c) | s}-\partial_{\tau}\, {}^3e_{(c)s})](\tau ,\vec 
\sigma )=\nonumber \\
&&=2\epsilon k [{}^3e ({}^3K^{rs}-{}^3e^r_{(c)}\, {}^3e^s_{(c)}\, {}^3K)
{}^3e_{(a)s}](\tau ,\vec \sigma ),\nonumber \\
&&{}\nonumber \\
&&\lbrace N(\tau ,\vec \sigma ),{\tilde \pi}^N(\tau ,{\vec \sigma}^{'} )
\rbrace = \delta^3(\vec \sigma ,{\vec \sigma}^{'}),\nonumber \\
&&\lbrace N_{(a)}(\tau ,\vec \sigma ),{\tilde \pi}^{\vec N}_{(b)}(\tau ,{\vec 
\sigma}^{'} )\rbrace =\delta_{(a)(b)} \delta^3(\vec \sigma ,{\vec \sigma}^{'}),
\nonumber \\
&&\lbrace \varphi_{(a)}(\tau ,\vec \sigma ),{\tilde \pi}^{\vec \varphi}_{(b)}
(\tau ,{\vec \sigma}^{'} )\rbrace = \delta_{(a)(b)} \delta^3(\vec \sigma ,
{\vec \sigma}^{'}),\nonumber \\
&&\lbrace {}^3e_{(a)r}(\tau ,\vec \sigma ),{}^3{\tilde \pi}^s_{(b)}(\tau ,
{\vec \sigma}^{'} )\rbrace =\delta_{(a)(b)} \delta^s_r \delta^3(\vec \sigma ,
{\vec \sigma}^{'}),\nonumber \\
&&{}\nonumber \\
&&\lbrace {}^3e^r_{(a)}(\tau ,\vec \sigma),{}^3{\tilde \pi}^s_{(b)}(\tau ,
{\vec \sigma}^{'})\rbrace =-{}^3e^r_{(b)}(\tau ,\vec \sigma )\, {}^3e^s
_{(a)}(\tau ,\vec \sigma ) \delta^3(\vec \sigma ,{\vec \sigma}^{'}),
\nonumber \\
&&\lbrace {}^3e(\tau ,\vec \sigma ), {}^3{\tilde \pi}^r_{(a)}(\tau ,{\vec 
\sigma}^{'})\rbrace ={}^3e(\tau ,\vec \sigma )\, {}^3e^r_{(a)}(\tau ,\vec 
\sigma )\, \delta^3(\vec \sigma ,{\vec \sigma}^{'}),
\label{III6}
\end{eqnarray}

\noindent where the Dirac delta distribution is a density of weight -1
[it behaves as $\sqrt{\gamma (\tau ,\vec \sigma )}$], because we have
the ${\vec \sigma}^{'}$-reparametrization invariant result $\int d^3\sigma^{'}
\delta^3(\vec \sigma ,{\vec \sigma}^{'}) f({\vec \sigma}^{'})=f(\vec \sigma )$].
The momentum ${}^3{\tilde \pi}^r_{(a)}$ is a density of weight -1.

Besides the seven primary constraints ${\tilde \pi}^{\vec \varphi}_{(a)}
(\tau ,\vec \sigma )\approx 0$, ${\tilde \pi}^N(\tau ,\vec \sigma )\approx
0$, ${\tilde \pi}^{\vec N}_{(a)}(\tau ,\vec \sigma )\approx 0$, there are the 
following three primary constraints (the generators of the inner rotations)

\begin{eqnarray}
{}^3{\tilde M}_{(a)}(\tau ,\vec \sigma )&=&\epsilon_{(a)(b)(c)}\, {}^3e_{(b)r}
(\tau ,\vec \sigma )\, {}^3{\tilde \pi}^r_{(c)}(\tau ,\vec \sigma )={1\over 2}
\epsilon_{(a)(b)(c)}\, {}^3{\tilde M}_{(b)(c)}(\tau ,\vec \sigma )\approx 0,
\nonumber \\
&\Rightarrow& {}^3{\tilde M}_{(a)(b)}(\tau ,\vec \sigma )=\epsilon_{(a)(b)(c)}
\, {}^3{\tilde M}_{(c)}(\tau ,\vec \sigma )=\nonumber \\
&=&{}^3e_{(a)r}(\tau ,\vec \sigma )\,
{}^3{\tilde \pi}^r_{(b)}(\tau ,\vec \sigma )-{}^3e_{(b)r}(\tau ,\vec \sigma )\,
{}^3{\tilde \pi}^r_{(a)}(\tau ,\vec \sigma )\approx 0.
\label{III7}
\end{eqnarray}

By using Eqs.(\ref{III5}) and (\ref{III6}) we get the following inversion

\begin{eqnarray}
{}^3e^r_{(a)}&(& N_{(b) | r}-\partial_{\tau}\, {}^3e_{(b)r})+{}^3e^r_{(b)} 
(N_{(a) | r}-\partial_{\tau}\, {}^3e_{(a)r})=\nonumber \\
&=&{{\epsilon N}\over {2k\, {}^3e}}\, 
{}^3G_{o(a)(b)(c)(d)}\, {}^3e_{(c)r}\, {}^3{\tilde \pi}^r_{(d)},
\label{III8}
\end{eqnarray}

\noindent so that, even if this equation cannot be solved for $\partial_{\tau}\,
{}^3e_{(a)r}$ [due to the degeneracy associated with the first class 
constraints], we can get the phase space expression of the extrinsic curvature
without using the Hamilton equations

\begin{eqnarray}
{}^3K_{rs}&=&{{\epsilon}\over {4k\, {}^3e}}\, {}^3G_{o(a)(b)(c)(d)}\, {}^3e
_{(a)r}\, {}^3e_{(b)s}\, {}^3e_{(c)u}\, {}^3{\tilde \pi}^u_{(d)},\nonumber \\
{}^3K&=& -{{\epsilon}\over {2k\sqrt{\gamma}}} {}^3{\tilde \Pi}=-{{\epsilon}
\over {4k\, {}^3e}} {}^3e_{(a)r}\, {}^3{\tilde \pi}^r_{(a)}.
\label{III9}
\end{eqnarray}

Since at the Lagrangian level the primary constraints are identically zero,
we have

\begin{eqnarray}
{}^3{\tilde \pi}^r_{(a)}&=&{}^3e^r_{(b)}\, {}^3e_{(b)s}\, {}^3{\tilde \pi}^s
_{(a)}={1\over 2}{}^3e^r_{(b)}[{}^3e_{(a)s}\, {}^3{\tilde \pi}^s_{(b)}+{}^3e
_{(b)s}\, {}^3{\tilde \pi}^s_{(a)}]-{1\over 2}{}^3{\tilde M}_{(a)(b)}\,
{}^3e^r_{(b)}\equiv \nonumber \\
&\equiv& {1\over 2}{}^3e^r_{(b)}[{}^3e_{(a)s}\, {}^3{\tilde \pi}^s_{(b)}+{}^3e
_{(b)s}\, {}^3{\tilde \pi}^s_{(a)}],\nonumber \\
&&{}\nonumber \\
{}^3{\tilde \pi}^r_{(a)} &\partial_{\tau}& {}^3e_{(a)r}\equiv
{1\over 2}[{}^3e_{(a)s}\, {}^3{\tilde \pi}^s_{(b)}+{}^3e
_{(b)s}\, {}^3{\tilde \pi}^s_{(a)}]{}^3e^r_{(b)}\, \partial_{\tau}\, {}^3e
_{(a)r}\equiv \nonumber \\
&\equiv& {}^3{\tilde \pi}^r_{(a)}\, N_{(a) | r}-{N\over {4k\, {}^3e}}\, 
{}^3G_{o(a)(b)(c)(d)}\,
{}^3e_{(a)s}\, {}^3{\tilde \pi}^s_{(b)}\, {}^3e_{(c)r}\, {}^3{\tilde \pi}^r
_{(d)},
\label{III11}
\end{eqnarray}

\noindent and the canonical Hamiltonian is

\begin{eqnarray}
{\hat H}_{(c)}&=&\int d^3\sigma [{\tilde \pi}^N\partial_{\tau}N+{\tilde
\pi}^{\vec N}_{(a)}\partial_{\tau}N_{(a)}+{\tilde \pi}^{\vec \varphi}_{(a)}
\partial_{\tau}\varphi_{(a)}+{}^3{\tilde \pi}^r_{(a)}\partial_{\tau}
{}^3e_{(a)r}](\tau ,\vec \sigma ) - {\hat L}_{ADMT}=\nonumber \\
&=&\int_{\Sigma_{\tau}}
d^3\sigma [\epsilon N\, (k\, {}^3e\, \epsilon_{(a)(b)(c)}\, {}^3e^r
_{(a)}\, {}^3e^s_{(b)}\, {}^3\Omega_{rs(c)}-\nonumber \\
&-&{1\over {8k\, {}^3e}}\, {}^3G_{o(a)(b)(c)(d)}\,
{}^3e_{(a)r}\, {}^3{\tilde \pi}^r_{(b)}\, {}^3e_{(c)s}\, {}^3{\tilde \pi}^s
_{(d)})-\nonumber \\
&-&
N_{(a)}\, {}^3{\tilde \pi}^r_{(a) | r}](\tau ,\vec \sigma )+\int_{\partial 
\Sigma_{\tau}}
d^2\Sigma_r[N_{(a)}\, {}^3{\tilde \pi}^r_{(a)}](\tau ,\vec \sigma ).
\label{III12}
\end{eqnarray}

\noindent In this paper we shall ignore the surface term.

The Dirac Hamiltonian is

\begin{equation}
{\hat H}_{(D)}={\hat H}_{(c)}+\int d^3\sigma [\lambda_N\, {\tilde 
\pi}^N+\lambda^{\vec N}_{(a)}\, {\tilde \pi}^{\vec N}_{(a)}+\lambda^{\vec
\varphi}_{(a)}\, {\tilde \pi}^{\vec \varphi}_{(a)}+\mu_{(a)}\, {}^3{\tilde
M}_{(a)}](\tau ,\vec \sigma ).
\label{III13}
\end{equation}

The $\tau$-constancy of the ten primary constraints generates four secondary
constraints [from $\partial_{\tau}\, {\tilde \pi}^N(\tau ,\vec \sigma )\approx
0$ and from $\partial_{\tau}\, {\tilde \pi}^{\vec N}_{(a)}(\tau ,\vec \sigma )
\approx 0$]

\begin{eqnarray}
{\hat {\cal H}}(\tau ,\vec \sigma )&=& \epsilon
\Big[ k\, {}^3e\, \epsilon_{(a)(b)(c)}\, {}^3e^r
_{(a)}\, {}^3e^s_{(b)}\, {}^3\Omega_{rs(c)}-\nonumber \\
&-&{1\over {8k\, {}^3e}}
{}^3G_{o(a)(b)(c)(d)}\, {}^3e_{(a)r}\, {}^3{\tilde \pi}^r_{(b)}\, {}^3e_{(c)s}\,
{}^3{\tilde \pi}^s_{(d)}\Big] (\tau ,\vec \sigma )=\nonumber \\
&=&\epsilon \Big[ k\, {}^3e\, {}^3R-{1\over {8k\, {}^3e}}
{}^3G_{o(a)(b)(c)(d)}\, {}^3e_{(a)r}\, {}^3{\tilde \pi}^r_{(b)}\, {}^3e_{(c)s}\,
{}^3{\tilde \pi}^s_{(d)}\Big] (\tau ,\vec \sigma )
\approx 0,\nonumber \\
{\hat {\cal H}}_{(a)}(\tau ,\vec \sigma )&=&[\partial_r\, {}^3{\tilde \pi}^r
_{(a)}-\epsilon_{(a)(b)(c)}\, {}^3\omega_{r(b)}\, {}^3{\tilde \pi}^r_{(c)}]
(\tau ,\vec \sigma )={}^3{\tilde \pi}^r_{(a)|r}(\tau ,\vec \sigma )
\approx 0,\nonumber \\
&&{}\nonumber \\
&\Rightarrow& {\hat H}_{(c)}= \int d^3\sigma [ N\, {\hat {\cal H}}-
N_{(a)}\, {\hat {\cal H}}_{(a)}](\tau ,\vec \sigma )\approx 0.
\label{III14}
\end{eqnarray}

It can be checked that the superhamiltonian constraint ${\hat {\cal H}}(\tau ,
\vec \sigma )\approx 0$ coincides with the ADM metric superhamiltonian one
${\tilde {\cal H}}(\tau ,\vec \sigma )\approx 0$ given in  Eqs.(\ref{c8}) of
Section V, where also the ADM metric supermomentum constraints will be
expressed in terms of the tetrad gravity constraints.

It is convenient to replace the constraints ${\hat {\cal H}}_{(a)}(\tau ,\vec 
\sigma )\approx 0$ [they are of the type of SO(3) Yang-Mills Gauss laws,
because they are the covariant
divergence of a vector density]  with the 3 constraints generating space
pseudodiffeomorphisms on the cotriads and their conjugate momenta

\begin{eqnarray}
{}^3{\tilde \Theta}_r(\tau ,\vec \sigma )&=&-[{}^3e_{(a)r}\, {\hat {\cal H}}
_{(a)}+{}^3\omega_{r(a)}\, {}^3{\tilde M}_{(a)}](\tau ,\vec \sigma )=
\nonumber \\
&=&[{}^3{\tilde \pi}^s_{(a)}\, \partial_r\, {}^3e_{(a)s}-\partial_s
({}^3e_{(a)r}\, {}^3{\tilde \pi}^s_{(a)})](\tau ,\vec \sigma )\approx 0,
\nonumber \\
&&{}\nonumber \\
{\hat {\cal H}}_{(a)}(\tau ,\vec \sigma )&=&-{}^3e^r_{(a)}(\tau ,\vec \sigma )
[{}^3{\tilde \Theta}_r + {}^3\omega_{r(b)}\, {}^3{\tilde M}_{(b)}]
(\tau ,\vec \sigma )\approx 0,\nonumber \\
&&{}\nonumber \\
&\Rightarrow& {\hat H}^{'}_{(D)}={\hat H}^{'}_{(c)}+\int d^3\sigma
[\lambda_N{\tilde \pi}^N+\lambda^{\vec N}_{(a)}{\tilde \pi}^{\vec N}_{(a)}+
\lambda^{\vec \varphi}_{(a)}{\tilde \pi}^{\vec \varphi}_{(a)}+{\hat \mu}_{(a)}
\, {}^3{\tilde M}_{(a)}](\tau ,\vec \sigma ),\nonumber \\
&& {\hat H}^{'}_{(c)}= \int d^3\sigma [ N {\hat {\cal H}}
+N_{(a)}\, {}^3e^r_{(a)}\, {}^3{\tilde \Theta}_r](\tau ,\vec \sigma ),
\nonumber \\
&&{}
\label{III15}
\end{eqnarray}

\noindent where we replaced $[\mu_{(a)}- N_{(b)}\, {}^3e^r_{(b)}\, 
{}^3\omega_{r(a)}](\tau ,\vec \sigma )$ with the new Dirac multipliers ${\hat
\mu}_{(a)}(\tau ,\vec \sigma )$.

All the constraints are first class because the only non-identically
vanishing Poisson brackets are

\begin{eqnarray}
\lbrace {}^3{\tilde M}_{(a)}(\tau ,\vec \sigma ),{}^3{\tilde M}_{(b)}
(\tau ,{\vec \sigma}^{'})\rbrace &=&\epsilon_{(a)(b)(c)}\, {}^3{\tilde M}_{(c)}
(\tau ,\vec \sigma ) \delta^3(\vec \sigma ,{\vec \sigma}^{'}),\nonumber \\
\lbrace {}^3{\tilde M}_{(a)}(\tau ,\vec \sigma ),{}^3{\tilde \Theta}_r
(\tau ,{\vec \sigma}^{'})\rbrace &=&{}^3{\tilde M}_{(a)}(\tau ,{\vec \sigma}
^{'})\, {{\partial \delta^3(\vec \sigma ,{\vec \sigma}^{'})}\over {\partial
\sigma^r}},\nonumber \\
\lbrace {}^3{\tilde \Theta}_r(\tau ,\vec \sigma ),{}^3{\tilde \Theta}_s
(\tau ,{\vec \sigma}^{'})\rbrace &=&[{}^3{\tilde \Theta}_r(\tau ,{\vec \sigma}
^{'}) {{\partial}\over {\partial \sigma^s}} +{}^3{\tilde \Theta}_s(\tau ,\vec 
\sigma ) {{\partial}\over {\partial \sigma^r}}] \delta^3(\vec \sigma ,{\vec 
\sigma}^{'}),\nonumber \\
\lbrace {\hat {\cal H}}(\tau ,\vec \sigma ),{}^3{\tilde \Theta}_r(\tau ,{\vec 
\sigma}^{'})\rbrace &=& {\hat {\cal H}}(\tau ,{\vec \sigma}^{'}) {{\partial
\delta^3(\vec \sigma ,{\vec \sigma}^{'})}\over {\partial \sigma^r}},
\nonumber \\
\lbrace {\hat {\cal H}}(\tau ,\vec \sigma ),{\hat {\cal H}}(\tau ,{\vec 
\sigma}^{'})\rbrace &=& [ {}^3e^r_{(a)}(\tau ,\vec \sigma )\, {\hat {\cal H}}
_{(a)}(\tau ,\vec \sigma ) +\nonumber \\
&+& {}^3e^r_{(a)}(\tau ,{\vec \sigma}^{'})\,
{\hat {\cal H}}_{(a)}(\tau ,{\vec \sigma}^{'}) ] {{\partial \delta^3(\vec \sigma
,{\vec \sigma}^{'})}\over {\partial \sigma^r}}=\nonumber \\
&=&\{ [{}^3e^r_{(a)}\, {}^3e^s_{(a)}\, [{}^3{\tilde \Theta}_s+{}^3\omega_{s(b)}
\, {}^3{\tilde M}_{(b)}]](\tau ,\vec \sigma ) + \nonumber \\
&+&[{}^3e^r_{(a)}\, {}^3e^s_{(a)}\, [{}^3{\tilde \Theta}_s+{}^3\omega_{s(b)}\,
{}^3{\tilde M}_{(b)}]](\tau ,{\vec \sigma}^{'}) \} \, {{\partial \delta^3(\vec 
\sigma ,{\vec \sigma}^{'})}\over {\partial \sigma^r}}.
\label{III16}
\end{eqnarray}

The Poisson brackets of the cotriads and of their conjugate momenta with the
constraints are [${}^3R=\epsilon_{(a)(b)(c)}\, {}^3e^r_{(a)}\, {}^3e^s_{(b)}\, 
{}^3\Omega_{rs(c)}$]

\begin{eqnarray}
&&\lbrace {}^3e_{(a)r}(\tau ,\vec \sigma ),{}^3{\tilde M}_{(b)}(\tau ,{\vec 
\sigma}^{'})\rbrace =\epsilon_{(a)(b)(c)}\, {}^3e_{(c)r}(\tau ,\vec \sigma )
\delta^3(\vec \sigma ,{\vec \sigma}^{'}),\nonumber \\
&&\lbrace {}^3e_{(a)r}(\tau ,\vec \sigma ),{}^3{\tilde \Theta}_s(\tau ,{\vec 
\sigma}^{'})\rbrace ={{\partial \, {}^3e_{(a)r}(\tau ,\vec \sigma )}\over
{\partial \sigma^s}}\delta^3(\vec \sigma ,{\vec \sigma}^{'})+{}^3e_{(a)s}
(\tau ,\vec \sigma ){{\partial \delta^3(\vec \sigma ,{\vec \sigma}^{'})}\over
{\partial \sigma^r}},\nonumber \\
&&\lbrace {}^3e_{(a)r}(\tau ,\vec \sigma ),{\hat {\cal H}}(\tau ,{\vec \sigma}
^{'})\rbrace =-{{\epsilon}\over {4k}}\Big[ {1\over {{}^3e}}\, 
{}^3G_{o(a)(b)(c)(d)}\, {}^3e_{(b)r}
\, {}^3e_{(c)s}\, {}^3{\tilde \pi}^s_{(d)}\Big] (\tau ,\vec \sigma )\delta^3
(\vec \sigma ,{\vec \sigma}^{'}),\nonumber \\
&&{}\nonumber \\
&&\lbrace {}^3{\tilde \pi}^r_{(a)}(\tau ,\vec \sigma ),{}^3{\tilde M}_{(b)}
(\tau ,{\vec \sigma}^{'})\rbrace =\epsilon_{(a)(b)(c)}\, {}^3{\tilde \pi}^r
_{(c)}(\tau ,\vec \sigma )\delta^3(\vec \sigma ,{\vec \sigma}^{'}),\nonumber \\
&&\lbrace {}^3{\tilde \pi}^r_{(a)}(\tau ,\vec \sigma ),{}^3{\tilde \Theta}_s
(\tau ,{\vec \sigma}^{'})\rbrace =- {}^3{\tilde \pi}^r_{(a)}(\tau ,{\vec \sigma}
^{'}){{\partial \delta^3(\vec \sigma ,{\vec \sigma}^{'})}\over {\partial
\sigma^{{'}s}}}+\delta^r_s {{\partial}\over {\partial \sigma^{{'}u}}} 
[{}^3{\tilde \pi}^u_{(a)}(\tau ,{\vec \sigma}^{'})\delta^3(\vec \sigma ,
{\vec \sigma}^{'})],\nonumber \\
&&\lbrace {}^3{\tilde \pi}^r_{(a)}(\tau ,\vec \sigma ),{\hat {\cal H}}
(\tau ,{\vec \sigma}^{'})\rbrace =\epsilon  \Big[ 2k\, {}^3e\,  ({}^3R^{rs}-
{1\over 2}\, {}^3g^{rs}\, {}^3R)\, {}^3e_{(a)s}+\nonumber \\
&+&{1\over {4k\, {}^3e}} 
{}^3G_{o(a)(b)(c)(d)}\, {}^3{\tilde \pi}^r_{(b)}\, {}^3e_{(c)s}\,
{}^3{\tilde \pi}^s_{(d)}-\nonumber \\
&-&{1\over {8k\, {}^3e}} {}^3e^r_{(a)}\, {}^3G_{o(b)(c)(d)(e)}\,
{}^3e_{(b)u}\, {}^3{\tilde \pi}^u_{(c)}\, {}^3e_{(d)v}\, {}^3{\tilde \pi}^v
_{(e)} \Big] (\tau ,\vec \sigma ) \delta^3(\vec \sigma ,{\vec \sigma}^{'})+
\nonumber \\
&+&2k\, {}^3e(\tau ,\vec \sigma )
\Big[ {}^3\Gamma^w_{uv}({}^3e^u_{(a)}\, {}^3g^{rv}-
{}^3e^r_{(a)}\, {}^3g^{uv})\Big] (\tau ,{\vec \sigma}^{'}){{\partial \delta^3
(\vec \sigma ,{\vec \sigma}^{'})}\over {\partial \sigma^w}}+\nonumber \\
&+&2k\, {}^3e(\tau ,\vec \sigma )\Big[ {}^3e^u_{(a)}\, {}^3g^{rv}-
{}^3e^r_{(a)}\, {}^3g^{uv})\Big] (\tau ,{\vec \sigma}^{'}){{\partial^2\delta
^3(\vec \sigma ,{\vec \sigma}^{'})}\over {\partial \sigma^u\partial \sigma^v}},
\label{III17}
\end{eqnarray}

\noindent where we used\hfill\break
 $\lbrace {}^3e(\tau ,\vec \sigma ){}^3R(\tau ,\vec 
\sigma ),{}^3{\tilde \pi}^r_{(a)}(\tau ,{\vec \sigma}^{'})\rbrace =
-2k\Big[ {}^3e({}^3R^{rs}-{1\over 2}{}^3g^{rs}\, {}^3R){}^3e_{(a)s}\Big] (\tau 
,\vec \sigma )\delta^3(\vec \sigma ,{\vec \sigma}^{'})+$\hfill\break
$+2k\, {}^3e(\tau ,\vec \sigma ) \Big[ {}^3\Gamma^w_{uv}({}^3e^u_{(a)}\, 
{}^3g^{rv}-{}^3e^r_{(a)}\, {}^3g^{uv})\Big]
(\tau ,{\vec \sigma}^{'}){{\partial \delta^3(\vec \sigma ,{\vec \sigma}^{'})}
\over {\partial \sigma^w}}+$\hfill\break
$+2k\, {}^3e(\tau ,\vec \sigma )\Big[ {}^3e^u_{(a)}\,
{}^3g^{rv}-{}^3e^r_{(a)}\, {}^3g^{uv}\Big] (\tau ,{\vec \sigma}^{'}){{\partial^2
\delta^3(\vec \sigma ,{\vec \sigma}^{'})}\over {\partial \sigma^u\partial
\sigma^v}}$.

The Hamilton equations associated with the Dirac Hamiltonian (\ref{III15}) are
[see Eqs.(\ref{II5a}) for ${}^3R^{uv}$]

\begin{eqnarray}
\partial_{\tau}N(\tau ,\vec \sigma ) &{\buildrel \circ \over =}& \lbrace
N(\tau ,\vec \sigma ),{\hat H}^{'}_{(D)}\rbrace =\lambda_N(\tau ,\vec 
\sigma ),\nonumber \\
\partial_{\tau}N_{(a)}(\tau ,\vec \sigma ) &{\buildrel \circ \over =}& \lbrace
N_{(a)}(\tau ,\vec \sigma ),{\hat H}^{'}_{(D)}\rbrace = \lambda^{\vec N}
_{(a)}(\tau ,\vec \sigma ),\nonumber \\
\partial_{\tau}\varphi_{(a)}(\tau ,\vec \sigma ) &{\buildrel \circ \over =}&
\lbrace \varphi_{(a)}(\tau ,\vec \sigma ),{\hat H}^{'}_{(D)}\rbrace =
\lambda^{\vec \varphi}_{(a)}(\tau ,\vec \sigma ),\nonumber \\
\partial_{\tau}\, {}^3e_{(a)r}(\tau ,\vec \sigma ) &{\buildrel \circ \over =}&
\lbrace {}^3e_{(a)r}(\tau ,\vec \sigma ),{\hat H}^{'}_{(D)}\rbrace =
\nonumber \\
&=&- {{\epsilon}\over {4k}}\Big[ {N\over{{}^3e}}\, {}^3G_{o(a)(b)(c)(d)}
\, {}^3e_{(b)r}\, {}^3e_{(c)s}\, {}^3{\tilde \pi}^s_{(d)}\Big] (\tau ,\vec 
\sigma )+\nonumber \\
&+&\Big[ N_{(b)}\, {}^3e^s_{(b)}{{\partial \, {}^3e_{(a)r}}\over {\partial 
\sigma^s}}+{}^3e_{(a)s} {{\partial}\over {\partial \sigma^r}} (N_{(b)}\,
{}^3e^s_{(b)})\Big] (\tau ,\vec \sigma )+\nonumber \\
&+&\epsilon_{(a)(b)(c)}\, {\hat \mu}_{(b)}(\tau ,\vec \sigma )\, {}^3e_{(c)r}
(\tau ,\vec \sigma ),\nonumber \\  
\partial_{\tau}\, {}^3{\tilde \pi}^r_{(a)}(\tau ,\vec \sigma ) &{\buildrel
\circ \over =}& \lbrace {}^3{\tilde \pi}^r_{(a)}(\tau ,\vec \sigma ),{\hat
H}^{'}_{(D)}\rbrace =\nonumber \\
&=&2k\epsilon \Big[ {}^3e N ({}^3R^{rs}-{1\over 2} {}^3g^{rs}\, {}^3R){}^3e
_{(a)s}+{}^3e (N^{|r|s}-{}^3g^{rs}\, N^{|u}{}_{|u}){}^3e_{(a)s}\Big] (\tau 
,\vec \sigma )-\nonumber \\
&-&\epsilon {{N(\tau ,\vec \sigma )}\over {8k}} 
\Big[ {1\over {{}^3e}}\, {}^3G_{o(a)(b)(c)(d)}\, {}^3{\tilde \pi}
^r_{(b)}\, {}^3e_{(c)s}\, {}^3{\tilde \pi}^s_{(d)}-\nonumber \\
&-&{2\over {{}^3e}}\, {}^3e^r_{(a)}\, {}^3G_{o(b)(c)(d)(e)}\, {}^3e_{(b)u}\,
{}^3{\tilde \pi}^u_{(c)}\, {}^3e_{(d)v}\, {}^3{\tilde \pi}^v_{(e)}\Big]
(\tau ,\vec \sigma )+\nonumber \\
&+&{{\partial}\over {\partial \sigma^s}} \Big[ N_{(b)}\, {}^3e^s_{(b)}\, 
{}^3{\tilde \pi}^r_{(a)}\Big] (\tau ,\vec \sigma )-{}^3{\tilde \pi}^u_{(a)}
(\tau ,\vec \sigma ){{\partial}\over {\partial \sigma^u}}\Big[ N_{(b)}\, {}^3e^r
_{(b)}\Big] (\tau ,\vec \sigma )+\nonumber \\
&+&\epsilon_{(a)(b)(c)}\, {\hat \mu}_{(b)}(\tau ,\vec \sigma )\, {}^3{\tilde
\pi}^r_{(c)}(\tau ,\vec \sigma ),\nonumber \\
&&\Downarrow \nonumber \\
\partial_{\tau}\, {}^3e^r_{(a)}(\tau ,\vec \sigma )
&=&-\Big[ {}^3e^r_{(b)}\, {}^3e^s_{(a)} \partial_{\tau}\, {}^3e_{(b)s}\Big]
(\tau ,\vec \sigma )\, {\buildrel \circ \over =}\, \nonumber \\
&{\buildrel \circ \over =}&\, 
{{\epsilon}\over {4k}}\Big[ {N\over {{}^3e}}\, 
{}^3G_{o(a)(b)(c)(d)}\, {}^3e_{(b)}^r\,
 {}^3e_{(c)s}\, {}^3{\tilde \pi}^s_{(d)}\Big] (\tau ,\vec \sigma )-
\nonumber \\
&-&{}^3e^s_{(a)}\Big[ N_{(c)}\, {}^3e^u_{(c)}\, {}^3e^r_{(b)}
{{\partial \, {}^3e_{(b)s}}\over {\partial 
\sigma^u}}+ {{\partial}\over {\partial \sigma^s}} (N_{(c)}\,
{}^3e^r_{(c)})\Big] (\tau ,\vec \sigma )+\nonumber \\
&+&\epsilon_{(a)(b)(c)}\, {\hat \mu}_{(b)}(\tau ,\vec \sigma )\, {}^3e_{(c)}^r
(\tau ,\vec \sigma ),\nonumber \\
\partial_{\tau}\, {}^3e(\tau ,\vec \sigma )
&=&\Big[ {}^3e\, {}^3e^r_{(a)} \partial_{\tau}\, {}^3e_{(a)r}\Big] (\tau ,\vec 
\sigma )\, {\buildrel \circ \over =}\nonumber \\
&{\buildrel \circ \over =}&\, {{\epsilon}\over {4k}}\Big[ N\, {}^3e_{(a)s}\,
{}^3{\tilde \pi}^s_{(a)}\Big] (\tau ,\vec \sigma )+\nonumber \\
&+&\Big( {}^3e\, \Big[ N_{(b)}\, {}^3e^s_{(b)}\, {}^3e^r_{(a)}\partial_s\,
{}^3e_{(a)r}+{}^3e^r_{(a)}\, {}^3e_{(a)s} \partial_r(N_{(b)}\, {}^3e^s_{(b)})
\Big] \Big)
(\tau ,\vec \sigma ).
\label{III18}
\end{eqnarray}

From the Hamilton equations and Eqs.(\ref{II5a}), (\ref{III9}), we  get

\begin{eqnarray}
\partial_{\tau}\, {}^3g_{rs}(\tau ,\vec \sigma )\, &{\buildrel \circ \over =}\,&
\Big[ N_{r|s}+N_{s|r}-2N\, {}^3K_{rs}\Big] (\tau ,\vec \sigma ),\nonumber \\
\partial_{\tau}\, {}^3K_{rs}(\tau ,\vec \sigma )
&{\buildrel \circ \over =}\,&{1\over {4k}}\, {}^3G_{o(a)(b)(c)(d)} \Big(
{{\epsilon}\over {{}^3e}} \Big[ \partial_v(N_{(m)}\, {}^3e^v_{(m)}\, {}^3e
_{(a)r}\, {}^3e_{(b)s}\, {}^3e_{(c)u}\, {}^3{\tilde \pi}^u_{(d)})+\nonumber \\
&+&{}^3e_{(c)u}\, {}^3{\tilde \pi}^u_{(d)}\Big[ 2k N ({}^3R^{uv}-{1\over 2}\, 
{}^3g^{uv}\, {}^3R)+\epsilon (N^{|u|v}-{}^3g^{uv} N^{|l}{}_{|l})\Big]
{}^3e_{(d)v}-\nonumber \\
&-&{N\over {4k\, {}^3e^2}}\Big[ {1\over 2}\, {}^3e_{(a)r}\, {}^3e_{(b)s}\,
{}^3e_{(c)u}\, {}^3G_{o(d)(e)(f)(g)}\, {}^3{\tilde \pi}^u_{(e)}\, {}^3e_{(f)v}\,
{}^3{\tilde \pi}^v_{(g)}-\nonumber \\
&-&{}^3e_{(a)r}\, {}^3e_{(b)s} \delta_{(c)(d)}\, {}^3G_{o(e)(f)(g)(h)}\,
{}^3e_{(e)u}\, {}^3{\tilde \pi}^u_{(f)}\, {}^3e_{(g)v}\, {}^3{\tilde \pi}^v
_{(h)}+\nonumber \\
&+&{}^3e_{(a)r}\, {}^3e_{(b)s} ({}^3e_{(m)v}\, {}^3{\tilde \pi}^v_{(m)}\, 
{}^3e_{(c)u}\, {}^3{\tilde \pi}^u_{(d)}+{}^3G_{o(c)(e)(f)(g)}\, {}^3e_{(e)u}\, 
{}^3{\tilde \pi}^u_{(d)}\, {}^3e_{(f)v}\, {}^3{\tilde \pi}^v_{(f)})+\nonumber \\
&+&({}^3e_{(a)r}\, {}^3G_{o(b)(e)(f)(g)}\, {}^3e_{(e)s}+{}^3e_{(b)s}\, {}^3G
_{o(a)(e)(f)(g)}\, {}^3e_{(e)r})\nonumber \\
&& {}^3e_{(f)u}\, {}^3{\tilde \pi}^u_{(g)}\,
{}^3e_{(c)v}\, {}^3{\tilde \pi}^v_{(d)}\Big] \Big) (\tau ,\vec \sigma )
,\nonumber \\
\partial_{\tau}\, {}^3K(\tau ,\vec \sigma )\, &{\buildrel \circ \over =}\,& 
\Big( {1\over 4} N\, {}^3R+ 4N^{|r}{}_{|r}+\nonumber \\
&+&{N\over {(4k\, {}^3e)^2}} \Big[ ({}^3e_{(a)r}\, {}^3{\tilde \pi}^r_{(a)})^2-
{3\over 2}\, {}^3G_{o(a)(b)(c)(d)}\, {}^3e_{(a)r}\, {}^3{\tilde \pi}^r_{(b)}\,
{}^3e_{(c)s}\, {}^3{\tilde \pi}^s_{(d)}\Big] -\nonumber \\
&-&{{\epsilon}\over {4k\, {}^3e}}\, {}^3{\tilde \pi}^r_{(a)}\Big[ N_{(m)}\,
{}^3e^u_{(m)}\partial_u\, {}^3e_{(a)r}+{}^3e_{(a)u}\partial_r(N_{(m)}\, {}^3e
^u_{(m)})\Big] \Big) (\tau ,\vec \sigma ),\nonumber \\
\partial_{\tau}\, {}^3\omega_{r(a)(b)}(\tau ,\vec \sigma )
&{\buildrel \circ \over =}\,&{{\epsilon N}\over {4k\, {}^3e}} \Big(
(\partial_r\, {}^3e_{(b)s}-\partial_s\, {}^3e_{(b)r}){}^3G_{o(a)(l)(m)(n)}\,
{}^3e^s_{(l)}\, +\nonumber \\
&+&(\partial_s\, {}^3e_{(a)r}-\partial_r\, {}^3e_{(a)s}){}^3G_{o(b)(l)(m)(n)}\,
{}^3e^s_{(l)}+\nonumber \\
&+&(\partial_v\, {}^3e_{(c)u}-\partial_u\, {}^3e_{(c)v})\Big[ {}^3e^v_{(b)}\, 
{}^3e_{(c)r}\, {}^3G_{o(a)(l)(m)(n)}\, {}^3e^u_{(l)}+\nonumber \\
&+&{}^3e^u_{(a)}\, {}^3e_{(c)r}\, {}^3G_{o(b)(l)(m)(n)}\, {}^3e^v_{(l)}-
\nonumber \\
&-&{}^3e^u_{(a)}\, {}^3e^v_{(b)}\, {}^3G_{o(c)(l)(m)(n)}\, {}^3e_{(l)r}\Big]
{}^3e_{(m)t}\, {}^3{\tilde \pi}^t_{(n)} \Big) -\nonumber \\
&-&{{\epsilon}\over {4k}}\Big( \Big[ {}^3e^s_{(a)}\, {}^3G_{o(b)(l)(m)(n)}-
{}^3e^s_{(b)}\, {}^3G_{o(a)(l)(m)(n)}\Big] \nonumber \\
&&\Big[ \partial_r({N\over {{}^3e}}\,
{}^3e_{(l)s}\, {}^3e_{(m)t}\, {}^3{\tilde \pi}^t_{(n)})-\partial_s({N\over 
{{}^3e}}\, {}^3e_{(l)r}\, {}^3e_{(m)t}\, {}^3{\tilde \pi}^t_{(n)})\Big]+
\nonumber \\
&+&{}^3e^u_{(a)}\, {}^3e^v_{(b)}\, {}^3e_{(c)r}\, {}^3G_{o(c)(l)(m)(n)}
\nonumber \\
&&\Big[
\partial_v({N\over {{}^3e}}\, {}^3e_{(l)u}\, {}^3e_{(m)t}\, {}^3{\tilde \pi}^t
_{(n)})-\partial_u({N\over {{}^3e}}\, {}^3e_{(l)v}\, {}^3e_{(m)t}\,
{}^3{\tilde \pi}^t_{(n)})\Big] \Big)-\nonumber \\
&-&\Big[ (\partial_r\, {}^3e_{(b)s}-\partial_s\, {}^3e_{(b)r}){}^3e^v_{(a)}-
(\partial_r\, {}^3e_{(a)s}-\partial_s\, {}^3e_{(a)r}){}^3e^v_{(b)}\Big]
\nonumber \\
&&\Big[N_{(w)}\, {}^3e^u_{(w)}\, {}^3e^s_{(l)}\partial_u(N_{(w)}\, {}^3e^s
_{(w)})\Big] -\nonumber \\
&-&(\partial_v\, {}^3e_{(c)u}-\partial_u\, {}^3e_{(c)v})\Big[ ({}^3e^v_{(b)}\, 
{}^3e^t_{(a)}+{}^3e^u_{(a)}\, {}^3e^t_{(b)}){}^3e_{(c)r}\nonumber \\
&&\Big[ N_{(m)}\, {}^3e^w_{(m)}\, {}^3e^u_{(l)}\partial_w\, {}^3e_{(l)t}+
\partial_t(N_{(w)}\, {}^3e^u_{(w)})\Big] +\nonumber \\
&+&{}^3e^u_{(a)}\, {}^3e^v_{(b)} \Big( N_{(m)}\, {}^3e^w_{(m)}\partial_w\,
{}^3e_{(c)r}+{}^3e_{(c)w}\partial_r(N_{(m)}\, {}^3e^w_{(m)}\Big) \Big] +
\nonumber \\
&+&{}^3e^s_{(a)}\Big( \partial_r(N_{(w)}\, {}^3e^u_{(w)}\partial_u\, {}^3e
_{(b)s}+{}^3e_{(b)u}\partial_s(N_{(w)}\, {}^3e^u_{(w)}) )-\nonumber \\
&-&\partial_s(
N_{(w)}\, {}^3e^u_{(w)}\partial_u\, {}^3e
_{(b)r}+{}^3e_{(b)u}\partial_r(N_{(w)}\, {}^3e^u_{(w)}) )\Big) -\nonumber \\
&-&{}^3e^s_{(b)}\Big( \partial_r(N_{(w)}\, {}^3e^u_{(w)}\partial_u\, {}^3e
_{(a)s}+{}^3e_{(a)u}\partial_s(N_{(w)}\, {}^3e^u_{(w)}) )-\nonumber \\
&-&\partial_s(N_{(w)}\, {}^3e^u_{(w)}\partial_u\, {}^3e
_{(a)r}+{}^3e_{(a)u}\partial_r(N_{(w)}\, {}^3e^u_{(w)}) )\Big) +\nonumber \\
&+&{}^3e^u_{(a)}\, {}^3e^v_{(b)}\, {}^3e_{(c)r}
\Big( \partial_v(N_{(w)}\, {}^3e^t_{(w)}\partial_t\, {}^3e
_{(c)u}+{}^3e_{(c)t}\partial_u(N_{(w)}\, {}^3e^t_{(w)}) )-\nonumber \\
&-&\partial_u(N_{(w)}\, {}^3e^t_{(w)}\partial_t\, {}^3e
_{(c)v}+{}^3e_{(c)t}\partial_v(N_{(w)}\, {}^3e^t_{(w)}) )\Big) +\nonumber \\
&+&\Big( \Big[ (\partial_r\, {}^3e_{(b)s}-\partial_s\, {}^3e_{(b)r})\epsilon
_{(a)(m)(n)}-(\partial_r\, {}^3e_{(a)s}-\partial_s\, {}^3e_{(a)r})\epsilon
_{(b)(m)(n)}\Big] {}^3e^s_{(n)}+\nonumber \\
&+&(\partial_v\, {}^3e_{(c)u}-\partial_u\, {}^3e_{(c)v})\Big[ {}^3e^v_{(b)}\, 
{}^3e_{(c)r}\epsilon_{(a)(m)(n)}\, {}^3e^u_{(n)}+\nonumber \\
&+&{}^3e^u_{(a)}\, {}^3e_{(c)r}\epsilon_{(b)(m)(n)}\, {}^3e^v_{(n)}+{}^3e^u
_{(a)}\, {}^3e^v_{(b)}\epsilon_{(c)(m)(n)}\, {}^3e_{(n)r}\Big] \Big) {\hat \mu}
_{(m)}+\nonumber \\
&+&\Big[ {}^3e^s_{(a)}\epsilon_{(b)(m)(n)}-{}^3e^s_{(b)}\epsilon_{(a)(m)(n)}
\Big] \Big[ \partial_r({\hat \mu}_{(m)}\, {}^3e_{(n)s})-\partial_s({\hat \mu}
_{(m)}\, {}^3e_{(n)r})\Big]+\nonumber \\
&+&{}^3e^u_{(a)}\, {}^3e^v_{(b)}\, {}^3e_{(c)r}\epsilon_{(c)(m)(n)}\Big[
\partial_v({\hat \mu}_{(m)}\, {}^3e_{(n)u})-\partial_u({\hat \mu}_{(m)}\,
{}^3e_{(n)v})\Big] .
\label{III18a}
\end{eqnarray}

\noindent
They  are needed in Appendix B, where there is the Hamiltonian version of the 
quantities given in Appendix A.

Let us consider the canonical transformation ${\tilde \pi}^N(\tau ,\vec \sigma )
\, dN(\tau ,\vec \sigma )+{\tilde \pi}^{\vec N}_{(a)}(\tau ,\vec \sigma )\,
dN_{(a)}(\tau ,\vec \sigma )+{\tilde \pi}^{\vec \varphi}_{(a)}(\tau ,\vec 
\sigma )\, d\varphi_{(a)}(\tau ,\vec \sigma )+{}^3{\tilde \pi}^r_{(a)}(\tau ,
\vec \sigma )\, d{}^3e_{(a)r}(\tau ,\vec \sigma )={}^4{\tilde \pi}^A_{(\alpha )}
(\tau ,\vec \sigma )\, d{}^4E^{(\alpha )}_A(\tau ,\vec \sigma )$, where
${}^4{\tilde \pi}^A_{(\alpha )}$ [$\lbrace {}^4E^{(\alpha )}_A(\tau ,\vec 
\sigma ),{}^4{\tilde \pi}^B_{(\beta )}(\tau ,{\vec \sigma}^{'})
\rbrace =\delta^B_A\delta^{(\alpha )}
_{(\beta )}\delta^3(\vec \sigma ,{\vec \sigma}^{'})$]
would be the canonical momenta if the ADM
action would be considered as a functional of the cotetrads ${}^4E^{(\alpha )}
_A={}^4E^{(\alpha )}_{\mu}\, b^{\mu}_A$ in the holonomic 
$\Sigma_{\tau}$-adapted basis, as essentially is done in Refs.\cite{hen3,hen2}.
If $\bar \gamma =\sqrt{1+\sum_{(c)}\varphi^{(c) 2}}$, we have

\begin{eqnarray}
{\tilde \pi}^N&=& (\bar \gamma \, {}^4{\tilde \pi}^{\tau}_{(o)}+\varphi^{(a)}\,
{}^4{\tilde \pi}^{\tau}_{(a)}),\nonumber \\
{\tilde \pi}^{\vec N}_{(a)}&=&-\epsilon
\varphi_{(a)}\, {}^4{\tilde \pi}^{\tau}_{(o)}+
[\delta_{(a)}^{(b)}-\epsilon {{\varphi_{(a)}\varphi^{(b)}}\over 
{1+\bar \gamma}}]\, 
{}^4{\tilde \pi}^{\tau}_{(b)},\nonumber \\
{\tilde \pi}^{\vec \varphi}_{(a)}&=&({{\epsilon N}\over {\bar \gamma}}\, 
\varphi_{(a)}-N_{(a)})\, {}^4{\tilde \pi}^{\tau}_{(o)}-\delta_{(a)}^{(b)}\, 
N\, {}^4{\tilde \pi}^{\tau}
_{(b)}-{}^3e_{(a)r}\, {}^4{\tilde \pi}^r_{(o)}-\nonumber \\
&-&{1\over {1+\bar \gamma}}(\delta_{(a)}^{(c)}\varphi^{(b)}+\delta_{(a)}^{(b)}
\varphi^{(c)}+\epsilon
{{\varphi_{(a)}\varphi^{(b)}\varphi^{(c)}}\over {\bar \gamma 
(1+\bar \gamma )}})
(N_{(c)}\, {}^4{\tilde \pi}^{\tau}_{(b)}+{}^3e_{(c)r}\, {}^4{\tilde \pi}^r
_{(b)}),\nonumber \\
{}^3{\tilde \pi}^r_{(a)}&=&-\epsilon \varphi_{(a)}\, {}^4{\tilde \pi}^r_{(o)}+
(\delta_{(a)}^{(b)}-\epsilon {{\varphi_{(a)}\varphi^{(b)}}\over 
{1+\bar \gamma}})\, 
{}^4{\tilde \pi}^r_{(b)},\nonumber \\
&&{}\nonumber \\
{}^4{\tilde \pi}^{\tau}_{(o)}&=&\bar \gamma {\tilde \pi}^N-\varphi^{(a)}\, 
{}^3{\tilde \pi}^{\vec N}_{(a)},\nonumber \\
{}^4{\tilde \pi}^{\tau}_{(a)}&=&\epsilon \varphi_{(a)} {\tilde \pi}^N+[\delta
_{(a)}^{(b)}-\epsilon
{{\varphi_{(a)}\varphi^{(b)}}\over {1+\bar \gamma}}] 
{\tilde \pi}^{\vec N}_{(b)},\nonumber \\
{}^4{\tilde \pi}^r_{(o)}&=&-\bar \gamma \, 
{}^3e^r_{(a)}[\delta_{(a)}^{(b)}-\epsilon
{{\varphi_{(a)}\varphi^{(b)}}\over {1+\bar \gamma}}] {\tilde \pi}^{\vec \varphi}
_{(b)}+\bar \gamma N_{(a)}\, {}^3e^r_{(a)}\, {\tilde \pi}^N+\nonumber \\
&+&{}^3e^r_{(a)} [-N\, \delta_{(a)}^{(b)}-(\delta_{(a)}^{(b)}\varphi^{(c)}-
\delta_{(a)}^{(c)}\varphi^{(b)}){{N_{(c)}}\over 
{1+\bar \gamma}}] {\tilde \pi}^{\vec N}
_{(b)}-\nonumber \\
&-&{1\over {1+\bar \gamma}}\, 
{}^3e^r_{(a)} [\delta^{(a)(b)}\varphi^{(c)}+{1\over
{\bar \gamma}} \delta^{(b)(c)}\varphi^{(a)}]{}^3e_{(c)s}\, 
{}^3{\tilde \pi}^s_{(b)},\nonumber \\
{}^4{\tilde \pi}^r_{(a)}&=&[\delta_{(a)}^{(b)}+\epsilon 
{{\varphi_{(a)}\varphi^{(b)}}\over
{\bar \gamma (1+\bar \gamma )}}] {}^3{\tilde \pi}^r_{(b)}-\epsilon 
\varphi_{(a)}\, {}^3e^r_{(b)}[\delta_{(b)}^{(c)}-\epsilon 
{{\varphi_{(b)}\varphi^{(c)}}\over {1+\bar \gamma}}]{\tilde \pi}
^{\vec \varphi}_{(c)}+\nonumber \\
&+&\epsilon \varphi_{(a)}\, {}^3e^r_{(b)} N_{(b)}{\tilde \pi}^N-\epsilon
\varphi_{(a)}\, {}^3e^r_{(b)}[N\delta^{(b)}_{(c)}+(\delta^{(b)}_{(c)}
\varphi^{(d)}-\delta_{(c)}^{(d)}
\varphi^{(b)}){{N_{(d)}}\over {1+\bar \gamma}}] {\tilde \pi}^{\vec N}_{(c)}-
\nonumber \\
&-&\epsilon
{{\varphi_{(a)}}\over {1+\bar \gamma}} [\varphi^{(b)}\delta^{(c)}_{(d)}+{1\over 
{\bar \gamma}} \varphi^{(c)} \delta^{(b)}_{(d)}]{}^3e^r_{(c)}\, {}^3e_{(b)s}\,
{}^3{\tilde \pi}^s_{(d)}.
\label{III19}
\end{eqnarray}

Our canonical transformation (\ref{III19}) allows to consider the metric ADM
Lagrangian as function of the cotetrads ${}^4E^{(\alpha )}_A={}^4E^{(\alpha )}
_{\mu}\, b_A^{\mu}$ and to find the conjugate momenta ${}^4{\tilde \pi}^A
_{(\alpha )}$. Eqs.(\ref{III19}) show that the four primary constraints, which
contain the informations ${\tilde \pi}^N\approx 0$ and ${\tilde \pi}^{\vec
N}_{(a)}\approx 0$, are ${}^4{\tilde \pi}^{\tau}_{(\alpha )}\approx 0$. 
The six primary constraints
(the generators of the local Lorentz transformations) ${}^4{\tilde M}_{(\alpha
)(\beta )}={}^4E^{(\gamma )}_A[{}^4\eta_{(\alpha )(\gamma )}\, {}^4{\tilde
\pi}^A_{(\beta )}-{}^4\eta_{(\beta )(\gamma )}\, {}^4{\tilde \pi}^A_{(\alpha )}]
\approx 0$ of this formulation have the following relation with ${\tilde \pi}
^{\vec \varphi}_{(a)}\approx 0$ and ${}^3{\tilde M}_{(a)}\approx 0$ 

\begin{eqnarray}
{}^4{\tilde M}_{(a)(b)}&=&-\epsilon {}^3{\tilde M}_{(a)(b)}+ (\varphi
_{(a)} {\tilde \pi}^{\vec \varphi}_{(b)}-\varphi_{(b)} {\tilde \pi}^{\vec
\varphi}_{(a)})+\epsilon (\varphi_{(a)}N_{(b)}-\varphi_{(b)}N_{(a)}) {\tilde 
\pi}^N-\nonumber \\
&-&(\delta^{(c)}_{(a)}\delta^{(d)}_{(b)}-\delta^{(c)}_{(b)}\delta^{(d)}_{(a)})
[\epsilon N \varphi_{(c)} \delta_{(d)(e)}+\nonumber \\
&+&(\delta_{(c)(f)}+{{\varphi_{(c)}\varphi_{(f)}}\over {1+\bar \gamma}})
(\delta_{(d)(e)}+{{\varphi_{(d)}\varphi_{(e)}}\over {1+\bar \gamma}}) N_{(f)}]
{\tilde \pi}^{\vec N}_{(e)}\approx 0,\nonumber \\
{}^4{\tilde M}_{(a)(o)}&=&-\epsilon \bar \gamma 
{\tilde \pi}^{\vec \varphi}_{(a)}-
{{1}\over {1+\bar \gamma}}\, {}^3{\tilde M}_{(a)(b)}\varphi_{(b)}-\epsilon
\bar \gamma (\delta_{(a)(b)}-{{\varphi_{(a)}\varphi_{(b)}}\over {\bar \gamma (1+
\bar \gamma )}})N_{(b)}{\tilde \pi}^N+\nonumber \\
&+& [-\epsilon \bar \gamma N(\delta_{(a)(b)}-{{\varphi_{(a)}\varphi_{(b)}}\over
{\bar \gamma (1+\bar \gamma )}})+
\varphi_{(c)}N_{(c)}\delta_{(a)(b)}-N_{(a)}\varphi
_{(b)}]{\tilde \pi}^{\vec N}_{(b)}\approx 0,\nonumber \\
&&{}\nonumber \\
{}^3{\tilde M}_{(a)(b)}&=&-\epsilon {}^4{\tilde M}_{(a)(b)}+{{\epsilon}\over 
{1+\bar \gamma}}
[\varphi_{(a)}\, {}^4{\tilde M}_{(b)(c)}-\varphi_{(b)}\, {}^4{\tilde
M}_{(a)(c)}]\varphi_{(c)}+\nonumber \\
&+& [\varphi_{(a)}\, {}^4{\tilde M}_{(b)(o)}-\varphi_{(b)}\, {}^4
{\tilde M}_{(a)(o)}]- [\varphi_{(a)}\, {}^4E^{\tau}_{(b)}-
\varphi_{(b)}\, {}^4E^{\tau}_{(a)}]{}^4{\tilde \pi}^{\tau}_{(o)}-\nonumber \\
&-&\epsilon [(\delta_{(a)(c)}\delta_{(d)(e)}-\delta_{(a)(e)}\delta_{(b)(c)})
(\delta_{(c)(d)}+{{\varphi_{(c)}\varphi_{(d)}}\over {1+\bar \gamma}})
{}^4E^{\tau}_{(c)}+\nonumber \\
&+&\epsilon ({}^4E^{\tau}_{(o)}+\epsilon 
{{\varphi_{(c)}\, {}^4E^{\tau}_{(c)}}\over {1+\bar \gamma }})
(\delta_{(a)(d)}\varphi_{(b)}-\delta_{(b)(d)}\varphi_{(a)})]{}^4{\tilde \pi}
^{\tau}_{(d)}\approx 0,\nonumber \\
{\tilde \pi}^{\vec \varphi}_{(a)}&=&\epsilon (\delta_{(a)(b)}-{{\varphi_{(a)}
\varphi_{(b)}}\over {\bar \gamma (1+\bar \gamma )}}) 
{}^4{\tilde M}_{(b)(o)}+{{1}
\over {1+\bar \gamma}}\, {}^4{\tilde M}_{(a)(b)} \varphi_{(b)}-\nonumber \\
&-&\epsilon (\delta_{(a)(b)}-{{\varphi_{(a)}\varphi_{(b)}}\over {\bar \gamma (1+
\bar \gamma )}})\, 
{}^4E^{\tau}_{(b)}\, {}^4{\tilde \pi}^{\tau}_{(o)}+\nonumber \\
&+&[\epsilon (\delta_{(a)(b)}-{{\varphi_{(a)}\varphi_{(b)}}\over {\bar \gamma 
(1+\bar \gamma )}})\, 
{}^4E^{\tau}_{(o)}-{{\varphi_{(c)}}\over {1+\bar \gamma}}(\delta
_{(c)(b)}\, {}^4E^{\tau}_{(a)}-\delta_{(c)(a)}\, {}^4E^{\tau}_{(b)})]
{}^4{\tilde \pi}^{\tau}_{(b)}\approx 0.
\label{III20}
\end{eqnarray}

Let us add a comment on the literature on tetrad gravity. The use of tetrads
(or vierbeins or local frames) started with Ref.\cite{weyl}, where vierbeins and
spin connections are used as independent variables in a Palatini form of the
Lagrangian. They were used by Dirac\cite{dirr} for the coupling of gravity to 
fermion fields (see also Ref.\cite{hen1}) and here $\Sigma_{\tau}$-adapted 
tetrads were introduced. In Ref.\cite{schw} the reduction of this theory at 
the Lagrangian level was done by introducing the so-called `time-gauge' ${}^4E
^{(o)}_r=0$ [or ${}^4E^o_{(a)}=0$], which distinguishes the time
coordinate $x^o=const.$ planes; in this paper there is also the coupling to
scalar fields, while in Ref.\cite{kib} the coupling to Dirac-Maiorana fields is
studied. In Ref.\cite{char}
there is a non-metric Lagrangian formulation, see Eq.(\ref{I15}), employing as
basic variables the cotetrads  ${}^4E^{(\alpha )}_{\mu}$,
which is different from our metric Lagrangian and has different primary
constraints; its Hamiltonian formulation is completely developed. See also
Ref.\cite{clay} for a study of the tetrad frame constraint algebra. In the
fourth of Refs.\cite{tetr} cotetrads ${}^4E^{(\alpha )}_{\mu}$ together with
the spin connection ${}^4\omega^{(\alpha )}_{\mu (\beta )}$ are used as
independent variables in a first order Palatini action [see also the
Nelson-Regge papers in Refs.\cite{tetr} for a different approach, the so-called
covariant canonical formalism], while in Ref.\cite{maluf} a first order
Lagrangian reformulation is done for Eq.(\ref{I15}) 
[in both these papers there is
a 3+1 decomposition of the tetrads different from our and, 
like in Ref.\cite{maluf}, use is done of the 
Schwinger time gauge to get free of three boost-like parameters].

Instead in most of Refs.\cite{tetr,hen2,hen3} one uses the space components
${}^4E^{(\alpha )}_r$ of cotetrads ${}^4E^{(\alpha )}_{\mu}$, together with 
the conjugate momenta ${}^4{\tilde \pi}^r_{(\alpha )}$ inside the ADM 
Hamiltonian, in which one puts ${}^3g_{rs}={}^4E_r^{(\alpha )}\, {}^4\eta
_{(\alpha )(\beta )}\, {}^4E^{(\beta )}_s$ and ${}^3{\tilde \Pi}^{rs}=
{1\over 4}{}^4\eta^{(\alpha )(\beta )}[{}^4E^r_{(\alpha )}\, {}^4{\tilde \pi}
^s_{(\beta )}+{}^4E^s_{(\alpha )}\, {}^4{\tilde \pi}^r_{(\beta )}]$. Lapse
and shift functions are treated as Hamiltonian multipliers and there is no
worked out Lagrangian formulation. In Ref.\cite{hen4} it is shown how to go from
the space components ${}^4E^{(\alpha )}_r$ to cotriads ${}^3e_{(a)r}$ by using 
the ``time gauge" on a surface $x^0=const.$; here it is introduced 
for the first time the concept of parameters of Lorentz boosts [if they are
put equal to zero, one recovers Schwinger's time gauge], which was our 
starting point to arrive at the identification of the Wigner boost parameters
$\varphi_{(a)}$. Finally in Ref.\cite{gold} there is a 3+1 decomposition of 
tetrads and cotetrads in which some boost-like parameters have been fixed (it 
is a Schwinger time gauge) so that one can arrive at a Lagrangian (different 
from ours) depending only on lapse, shift and cotriads.

In Ref.\cite{hen4} there is another canonical transformation from cotriads
and their conjugate momenta to a new canonical basis containing densitized
triads and their conjugate momenta

\begin{eqnarray}
(\, {}^3e_{(a)r}&,& {}^3{\tilde \pi}^r_{(a)}\, ) \mapsto 
(\,\,\, {}^3{\tilde h}^r_{(a)}
={}^3e\, {}^3e^r_{(a)},\nonumber \\
&&2\, {}^3K_{(a)r}=
2[{}^3e^s_{(a)}\, {}^3K_{sr}+{1\over {4\, {}^3e}} {}^3{\tilde M}_{(a)(b)}\,
{}^3e_{(b)r}]=\nonumber \\
&=&{1\over 2}[{1\over k}{}^3G_{o(a)(b)(c)(d)}\, {}^3e_{(b)r}\, 
{}^3e_{(c)u}\, {}^3{\tilde \pi}^u_{(d)}+{1\over {{}^3e}}{}^3{\tilde M}
_{(a)(b)}\, {}^3e_{(b)r}]\, ),
\label{III21}
\end{eqnarray}

\noindent which is used to make the transition to the complex Ashtekar 
variables \cite{ash}

\begin{equation}
(\, {}^3{\tilde h}^r_{(a)},\quad\quad {}^3A_{(a)r}=2\,\, {}^3K_{(a)r}+i 
\, {}^3\omega_{r(a)}\, ),
\label{III22}
\end{equation}

\noindent where ${}^3A_{(a)r}$ is a zero density whose real part (in this 
notation) can be considered the
gauge potential of the Sen connection  and plays an
important role in the simplification of the functional form of the constraints
present in this approach; the
conjugate variable is a density 1 SU(2) soldering form.

\vfill\eject

\section{Comparison with ADM Canonical Metric Gravity.}

In this Section  we give a brief review of the Hamiltonian formulation
of ADM metric gravity [see Refs.\cite{witt,mtw,ish,ro,fm}] to express its
constraints in terms of those of Section IV.

Let us rewrite Eq.(\ref{I10})  in terms of the independent variables
N, $N_r={}^3g_{rs}N^s$, ${}^3g_{rs}$ as $S_{ADM}=\int d\tau \, L_{ADM}(\tau )=
\int d\tau d^3\sigma {\cal L}_{ADM}(\tau ,\vec \sigma )=-
\epsilon k\int_{\triangle \tau}d\tau  \, \int d^3\sigma \, \lbrace 
\sqrt{\gamma} N\, [{}^3R+{}^3K_{rs}\, {}^3K^{rs}-({}^3K)^2]\rbrace (\tau ,\vec 
\sigma )$. Since $\delta_o(\sqrt{\gamma}
{}^3R)=\sqrt{\gamma} ({}^3R_{rs}-{1\over 2}\, {}^3g_{rs}\, {}^3R) \delta_o\, 
{}^3g^{rs} +\sqrt{\gamma} ({}^3g_{rs}\delta_o\, {}^3g^{rs |u}-\delta_o\, 
{}^3g^{ru}{}_{|r})_{|u}$, we get\hfill\break
$\delta_oS_{ADM}=\int d\tau d^3\sigma \Big( L_N \delta_oN +L^r_{\vec N} \delta
_o N_r +L^{rs}_g \delta_o\, {}^3g_{rs}$\hfill\break
$+\partial_{\tau}\Big[ \epsilon k \sqrt{\gamma}({}^3K^{rs}-{}^3g^{rs}\, {}^3K)
\delta_o\, {}^3g_{rs}\Big] -\partial_r\Big[ 2\epsilon k \sqrt{\gamma}({}^3K^{rs}
-{}^3g^{rs}\, {}^3K)\delta_o\, N_s\Big] $\hfill\break
$+\partial_r\Big[ \epsilon k \sqrt{\gamma} (N[{}^3g^{uv}\delta_o\, 
{}^3g_{uv}{}^{|r}-{}^3g^{ur}\delta_o\, {}^3g_{uv}{}^{|v}] +N^{|u}\, {}^3g^{rs}
\delta_o\, {}^3g_{us}-N^{|r}\, {}^3g^{uv}\delta_o\, {}^3g_{uv}\Big] \Big) ,$
\hfill\break
so that the Euler-Lagrange equations are

\begin{eqnarray}
L_N&=&{{\partial {\cal L}_{ADM}}\over {\partial N}}-\partial_{\tau}
{{\partial {\cal L}_{ADM}}\over {\partial \partial_{\tau}N}}-\partial_r
{{\partial {\cal L}_{ADM}}\over {\partial \partial_rN}}=\nonumber \\
&=&-\epsilon k
\sqrt{\gamma} [{}^3R-{}^3K_{rs}\, {}^3K^{rs}+({}^3K)^2]=-2\epsilon k\, 
{}^4{\bar G}_{ll}\, {\buildrel \circ 
\over =}\, 0,\nonumber \\
L^r_{\vec N}&=&{{\partial {\cal L}_{ADM}}\over {\partial N_r}}-\partial_{\tau}
{{\partial {\cal L}_{ADM}}\over {\partial \partial_{\tau}N_r}}-\partial_s
{{\partial {\cal L}_{ADM}}\over {\partial \partial_sN_r}}=\nonumber \\
&=&2\epsilon k
[\sqrt{\gamma}({}^3K^{rs}-{}^3g^{rs}\, {}^3K)]_{\, |s}=2k\, {}^4{\bar G}_l{}^r
\, {\buildrel \circ \over =}\, 0,\nonumber \\
L_g^{rs}&=& -\epsilon k \Big[
{{\partial}\over {\partial \tau}}[\sqrt{\gamma}({}^3K^{rs}-{}^3g
^{rs}\, {}^3K)]\, - N\sqrt{\gamma}({}^3R^{rs}-
{1\over 2} {}^3g^{rs}\, {}^3R)+\nonumber \\
&&+2N\, \sqrt{\gamma}({}^3K^{ru}\, {}^3K_u{}^s-{}^3K\, {}^3K^{rs})+{1\over 2}N 
\sqrt{\gamma}[({}^3K)^2-{}^3K_{uv}\, {}^3K^{uv}){}^3g^{rs}+\nonumber \\
&&+\sqrt{\gamma} ({}^3g^{rs} N^{|u}{}_{|u}-N^{|r |s})\Big] =-\epsilon kN
\sqrt{\gamma}\, {}^4{\bar G}^{rs}\, {\buildrel \circ \over =}\,0,
\label{V0}
\end{eqnarray}

\noindent and correspond to the Einstein equations in the form ${}^4{\bar G}
_{ll}\, {\buildrel \circ \over =}\, 0$, ${}^4{\bar G}_{lr}\, {\buildrel \circ 
\over =}\, 0$, ${}^4{\bar G}_{rs}\, {\buildrel \circ \over =}\, 0$, 
respectively. As shown after Eq.(\ref{I6}) there are four contracted Bianchi
identities implying that only two of the equations $L^{rs}_g\, {\buildrel \circ
\over =}\, 0$ are independent.

The canonical momenta (densities of weight -1) are

\begin{eqnarray}
&&{\tilde \Pi}^N(\tau ,\vec \sigma )={{\delta S_{ADM}}\over {\delta
\partial_{\tau}N(\tau ,\vec \sigma )}} =0,\nonumber \\
&&{\tilde \Pi}^r_{\vec N}(\tau ,\vec \sigma )={{\delta S_{ADM}}\over
{\delta \partial_{\tau} N_r(\tau ,\vec \sigma )}} =0,\nonumber \\
&&{}^3{\tilde \Pi}^{rs}(\tau ,\vec \sigma )={{\delta S_{ADM}}\over
{\delta \partial_{\tau} {}^3g_{rs}(\tau ,\vec \sigma )}}=\epsilon k\, [
\sqrt{\gamma}({}^3K^{rs}-{}^3g^{rs}\, {}^3K)](\tau ,\vec \sigma ),\nonumber \\
&&{}\nonumber \\
&&{}^3K_{rs}={{\epsilon}\over {k\sqrt{\gamma}}} [{}^3{\tilde \Pi}_{rs}-{1\over 
2}{}^3g_{rs}\, {}^3{\tilde \Pi}],\quad\quad {}^3\tilde \Pi ={}^3g_{rs}\, 
{}^3{\tilde \Pi}^{rs}=-2\epsilon k\sqrt{\gamma}\, {}^3K,
\label{c1}
\end{eqnarray}

\noindent and satisfy the Poisson brackets

\begin{eqnarray}
&&\lbrace N(\tau ,\vec \sigma ),{\tilde \Pi}^N(\tau ,{\vec \sigma}^{'} )
\rbrace =\delta^3(\vec \sigma ,{\vec \sigma}^{'}),\nonumber \\
&&\lbrace N_r(\tau ,\vec \sigma ),{\tilde \Pi}^s_{\vec N}(\tau ,{\vec \sigma}
^{'} )\rbrace =\delta^s_r \delta^3(\vec \sigma ,{\vec \sigma}^{'}),\nonumber \\
&&\lbrace {}^3g_{rs}(\tau ,\vec \sigma ),{}^3{\tilde \Pi}^{uv}(\tau ,{\vec 
\sigma}^{'}\rbrace = {1\over 2} (\delta^u_r\delta^v_s+\delta^v_r\delta^u_s)
\delta^3(\vec \sigma ,{\vec \sigma}^{'}).
\label{c2}
\end{eqnarray}

Let us introduce a new tensor, the Wheeler- DeWitt supermetric

\begin{equation}
{}^3G_{rstw}(\tau ,\vec \sigma )=[{}^3g_{rt}\, {}^3g_{sw}+{}^3g_{rw}\, {}^3g
_{st}-{}^3g_{rs}\, {}^3g_{tw}](\tau ,\vec \sigma ),
\label{c3}
\end{equation}

\noindent whose inverse is defined by the equations

\begin{eqnarray}
{}&&{1\over 2} {}^3G_{rstw}\, {1\over 2} {}^3G^{twuv} ={1\over 2}(\delta^u_r
\delta^v_s+\delta^v_r\delta^u_s),\nonumber \\
{}^3G^{twuv}(\tau ,\vec \sigma )&=&[{}^3g^{tu}\, {}^3g^{wv}+{}^3g^{tv}\, {}^3g
^{wu}-2\, {}^3g^{tw}\, {}^3g^{uv}](\tau ,\vec \sigma ),
\label{c4}
\end{eqnarray}

\noindent so that we get

\begin{eqnarray}
{}^3{\tilde \Pi}^{rs}(\tau ,\vec \sigma )&=&{1\over 2}\epsilon k \sqrt{\gamma}\,
{}^3G^{rsuv}(\tau ,\vec \sigma )\, {}^3K_{uv}(\tau ,\vec \sigma ),\nonumber \\
{}^3K_{rs}(\tau ,\vec \sigma )&=&{{\epsilon}\over {2k\sqrt{\gamma}}}\,
{}^3G_{rsuv}(\tau ,\vec \sigma )\, {}^3{\tilde \Pi}^{uv}(\tau ,\vec \sigma ),
\nonumber \\
&&[{}^3K^{rs}\, {}^3K_{rs}-({}^3K)^2](\tau ,\vec \sigma )=\nonumber \\
&&=k^{-2}[\gamma^{-1}({}^3{\tilde \Pi}^{rs}\, {}^3{\tilde \Pi}_{rs}-{1\over 2}
({}^3{\tilde \Pi})^2](\tau ,\vec \sigma )=(2k)^{-1}[\gamma^{-1}\, {}^3G_{rsuv}
\, {}^3{\tilde \Pi}^{rs}\, {}^3{\tilde \Pi}^{uv}](\tau ,\vec \sigma ),
\nonumber \\
\partial_{\tau}\, {}^3g_{rs}(\tau ,\vec \sigma )&=&[N_{r|s}+N_{s|r}-{{\epsilon
N}\over {k\sqrt{\gamma}}}\, {}^3G_{rsuv}\, {}^3{\tilde \Pi}^{uv}](\tau ,
\vec \sigma ).
\label{c5}
\end{eqnarray}

Since ${}^3{\tilde \Pi}^{rs}\partial_{\tau}\, {}^3g_{rs}=$${}^3{\tilde \Pi}
^{rs} [N_{r | s}+N_{s | r}-{{\epsilon N}\over {k\sqrt{\gamma}}} {}^3G_{rsuv}
{}^3{\tilde \Pi}^{uv}]=$$-2 N_r {}^3{\tilde \Pi}^{rs}{}_{| s}-{{\epsilon N}\over
{k\sqrt{\gamma}}}\, {}^3G_{rsuv}\, {}^3{\tilde \Pi}^{rs} {}^3{\tilde \Pi}^{uv}+
(2N_r\, {}^3{\tilde \Pi}^{rs})_{| s}$, we obtain the canonical Hamiltonian
[since $N_r\, {}^3{\tilde \Pi}^{rs}$ is a vector density of weight -1, we have
${}^3\nabla_s(N_r\, {}^3{\tilde \Pi}^{rs})=\partial_s(N_r\, {}^3{\tilde \Pi}
^{rs})$]

\begin{eqnarray}
H_{(c)ADM}&=& \int_Sd^3\sigma \, [{\tilde \Pi}^N\partial_{\tau}N+{\tilde
\Pi}^r_{\vec N}\partial_{\tau}N_r+{}^3{\tilde \Pi}^{rs}\partial_{\tau}
{}^3g_{rs}](\tau ,\vec \sigma ) -L_{ADM}=\nonumber \\
&=&\int_Sd^3\sigma \, [\epsilon N(k\sqrt{\gamma}\, {}^3R-
{1\over {2k\sqrt{\gamma}}} {}^3G_{rsuv}{}^3{\tilde \Pi}^{rs} {}^3{\tilde
\Pi}^{uv})-2N_r\, {}^3{\tilde \Pi}^{rs}{}_{| s}](\tau ,\vec \sigma )+
\nonumber \\
&+&2\int_{\partial S}d^2\Sigma_s [N_r\, {}^3{\tilde \Pi}^{rs}\,\, 
](\tau ,\vec \sigma ),
\label{c6}
\end{eqnarray}

\noindent In the following discussion we shall omit the surface term.

The Dirac Hamiltonian is [the $\lambda (\tau ,\vec \sigma )$'s are arbitrary
Dirac multipliers]

\begin{equation}
H_{(D)ADM}=H_{(c)ADM}+\int d^3\sigma \, [\lambda_N\, {\tilde \Pi}^N + \lambda
^{\vec N}_r\, {\tilde \Pi}^r_{\vec N}](\tau ,\vec \sigma ).
\label{c7}
\end{equation}

The $\tau$-constancy of the primary constraints [$\partial_{\tau} {\tilde
\Pi}^N(\tau ,\vec \sigma )=\lbrace {\tilde \Pi}^N(\tau ,\vec \sigma ),H_{(D)
ADM}\rbrace \approx 0$, $\partial_{\tau} {\tilde \Pi}^r_{\vec N}(\tau ,\vec 
\sigma )=\lbrace {\tilde \Pi}^r_{\vec N}(\tau ,\vec \sigma ),H_{(D)ADM}
\rbrace \approx 0$] generates four secondary constraints [all 4 are densities
of weight -1] which correspond to the Einstein equations ${}^4{\bar G}_{ll}
(\tau ,\vec \sigma )\, {\buildrel \circ \over =}\, 0$, ${}^4{\bar G}_{lr}
(\tau ,\vec \sigma )\, {\buildrel \circ \over =}\, 0$ [see after Eqs.(\ref{I6})]

\begin{eqnarray}
{\tilde {\cal H}}(\tau ,\vec \sigma )&=&\epsilon
[k\sqrt{\gamma}\, {}^3R-{1\over {2k
\sqrt{\gamma}}} {}^3G_{rsuv}\, {}^3{\tilde \Pi}^{rs}\, {}^3{\tilde \Pi}^{uv}]
(\tau ,\vec \sigma )=\nonumber \\
&=&\epsilon [\sqrt{\gamma}\, {}^3R-{1\over {k\sqrt{\gamma}}}({}^3{\tilde \Pi}
^{rs}\, {}^3{\tilde \Pi}_{rs}-{1\over 2}({}^3\tilde \Pi )^2)](\tau ,\vec 
\sigma )=\nonumber \\
&=&\epsilon k \{ \sqrt{\gamma} [{}^3R-({}^3K_{rs}\, {}^3K^{rs}-({}^3K)^2 )]\}
(\tau ,\vec \sigma )\approx 0,
\nonumber \\
{}^3{\tilde {\cal H}}^r(\tau ,\vec \sigma )&=&-2\, {}^3{\tilde \Pi}^{rs}{}_{| s}
(\tau ,\vec \sigma )=-2[\partial_s\, {}^3{\tilde \Pi}^{rs}+{}^3\Gamma^r_{su}
{}^3{\tilde \Pi}^{su}](\tau ,\vec \sigma )=\nonumber \\
&=&-2\epsilon k \{ \partial_s[\sqrt{\gamma}({}^3K^{rs}-{}^3g^{rs}\, {}^3K)]+
{}^3\Gamma^r_{su}\sqrt{\gamma}({}^3K^{su}-{}^3g^{su}\, {}^3K) \}
(\tau ,\vec \sigma )\approx 0,
\label{c8}
\end{eqnarray}

\noindent so that we have

\begin{equation}
H_{(c)ADM}= \int d^3\sigma [N\, {\tilde {\cal H}}+N_r\, {}^3{\tilde
{\cal H}}^r](\tau ,\vec \sigma )\approx 0,
\label{c9}
\end{equation}

\noindent with ${\tilde {\cal H}}(\tau ,\vec \sigma )\approx 0$ called the
superhamiltonian constraint and ${}^3{\tilde {\cal H}}^r(\tau ,\vec \sigma )
\approx 0$ called the supermomentum constraints. See Ref.\cite{tei} for
their interpretation as the generators of the change of the canonical data 
${}^3g_{rs}$, ${}^3{\tilde \Pi}^{rs}$, under the normal and tangent 
deformations of the spacelike hypersurface $\Sigma_{\tau}$ which generate 
$\Sigma_{\tau +d\tau}$ [one thinks to $\Sigma_{\tau}$ as determined by a cloud
of observers, one per space point; the idea of bifurcation and reencounter of 
the observers is expressed by saying that the data on $\Sigma_{\tau}$ (where
the bifurcation took place) are propagated to some final $\Sigma_{\tau +
d\tau}$ (where the reencounter arises) along different intermediate paths, each
path being a monoparametric family of surfaces that fills the sandwich in 
between the two surfaces; embeddability of $\Sigma_{\tau}$ in $M^4$ becomes the
synonymous with path independence; see also Ref.\cite{gr13} for the connection
with the theorema egregium of Gauss).

In ${\tilde {\cal H}}(\tau ,\vec \sigma )\approx 0$ one can say that the term 
$-\epsilon k \sqrt{\gamma}({}^3K_{rs}\, {}^3K^{rs}-{}^3K^2)$ is the kinetic 
energy and $\epsilon k\sqrt{\gamma}\, {}^3R$ the potential energy: in any Ricci 
flat spacetime
(i.e. one satisfying Einstein's empty-space equations) the extrinsic and
intrinsic scalar curvatures of any spacelike hypersurface $\Sigma_{\tau}$
are both equal to zero (also the converse is true\cite{jawh}).

All the constraints are first class, because the only non-identically zero
Poisson brackets correspond to the so called universal Dirac algebra 
\cite{dirac}:

\begin{eqnarray}
\lbrace {}^3{\tilde {\cal H}}_r(\tau ,\vec \sigma ),{}^3{\tilde {\cal H}}_s
(\tau ,{\vec \sigma}^{'})\rbrace &=&{}\nonumber \\
&=&{}^3{\tilde {\cal H}}_r(\tau ,{\vec 
\sigma}^{'} )\, {{\partial \delta^3(\vec \sigma ,{\vec \sigma}^{'})}\over
{\partial \sigma^s}} + {}^3{\tilde {\cal H}}_s(\tau ,\vec \sigma ) {{\partial
\delta^3(\vec \sigma ,{\vec \sigma}^{'})}\over {\partial \sigma^r}},
\nonumber \\
\lbrace {\tilde {\cal H}}(\tau ,\vec \sigma ),{}^3{\tilde {\cal H}}_r(\tau ,
{\vec \sigma}^{'})\rbrace &=& {\tilde {\cal H}}(\tau ,\vec \sigma )
{{\partial \delta^3(\vec \sigma ,{\vec \sigma}^{'})}\over {\partial \sigma^r}},
\nonumber \\
\lbrace {\tilde {\cal H}}(\tau ,\vec \sigma ),{\tilde {\cal H}}(\tau ,{\vec 
\sigma}^{'})\rbrace &=&[{}^3g^{rs}(\tau ,\vec \sigma ) {}^3{\tilde {\cal H}}_s
(\tau ,\vec \sigma )+\nonumber \\
&+&{}^3g^{rs}(\tau ,{\vec \sigma}^{'}){}^3{\tilde
{\cal H}}_s(\tau ,{\vec \sigma}^{'})]{{\partial \delta^3(\vec \sigma ,{\vec 
\sigma}^{'})}\over {\partial \sigma^r}},
\label{c10}
\end{eqnarray}

\noindent with ${}^3{\tilde {\cal H}}_r={}^3g_{rs}\, {}^3{\tilde {\cal H}}^r$
as the combination of the supermomentum constraints satisfying the algebra of
3-diffeomorphisms. In Ref.\cite{tei} it is shown that Eqs.(\ref{c10}) are 
sufficient conditions for the embeddability of $\Sigma_{\tau}$ into $M^4$.
In the second paper in Ref.\cite{kuchar} it is shown that the last two lines 
of the Dirac algebra are the equivalent in phase space of the Bianchi 
identities  ${}^4G^{\mu\nu}{}_{;\nu}\equiv 0$.

The Hamilton-Dirac equations are 

\begin{eqnarray}
\partial_{\tau}N(\tau ,\vec \sigma )\, &{\buildrel \circ \over =}\,& 
\lbrace N(\tau ,\vec \sigma ),H_{(D)ADM}
\rbrace =\lambda_N(\tau ,\vec \sigma ),\nonumber \\
\partial_{\tau}N_r(\tau ,\vec \sigma )\, &{\buildrel \circ \over =}\,&
\lbrace N_r(\tau ,\vec \sigma ),
H_{(D)ADM}\rbrace =\lambda^{\vec N}_r(\tau ,\vec \sigma ),\nonumber \\
\partial_{\tau}\, {}^3g_{rs}(\tau ,\vec \sigma )\, &{\buildrel \circ \over =}\,&
\lbrace {}^3g_{rs}(\tau ,
\vec \sigma ),H_{(D)ADM}\rbrace =[N_{r | s}+N_{s | r}-{{2\epsilon N}\over
{k\sqrt{\gamma}}}({}^3{\tilde \Pi}_{rs}-{1\over 2}{}^3g_{rs}\, {}^3{\tilde 
\Pi})](\tau ,\vec \sigma )=\nonumber \\
&=&[N_{r|s}+N_{s|r}-2N\, {}^3K_{rs}](\tau ,\vec \sigma ),\nonumber \\
\partial_{\tau}\, {}^3{\tilde \Pi}^{rs}(\tau ,\vec \sigma )
\, &{\buildrel \circ \over =}\,& \lbrace {}^3
{\tilde \Pi}^{rs}(\tau ,\vec \sigma ),H_{(D)ADM}\rbrace =\epsilon [N\, 
k\sqrt{\gamma} ({}^3R^{rs}-{1\over 2}{}^3g^{rs}\, {}^3R)](\tau ,\vec \sigma )-
\nonumber \\
&-&2\epsilon [{N\over {k\sqrt{\gamma}}}({1\over 2}{}^3{\tilde \Pi}\, {}^3{\tilde
\Pi}^{rs}-{}^3{\tilde \Pi}^r{}_u\, {}^3{\tilde \Pi}^{us})(\tau ,\vec \sigma )-
\nonumber \\
&-&{{\epsilon N}\over 2}
{{{}^3g^{rs}}\over {k\sqrt{\gamma}}}({1\over 2}{}^3{\tilde \Pi}^2-{}^3{\tilde
\Pi}_{uv}\, {}^3{\tilde \Pi}^{uv})](\tau ,\vec \sigma )+\nonumber \\
&+&{\cal L}_{\vec N}\, {}^3{\tilde \Pi}^{rs}(\tau ,\vec \sigma )+\epsilon
[k\sqrt{\gamma} (N^{| r | s}-{}^3g^{rs}\, N^{| u}{}_{| u})](\tau ,\vec \sigma ),
\nonumber \\
&&with\quad\quad {\cal L}_{\vec N}\, {}^3{\tilde \Pi}^{rs}=\epsilon \Big[ 
({}^3{\tilde \pi}^{rs}N^u)_{|u}-N^r{}_{|u}\, {}^3{\tilde \pi}^{us}-N^s{}_{|u}
\, {}^3{\tilde \pi}^{ur}\Big] ,\nonumber \\
&&\Downarrow \nonumber \\
\partial_{\tau}\, {}^3K_{rs}(\tau ,\vec \sigma )\, &{\buildrel \circ \over =}\,&
\Big( N [{}^3R_{rs}+{}^3K\, {}^3K_{rs}-2\, {}^3K_{ru}\, {}^3K^u{}_s]-
\nonumber \\
&-&N_{|s|r}+N^u{}_{|s}\, {}^3K_{ur}+N^u{}_{|r}\, {}^3K_{us}+N^u\, {}^3K_{rs|u}
\Big) (\tau ,\vec \sigma ),\nonumber \\
\partial_{\tau}\, \gamma (\tau ,\vec \sigma )\, &{\buildrel \circ \over =}\,&
\Big( 2 \gamma [-N\, {}^3K +N^u{}_{|u}]\Big) (\tau ,\vec \sigma ),
\nonumber \\
\partial_{\tau}\, {}^3K(\tau ,\vec \sigma )\, &{\buildrel \circ \over =}\,&
\Big( N [{}^3g^{rs}\, {}^3R_{rs} +({}^3K)^2] -N_{|u}{}^{|u}+N^u\, {}^3K_{|u}
\Big) (\tau ,\vec \sigma ),
\label{c11}
\end{eqnarray}

\noindent with \hfill\break
\hfill\break
${\cal L}_{\vec N}\, {}^3{\tilde \Pi}^{rs}=-\sqrt{\gamma}\,
{}^3\nabla_u({{N^u}\over {\sqrt{\gamma}}} {}^3{\tilde \Pi}^{rs})+
{}^3{\tilde \Pi}^{ur}\, {}^3\nabla_u N^s+{}^3{\tilde \Pi}^{us}\, {}^3\nabla_u 
N^r$.\hfill\break
\hfill\break
We have also used \hfill\break
$\delta (\sqrt{\gamma}\, {}^3R)(\tau ,\vec \sigma )=
\int d^3\sigma_1 \lbrace (\sqrt{\gamma}\, {}^3R)(\tau ,\vec \sigma ), 
{}^3{\tilde \Pi}^{rs}(\tau ,{\vec \sigma}_1) \rbrace \delta \, {}^3g_{rs}
(\tau ,{\vec \sigma}_1)=\int d^3\sigma_1 \delta \, {}^3g_{rs}(\tau ,{\vec 
\sigma}_1) \{ [-\sqrt{\gamma} ({}^3R^{rs}-{1\over 2} {}^3g^{rs}\, {}^3R)](\tau 
,\vec \sigma ) \delta^3(\vec \sigma ,{\vec \sigma}_1)+[\sqrt{\gamma}\, 
{}^3\Gamma^n_{lm}({}^3g^{rl}\, {}^3g^{sm}-{}^3g^{rs}\, {}^3g^{lm})](\tau ,{\vec 
\sigma}_1){{\partial \delta^3(\vec \sigma ,{\vec \sigma}_1)}\over {\partial
\sigma^n}}+[\sqrt{\gamma} ({}^3g^{rl}\, {}^3g^{sm}-{}^3g^{rs}\, {}^3g^{lm})]
(\tau ,{\vec \sigma}_1) {{\partial^2 \delta^3(\vec \sigma ,{\vec \sigma}_1)}
\over {\partial \sigma^l\partial \sigma^m}} \}$.

The above equation for $\partial_{\tau}\, {}^3g_{rs}(\tau ,\vec \sigma )$
shows that the generator of space diffeomorphisms $\int d^3\sigma N_r(\tau
,\vec \sigma )\, {}^3{\tilde {\cal H}}^r(\tau ,\vec \sigma )$ produces a
variation, tangent to $\Sigma_{\tau}$, $\delta_{tangent} {}^3g_{rs}={\cal L}
_{\vec N}\, {}^3g_{rs}=N_{r|s}+N_{s|r}$ in accord with the infinitesimal
pseudodiffeomorphisms in $Diff\, \Sigma_{\tau}$. Instead, the superhamiltonian
generator $\int d^3\sigma N(\tau ,\vec \sigma )\, {\tilde {\cal H}}(\tau ,\vec 
\sigma )$ does not reproduce the infinitesimal diffeomorphisms in $Diff\, M^4$
normal to $\Sigma_{\tau}$ (see also Ref.\cite{wa}). In Ref.\cite{anton}
there is a study of the assumptions hidden in the ADM formulation
(essentially the embedding of the model $\Sigma$ hypersurface in $M^4$ is
fixed and not variable), whose relaxation allows to turn an arbitrary normal
deformation to $\Sigma_{\tau}$ (as an element of $Diff\, M^4$) into the
deformation $-2N(\tau ,\vec \sigma )\, {}^3K_{rs}(\tau ,\vec \sigma )$
generated by the superhamiltonian constraint.

Let us remark that the canonical transformation [${}^4g_{AB}$ and ${}^4g^{AB}$
are given in Eqs.(\ref{I1})] ${\tilde \Pi}^N\, dN+{\tilde \Pi}^r_{\vec N}\,
dN_r+{}^3{\tilde \Pi}^{rs}\, d{}^3g_{rs} = {}^4{\tilde \Pi}^{AB}\, d{}^4g
_{AB}$ defines the following momenta conjugated to ${}^4g_{AB}$

\begin{eqnarray}
{}^4{\tilde \Pi}^{\tau\tau}&=&{{\epsilon}\over {2N}} {\tilde \Pi}^N,\nonumber \\
{}^4{\tilde \Pi}^{\tau r}&=&{{\epsilon}\over 2} ({{N^r}\over N} {\tilde \Pi}
^N-{\tilde \Pi}^r_{\vec N}),\nonumber \\
{}^4{\tilde \Pi}^{rs}&=&\epsilon ({{N^rN^s}\over {2N}} {\tilde \Pi}^N-
{}^3{\tilde \Pi}^{rs}),
\nonumber \\
&&{}\nonumber \\
&&\lbrace {}^4g_{AB}(\tau ,\vec \sigma ),{}^4{\tilde \Pi}^{CD}(\tau ,{\vec 
\sigma}^{'} )\rbrace ={1\over 2}(\delta^C_A\delta^D_B+\delta^D_A\delta^C_B)
\delta^3(\vec \sigma ,{\vec \sigma}^{'}),\nonumber \\
&&{}\nonumber \\
{\tilde \Pi}^N&=& {{2\epsilon}\over {\sqrt{\epsilon {}^4g^{\tau\tau}}}}
{}^4{\tilde \Pi}^{\tau\tau},\nonumber \\
{\tilde \Pi}^r_{\vec N}&=&2\epsilon {{{}^4g^{\tau r}}\over {{}^4g^{\tau\tau}}}
{}^4{\tilde \Pi}^{\tau\tau}-2\epsilon {}^4{\tilde \Pi}^{\tau r},\nonumber \\
{}^3{\tilde \Pi}^{rs}&=&\epsilon {{{}^4g^{\tau r} {}^4g^{\tau S}}\over
{({}^4g^{\tau\tau})^2}}\, {}^4{\tilde \Pi}^{\tau\tau} -\epsilon {}^4{\tilde
\Pi}^{rs},
\label{c12}
\end{eqnarray}

\noindent which would emerge if the ADM action would be considered function
of ${}^4g_{AB}$ instead of N, $N_r$ and ${}^3g_{rs}$.

The standard ADM momenta ${}^3{\tilde \Pi}^{rs}$, defined in Eq.
(\ref{c1}), may now be expressed in terms of the cotriads and their conjugate 
momenta of the canonical formulation of tetrad gravity given in Section IV:

\begin{eqnarray}
{}^3{\tilde \Pi}^{rs}&=&\epsilon k\sqrt{\gamma} ({}^3K^{rs}-{}^3g^{rs}\, {}^3K)=
{{1}\over 4} [{}^3e^r_{(a)}\, {}^3{\tilde \pi}^s_{(a)}+{}^3e
^s_{(a)}\, {}^3{\tilde \pi}^r_{(a)}],\nonumber \\
&\Rightarrow& {}^3{\tilde \Pi}={}^3{\tilde \Pi}^{rs}\, {}^3g_{rs}=
-2\epsilon k\sqrt{\gamma} \, {}^3K=
{{1}\over 2} {}^3e_{(a)r}\, {}^3{\tilde \pi}^r_{(a)},\nonumber \\
&&{}\nonumber \\
&&\lbrace {}^3g_{rs}(\tau ,\vec \sigma )=
{}^3e_{(a)r}(\tau ,\vec \sigma )\, {}^3e_{(a)s}(\tau ,\vec \sigma ),
{}^3{\tilde \Pi}^{uv}(\tau ,{\vec \sigma}^{'})\rbrace ={1\over 2}(\delta^u_r
\delta^v_s+\delta^u_s \delta^v_r)\delta^3(\vec \sigma ,{\vec \sigma}^{'}),
\nonumber \\
&&\lbrace {}^3{\tilde \Pi}^{rs}(\tau ,\vec \sigma ),{}^3{\tilde \Pi}^{uv}(\tau ,
{\vec \sigma}^{'})\rbrace ={1\over {8}} \delta^3(\vec \sigma ,{\vec \sigma}
^{'}) \times \nonumber \\
&&{} [{}^3g^{ru}\, {}^3e^v_{(a)}\, {}^3e^s_{(b)}+{}^3g^{rv}\, {}^3e^u_{(a)}\,
{}^3e^s_{(b)}+{}^3g^{su}\, {}^3e^v_{(a)}\, {}^3e^r_{(b)}+{}^3g^{sv}\,
{}^3e^u_{(a)}\, {}^3e^r_{(b)}](\tau ,\vec \sigma )\cdot \nonumber \\
&& \cdot {}^3{\tilde M}_{(a)(b)}(\tau ,\vec \sigma )\approx 0.\nonumber \\
&&{}
\label{III10}
\end{eqnarray}

\noindent The fact that in tetrad gravity the last Poisson brackets is only 
weakly zero has been noted in Ref.\cite{hen2}.

Let us now consider the expression of the ADM supermomentum constraints in
tetrad gravity.
Since ${}^3e_{(b)u}\, {}^3{\tilde \Pi}^{us}={{1}\over 4}{}^3e_{(b)u}
[{}^3e^u_{(a)}\, {}^3{\tilde \pi}^s_{(a)}+{}^3e^s_{(a)}\, {}^3{\tilde \pi}^u
_{(a)}]={{1}\over 4}[{}^3{\tilde \pi}^s_{(b)}+{}^3e^s_{(a)}\, {}^3e
_{(b)u}\, {}^3{\tilde \pi}^u_{(a)}]={{1}\over 4}[{}^3{\tilde \pi}^s_{(b)}
+{}^3e^s_{(a)} ({}^3e_{(a)u}\, {}^3{\tilde \pi}^u_{(b)}+{}^3{\tilde M}_{(b)(a)}
)]={{1}\over 4}[2\, {}^3{\tilde \pi}^s_{(b)}-{}^3e^s_{(a)}\, 
{}^3{\tilde M}_{(a)(b)}]$, we have

\begin{eqnarray}
&&{}^3{\tilde \Pi}^{rs}{}_{|s}=\partial_s\, {}^3{\tilde \Pi}^{rs}+{}^3\Gamma^r
_{su}\, {}^3{\tilde \Pi}^{us}=\nonumber \\
&&=\partial_s\, {}^3{\tilde \Pi}^{rs}+[\epsilon_{(a)(b)(c)}\, {}^3e^r_{(a)}\, 
{}^3\omega_{s(c)}-\partial_s\, {}^3e^r_{(b)}] {}^3e_{(b)u}\, {}^3{\tilde \Pi}
^{us}=\nonumber \\
&&={{1}\over 4} ( \partial_s[{}^3e^r_{(a)}\, {}^3{\tilde \pi}^s_{(a)}
+{}^3e^s_{(a)}\, {}^3{\tilde \pi}^r_{(a)}]-\nonumber \\
&&-[\epsilon_{(a)(c)(b)}\, {}^3e^r_{(a)}\, {}^3\omega_{s(c)}+\partial_s\,
{}^3e^r_{(b)}]\cdot [2\, {}^3{\tilde \pi}^s_{(b)}-{}^3e^s_{(d)}\, {}^3{\tilde M}
_{(d)(b)}]\, )=\nonumber \\
&&={{1}\over 4} \{ \, {}^3e^r_{(a)} [\partial_s\, {}^3{\tilde \pi}^s_{(a)}-2
\epsilon_{(a)(b)(c)}\, {}^3\omega_{s(b)}\, {}^3{\tilde \pi}^s_{(c)}]-
{}^3{\tilde \pi}^s_{(a)} \partial_s\, {}^3e^r_{(a)}+\nonumber \\
&&+\partial_s({}^3e^s_{(a)}\, {}^3{\tilde \pi}^r_{(a)})+[\epsilon_{(a)(c)(b)}\,
{}^3e^r_{(a)}\, {}^3\omega_{s(c)}+\partial_s\, {}^3e^r_{(b)}]{}^3e^s_{(d)}\,
{}^3{\tilde M}_{(d)(b)}\, \} =\nonumber \\
&&={{1}\over 4} \{ 2\, {}^3e^r_{(a)} {\hat {\cal H}}_{(a)}+\partial_s
[{}^3e^s_{(a)}\, {}^3{\tilde \pi}^r_{(a)}-{}^3e^r_{(a)}\, {}^3{\tilde \pi}^s
_{(a)}]-\nonumber \\
&&-[\epsilon_{(a)(b)(c)}\, {}^3e^r_{(a)}\, {}^3\omega_{s(b)}+\partial_s\, {}^3e
^r_{(c)}]{}^3e^s_{(d)}\, {}^3{\tilde M}_{(c)(d)} \} .
\label{aaa1}
\end{eqnarray}

\noindent Since ${}^3{\tilde \pi}^r_{(a)}={1\over 2} {}^3e^r_{(b)} [{}^3e
_{(b)u}\, {}^3{\tilde \pi}^u_{(a)}+{}^3e_{(a)u}\, {}^3{\tilde \pi}^u_{(b)}]-
{1\over 2} {}^3{\tilde M}_{(a)(b)}\, {}^3e^r_{(b)}$, we get $\partial_s [{}^3e
^s_{(a)}\, {}^3{\tilde \pi}^r_{(a)}- {}^3e^r_{(a)}\, {}^3{\tilde \pi}^s_{(a)}]
=\partial_s [{1\over 2} ({}^3e^s_{(a)}\, {}^3e^r_{(b)}-{}^3e^r_{(a)}\,
{}^3e^s_{(b)})({}^3e_{(b)u}\, {}^3{\tilde \pi}^u_{(a)}+{}^3e_{(a)u}\, 
{}^3{\tilde \pi}^u_{(b)})-{}^3e^s_{(a)}\, {}^3e^r_{(b)}\, {}^3{\tilde M}
_{(a)(b)}]=-\partial_s [{}^3e^s_{(a)}\, {}^3e^r_{(b)}\, {}^3{\tilde M}_{(a)(b)}
]$,  the ADM metric supermomentum constraints (\ref{c8}) are satisfied 
in the following form

\begin{eqnarray}
{}^3{\tilde {\cal H}}^r&=&-2{}^3{\tilde \Pi}^{rs}{}_{|s}={{1}\over 2} 
\{ \, -2\, {}^3e^r_{(a)} {\hat {\cal H}}_{(a)} +\partial_s [{}^3e^s_{(a)}\, 
{}^3e^r_{(b)}\, {}^3{\tilde M}_{(a)(b)}]+\nonumber \\
&+&[\partial_s\, {}^3e^r_{(c)}-\epsilon_{(c)(b)(a)}\,  {}^3\omega_{s(b)}\,
{}^3e^r_{(a)}]{}^3e^s_{(d)}\, {}^3{\tilde M}_{(c)(d)} \, \} =\nonumber \\
&=&{{1}\over 2} \{ \, 2\, {}^3e^r_{(a)}\, {}^3e^s_{(a)}\, {}^3{\tilde \Theta}
_s+[{}^3e^r_{(a)}\, {}^3\omega_{s(b)}-{}^3e^r_{(b)}\, {}^3\omega_{s(a)}]{}^3e
^s_{(a)}\, {}^3{\tilde M}_{(b)}+\nonumber \\
&+&\epsilon_{(a)(b)(c)}\, {}^3e^r_{(b)} \partial_s[{}^3e^s_{(a)}\, {}^3{\tilde 
M}_{(c)}] \, \} \approx 0.
\label{III15a}
\end{eqnarray}

Let us add a comment on the structure of gauge-fixings for metric gravity;
the same results hold for tetrad gravity. As said in Refs.\cite{giap,lusa},
in a system with only primary and secondary first class constraints (like
electromagnetism, Yang-Mills theory and both metric and tetrad gravity) the
Dirac Hamiltonian $H_D$ contains only the arbitrary Dirac multipliers associated
with the primary first class constraints. The secondary first class constraints
are already contained in the canonical Hamiltonian with well defined
coefficients [the temporal components $A_{ao}$ of the gauge potential in
Yang-Mills theory; the lapse and shift functions in metric and tetrad gravity;
in both cases, through the first half of the Hamilton equations, the Dirac 
multipliers turn out to be equal to the $\tau$-derivatives of these quantities, 
which, therefore, inherit an induced arbitrariness]. See the second paper in 
Ref.\cite{sha} for a discussion of this point and for a refusal of Dirac's 
conjecture\cite{dirac} according to which also the secondary first class 
constraints must have arbitrary Dirac multipliers (in such a case one does not
recover the original Lagrangian by inverse Legendre transformation and one
obtains a different "off-shell" theory). In these cases one must adopt the
following gauge-fixing strategy: i) add gauge-fixing constraints
$\chi_a\approx 0$ to the secondary constraints; ii) their time constancy,
$\partial_{\tau} \chi_a \, {\buildrel \circ \over =}\, \lbrace \chi_a,H_D
\rbrace =g_a \approx 0$, implies the appearance of gauge-fixing constraints 
$g_a\approx 0$ for the primary constraints; iii) the time constancy of the
constraints $g_a\approx 0$, $\partial_{\tau} g_a\, {\buildrel \circ \over =}\,
\lbrace g_a,H_D\rbrace \approx 0$, determines the Dirac multipliers in front 
of the primary constraints.

As shown in the second paper of Ref.\cite{lusa} for the electromagnetic case,
this method works also with covariant gauge-fixings: the electromagnetic
Lorentz gauge $\partial^{\mu}A_{\mu}(x) \approx 0$ may be rewritten in phase
space as a gauge-fixing constraint depending upon the Dirac multiplier; its
time constancy gives a multiplier-dependent gauge-fixing for $A_o(x)$ and the
time constancy of this new constraint gives the elliptic equation for the 
multiplier with the residual gauge freedom connected with the kernel of the
elliptic operator.

In metric gravity, the covariant gauge-fixings analogous to the Lorentz
gauge are those determining the harmonic coordinates
(harmonic or DeDonder gauge): $\chi^B={1\over
{\sqrt{{}^4g}}} \partial_A(\sqrt{{}^4g}\, {}^4g^{AB}) \approx 0$ in the
$\Sigma_{\tau}$-adapted holonomic coordinate basis. More explicitly, they are:
\hfill\break
\hfill\break
i) for $B=\tau$: $N \partial_{\tau} \gamma -\gamma \partial_{\tau} N -N^2
\partial_r({{\gamma N^r}\over N}) \approx 0$; \hfill\break
ii) for $B=s$: $N N^s \partial
_{\tau}\gamma +\gamma (N \partial_{\tau}N^s-N^s \partial_{\tau}N)+N^2
\partial_r[N\gamma ({}^3g^{rs}-{{N^rN^s}\over {N^2}})] \approx 0$.\hfill\break
\hfill\break 
From Eqs.(\ref{c11}) we get $\partial_{\tau}N\, {\buildrel \circ \over =}\, 
\lambda_N$, $\partial_{\tau} N_r\, {\buildrel \circ \over =}\, \lambda
^{\vec N}_r$ and $\partial_{\tau} \gamma ={1\over 2}\gamma \, {}^3g^{rs}
\partial_{\tau}\, {}^3g_{rs}\, {\buildrel \circ \over =}\, {1\over 2}\gamma 
[{}^3g^{rs}(N_{r|s}+N_{s|r})-{{5\epsilon N}\over {k\sqrt{\gamma}}}\,
{}^3{\tilde \Pi}].$

Therefore, in phase space the harmonic coordinate gauge-fixings take the form 
$\chi^B={\bar \chi}^B(N,N_r,N_{r|s}, {}^3g_{rs}, {}^3{\tilde \Pi}^{rs}, \lambda
_N, \lambda^{\vec N}_r) \approx 0$ and have to be associated with the secondary
superhamiltonian and supermomentum constraints. The conditions $\partial_{\tau}
{\bar \chi}^B\, {\buildrel \circ \over =}\, \lbrace {\bar \chi}^B,H_D\rbrace
=g^B\approx 0$ give the gauge-fixings for the primary constraints ${\tilde 
\Pi}^N\approx 0$, ${\tilde \Pi}^r_{\vec N}\approx 0$. The conditions $\partial
_{\tau} g^B\, {\buildrel \circ \over =}\, \lbrace g^B,H_D\rbrace \approx 0$
are partial differential equations for the Dirac multipliers $\lambda_N$,
$\lambda_r^{\vec N}$, implying a residual gauge freedom like it happens for
the electromagnetic Lorentz gauge.

\vfill\eject

\section{Conclusions.}

Motivated by the attempt to get a unified description and a canonical
reduction of the four interactions in the framework of Dirac-Bergmann theory
of constraint (the presymplectic approach), we begin an investigation of
general relativity along these lines. A complete analysis of this theory
along these lines is still lacking, probably due to the fact that it does not
respect the requirement of manifest general covariance. Instead, the
presymplectic approach is the natural one to get an explicit control on the
degrees of freedom of theories described by singular Lagrangians at the
Hamiltonian level. After the completion of the canonical reduction along
these lines, one will come back to the interpretational problems connected with
general covariance, which are deeply different from those of ordinary gauge
theories like Yang-Mills one.

In this first paper we have reviewed the kinematical framework for tetrad 
gravity (natural for the coupling to fermion fields)
on globally hyperbolic, asymptotically flat at spatial infinity
spacetimes whose 3+1 decomposition may be obtained with simultaneity spacelike
hypersurfaces $\Sigma_{\tau}$ diffeomorphic to $R^3$ (they are the Cauchy
surfaces).

Then, we have given a new parametrization of arbitrary cotetrads in terms of 
lapse and
shift functions, of cotriads on $\Sigma_{\tau}$ and of three boost parameters.
Such parametrized cotetrads are put in the ADM action for metric gravity to
obtain the new Lagrangian for tetrad gravity. In the Hamiltonian
formulation, we obtain 14 first class constraints, ten primary and four 
secondary ones, whose algebra is studied.

A comparison with other formulations of tetrad gravity and with the Hamiltonian
ADM metric gravity has been done.

In the next paper \cite{russo2}, 
we shall study the Hamiltonian group of gauge transformations
induced by the first class constraints. Then, the multitemporal equations
associated with the constraints generating space rotations and space
diffeomorphisms on the cotriads will be studied and solved. The Dirac
observables with respect to thirteen of the fourteen constraints will be found 
in 3-orthogonal coordinates on $\Sigma_{\tau}$ and the associated Shanmugadhasan
canonical transformation will be done. The only left constraint to be
studied will be the superhamiltonian one. Some interpretational problems
(Dirac observables versus general covariance) \cite{rove,be} will be faced.

\vfill\eject

\appendix

\section{4-Tensors in the $\Sigma_{\tau}$-adapted Holonomic Coordinates.}

The connection coefficients ${}^4\Gamma^B_{AC}={1\over 2}\, {}^4g^{BD}
(\partial_A\, {}^4g_{CD}+\partial_C\, {}^4g_{AD}-\partial_D\, {}^4g_{AC})=
{}^4\Gamma^B_{CA}$ in the $\Sigma_{\tau}$-adapted  coordinate basis
associated with ${}^4g_{AB}$ and ${}^4g^{AB}$
of Eqs.(\ref{II13}) and (\ref{II14})
are independent from the boost parameters $\varphi_{(a)}$ and have the
following expression [use is made of  $N_{v|r}=N_{(a)|r}\, {}^3e_{(a)v}$]

\begin{eqnarray}
{}^4\Gamma^{\tau}_{\tau\tau}&=&{1\over {N}} \Big[ \partial_{\tau}N+N^r\partial
_rN-N^rN^s\, {}^3K_{rs}\Big] =\nonumber \\
&=&{1\over N}\Big[ \partial_{\tau}N+{}^3e^r_{(a)}N_{(a)}\partial_rN+{{N_{(a)}N
_{(b)}}\over N}\, {}^3e^r_{(a)}(\partial_{\tau}\, {}^3e_{(b)r}-N_{(b)|r})\Big]
,\nonumber \\
{}^4\Gamma^{\tau}_{r\tau}&=&{}^4\Gamma^{\tau}_{\tau r}={1\over {N}}
\Big[ \partial_rN -{}^3K_{rs} N^s]=\nonumber \\
&=&{1\over N}[\partial_rN+{{N_{(a)}}\over {2N}}(\delta^u_r\delta^v_s+\delta^u_s
\delta^v_r){}^3e^s_{(a)}\, {}^3e_{(b)u}(\partial_{\tau}\, {}^3e_{(b)v}-N_{(b)|v}
)\Big],\nonumber \\
{}^4\Gamma^{\tau}_{rs}&=&{}^4\Gamma^{\tau}_{sr}=-{1\over N} {}^3K_{rs}
=\nonumber \\
&=&{1\over {2N^2}}(\delta^u_r\delta^v_s+\delta^u_s\delta^v_r)(\partial_{\tau}\,
{}^3e_{(a)u}-N_{(a)|u}){}^3e_{(a)v},\nonumber \\
{}^4\Gamma^u_{\tau\tau}&=&\partial_{\tau} N^u-{{N^u}\over N}\partial_{\tau}N+
({}^3g^{uv}-{{N^uN^v}\over {N^2}}) N\partial_vN+N^u{}_{|v}\, N^v-\nonumber \\
&-&2N ({}^3g^{uv}-{{N^uN^v}\over {2N^2}}) {}^3K_{vr} N^r=\nonumber \\
&=&
{}^3e^u_{(a)} (\partial_{\tau} N_{(a)}-{{N_{(a)}}\over N}\partial_{\tau} N)+
N_{(a)} \partial_{\tau}\, {}^3e^u_{(a)}+\nonumber \\
&+&N (\delta_{(a)(b)}-{{N_{(a)}N_{(b)}}\over {N^2}}){}^3e^u_{(a)}\, {}^3e^v
_{(b)} \partial_vN+  {}^3e^v_{(b)} N_{(b)} ({}^3e^u_{(a)}\, N_{(a)}){|}_{|v}-
\nonumber \\
&-&(\delta_{(a)(b)}-{{N_{(a)}N_{(b)}}\over {2N^2}})({}^3e^v_{(c)}\delta_{(b)(d)}
+{}^3e^v_{(b)}\delta_{(c)(d)})\, {}^3e^u_{(a)}(N_{(d)|v}-\partial_{\tau}\,
{}^3e_{(d)v}) N_{(c)},\nonumber \\
{}^4\Gamma^u_{r\tau}&=&{}^4\Gamma^u_{\tau r}=N^u{}_{|r}-{{N^u}\over N}
\partial_rN -N ({}^3g^{uv}-{{N^uN^v}\over {N^2}}) {}^3K_{vr}=\nonumber \\
&=&{}^3e^u_{(a)} (N_{(a)|r}-{{N_{(a)}}\over N} \partial_rN)-\nonumber \\
&-&{1\over 2}\, {}^3e^u_{(a)}(\delta^u_r\delta_{(b)(c)}+{}^3e^s_{(b)}\,
{}^3e_{(c)r})(\delta_{(a)(b)}-{{N_{(a)}N_{(b)}}\over {N^2}}) (N_{(c)|s}-
\partial_{\tau}\, {}^3e_{(c)s}),\nonumber \\
{}^4\Gamma^u_{rs}&=&{}^4\Gamma^u_{sr}={}^3\Gamma^u_{rs}-{{N^u}\over N}
{}^3K_{rs}=\nonumber \\
&=&{}^3\Gamma^u_{rs}+{{N_{(a)}}\over {2N^2}}(\delta^m_r\delta^v_s+\delta^m_s
\delta^v_r) {}^3e^u_{(a)}\, {}^3e_{(b)m} (\partial_{\tau}\, {}^3e_{(b)v}-
N_{(b)|v}).
\label{d1}
\end{eqnarray}

\noindent In these equations we use the 3-dimensional Christoffel symbols
${}^3\Gamma^u_{rs}$, whose associated spin connection is
${}^3\omega_{r(a)(b)}$ of Eqs.(\ref{II5a}).

The spacetime spin connection ${}^4\omega_{A(\alpha )(\beta )}={}^4\eta
_{(\alpha )(\gamma )}\, {}^4\omega_A{}^{(\gamma )}{}_{(\beta )}$ 

\begin{eqnarray}
{}^4\omega_A{}^{(\alpha )}{}_{(\beta )}&=&{}^4E^{(\alpha )}_B[\partial_A\,
{}^4E^B_{(\beta )}+{}^4\Gamma^B_{AC}\, {}^4E^C_{(\beta )}]=\nonumber \\
&=&[\Lambda (\varphi_{(a)}(\sigma ))\, {}^4{\buildrel \circ \over {\omega}}_A
\, \Lambda^{-1}(\varphi_{(a)}(\sigma ))+\partial_A\Lambda (\varphi_{(a)}
(\sigma ))\, \Lambda^{-1}(\varphi_{(a)}(\sigma ))]{}^{(\alpha )}{}_{(\beta )},
\label{d2}
\end{eqnarray}

\noindent is expressed
in terms of the boost parameter independent spin connection
[the Christoffel symbols are invariant under the local Lorentz rotation]

\begin{equation}
{}^4{\buildrel \circ \over {\omega}}_A{}^{(\alpha )}{}_{(\beta )}={}^4
_{(\Sigma )}{\check {\tilde E}}^{(\alpha )}_B[\partial_A\, {}^4_{(\Sigma )}
{\check {\tilde E}}^B_{(\beta )}+{}^4\Gamma^B_{AC}\, {}^4_{(\Sigma )}{\check
{\tilde E}}^C_{(\beta )}].
\label{d3}
\end{equation}

\noindent Analogously we have ${}^4\Omega_{AB}{}^{(\alpha )}{}_{(\beta )}=
[\Lambda (\varphi_{(a)}(\sigma ))\, {}^4{\buildrel \circ \over {\Omega}}_{AB}
\, \Lambda^{-1}(\varphi_{(a)}(\sigma ))]{}^{(\alpha )}{}_{(\beta )}$ for the
associated field strengths.

For the spacetime spin connection ${}^4{\buildrel \circ \over {\omega}}
_{A(\alpha )(\beta )}={}^4\eta_{(\alpha )(\gamma )}\, {}^4{\buildrel \circ 
\over {\omega}}_A{}^{(\gamma )}{}_{(\beta )}$, also using Eqs.(\ref{III2}), 
(\ref{II5a}) we have [see also 
Refs.\cite{hen1}] 

\begin{eqnarray}
{}^4{\buildrel \circ \over {\omega}}_{\tau (o)(a)}&=&-{}^4{\buildrel \circ 
\over {\omega}}_{\tau (a)(o)}=-\epsilon [\partial_rN+{}^3K_{rs}\, {}^3e^s_{(b)}
N_{(b)}]{}^3e^r_{(a)},\nonumber \\
{}^4{\buildrel \circ \over {\omega}}_{\tau (a)(b)}&=&-{}^4{\buildrel \circ 
\over {\omega}}_{\tau (b)(a)}=\nonumber \\
&=&-\epsilon \, {}^3e_{(a)r}\Big[ \partial_{\tau}\, {}^3e^r_{(b)}-N_{(c)}\,
{}^3e^s_{(c)}\partial_s\, {}^3e^r_{(b)}+{}^3e^s_{(b)}\partial_s(N_{(c)}\,
{}^3e^r_{(c)})\Big] -\nonumber \\
&-&\epsilon [N\, {}^3e^r_{(a)}\, {}^3K_{rs}\, {}^3e^s_{(b)}+N_{(c)}\, {}^3e^r
_{(c)}\, {}^3\omega_{r(a)(b)}]=\nonumber \\
&=&-\epsilon \, {}^3\omega_{r(a)(b)}\, {}^3e^r_{(c)}N_{(c)}-{{\epsilon}\over 2}
\Big( {}^3e^r_{(a)}\partial_{\tau}\, {}^3e_{(b)r}-{}^3e^r_{(b)}\partial_{\tau}\,
{}^3e_{(a)r}\Big) +\nonumber \\
&+&{{\epsilon}\over 2}N_{(c)}\, {}^3e^s_{(c)}\Big( {}^3e_{(a)r}\partial_s\, 
{}^3e^r_{(b)}-{}^3e_{(b)r}\partial_s\, {}^3e^r_{(a)}\Big) -\nonumber \\
&-&{{\epsilon}\over 2}\Big( {}^3e_{(a)s}\, {}^3e^r_{(b)}-{}^3e_{(b)s}\, {}^3e^r
_{(a)}\Big) \partial_r(N_{(c)}\, {}^3e^s_{(c)}),\nonumber \\
{}^4{\buildrel \circ \over {\omega}}_{r(o)(a)}&=&-{}^4{\buildrel \circ \over 
{\omega}}_{r(a)(o)}=-\epsilon \, {}^3K_{rs}\, {}^3e^s_{(a)}=\nonumber \\
&=&-{{\epsilon}\over {2N}} [\delta_{(a)(b)}\delta^w_r+{}^3e^w_{(a)}\, {}^3e
_{(b)r}][N_{(b)|w}-\partial_{\tau}\, {}^3e_{(b)w}],\nonumber \\
{}^4{\buildrel \circ \over {\omega}}_{r(a)(b)}&=&-{}^4{\buildrel \circ \over 
{\omega}}_{r(b)(a)}=-\epsilon {}^3\omega_{r(a)(b)}=
{{\epsilon}\over 2}\Big[ {}^3e^s_{(a)}(\partial_r\, {}^3e_{(b)s}-
\partial_s\, {}^3e_{(b)r})+\nonumber \\
&+&{}^3e^s_{(b)}(\partial_s\, {}^3e_{(a)r}-\partial_r\, {}^3e_{(a)s})+{}^3e^u
_{(a)}\, {}^3e^v_{(b)}\, {}^3e_{(c)r}(\partial_v\, {}^3e_{(c)u}-\partial_u\,
{}^3e_{(c)v}) \Big] .
\label{d4}
\end{eqnarray}

The field stregth
${}^4{\buildrel \circ \over {\Omega}}_{AB(\alpha )(\beta )}=
{}^4_{(\Sigma )}{\check {\tilde E}}^C_{(\alpha )}\, {}^4_{(\Sigma )}{\check
{\tilde E}}^D_{(\beta )}\, {}^4R_{CDAB}=\partial_A\, {}^4{\buildrel \circ \over
{\omega}}_{B(\alpha )(\beta )}-\partial_B\, {}^4{\buildrel \circ \over
{\omega}}_{A(\alpha )(\beta )}+{}^4{\buildrel \circ \over {\omega}}_{A(\alpha
)(\gamma )}\, {}^4{\buildrel \circ \over {\omega}}_{B(\beta )}^{(\gamma )}-
{}^4{\buildrel \circ \over {\omega}}_{B(\alpha )(\gamma )}\,
{}^4{\buildrel \circ \over {\omega}}_{A(\beta )}^{(\gamma )}$ 
is obtained starting from the 
spin connection ${}^4{\buildrel \circ \over {\omega}}_{A(\alpha )(\beta )}=
{}^4\eta_{(\alpha )(\gamma )}\, {}^4{\buildrel \circ \over {\omega}}_{A(\beta )}
^{(\gamma )}$. We have [see Eqs.(\ref{II5a}) for ${}^3\Omega_{rs(a)(b)}$]

\begin{eqnarray}
{}^4{\buildrel \circ \over {\Omega}}_{rs(a)(b)}&=&{}^3e^u_{(a)}\, {}^3e^v
_{(b)}\, {}^4R_{uvrs}=-\epsilon \Big[ {}^3\Omega_{rs(a)(b)}+\nonumber \\
&+&({}^3K_{ru}\, {}^3K_{sv}-{}^3K_{su}\, {}^3K_{rv}) {}^3e^u_{(a)}\,
{}^3e^v_{(b)}\Big] ,\nonumber \\
{}^4{\buildrel \circ \over {\Omega}}_{rs(o)(a)}&=&{1\over N}{}^3e^v_{(a)}
({}^4R_{\tau vrs}-N^u\, {}^4R_{uvrs})=\nonumber \\
&=&\epsilon ({}^3K_{ru|s}-{}^3K_{su|r})\, {}^3e^u_{(a)},\nonumber \\
{}^4{\buildrel \circ \over {\Omega}}_{\tau r(a)(b)}&=&{}^3e^u_{(a)}\, {}^3e^v
_{(b)}\, {}^4R_{uv\tau r}=-\epsilon \Big( \partial_{\tau} \,
{}^3\omega_{r(a)(b)}+\nonumber \\
&+&{1\over 2} (\epsilon_{(a)(b)(c)}\epsilon_{(d)(e)(f)}-
\epsilon_{(a)(b)(d)}\epsilon_{(c)(e)(f)})\cdot \nonumber \\
&&{}^3e^s_{(c)} \Big[ \partial_{\tau}\, {}^3e_{(d)s}-
\Big( N_{(g)}\, {}^3e^u_{(g)}
\partial_u\, {}^3e_{(d)s}+{}^3e_{(d)u} \partial_s(N_{(g)}\, {}^3e^u_{(g)})\Big)
\Big] \, {}^3\omega_{r(e)(f)}+\nonumber \\
&+&N_{(c)}\, {}^3e^s_{(c)} [{}^3\omega_s, {}^3\omega_r]_{(a)(b)}+\nonumber \\
&+&{}^3K_{rs} \Big( {}^3e^s_{(a)}\, {}^3e^u_{(b)}-{}^3e^u_{(a)}\, {}^3e^s_{(b)}
\Big) \partial_u N+\nonumber \\
&+&({}^3K_{sv}\, {}^3K_{ru}-{}^3K_{uv}\, {}^3K_{rs}) {}^3e^u_{(a)}\,
{}^3e^s_{(b)} N_{(c)}\, {}^3e^v_{(c)}\, \Big)  ,\nonumber \\
{}^4{\buildrel \circ \over {\Omega}}_{\tau r(o)(a)}&=&{1\over N}\, {}^3e^u
_{(a)}({}^4R_{\tau u\tau r}-N^s\, {}^4R_{su\tau r})=-\epsilon \Big[ \partial
_{\tau}\, {}^3K_{rs}-\nonumber \\
&-&{}^3K_{ru}\, (N_{(b)}\, {}^3e^u_{(b)})_{|s} -{}^3K_{su}\, (N_{(b)}\,
{}^3e^u_{(b)})_{|r}-N_{(b)}\, {}^3e^u_{(b)}\, {}^3K_{su|r}+
N_{|s|r}\Big] {}^3e^s_{(a)}.
\label{d5}
\end{eqnarray}

The Riemann tensor ${}^4R_{ABCD}={}^4_{(\Sigma )}{\check {\tilde E}}_C
^{(\alpha )}\, {}^4_{(\Sigma )}{\check {\tilde E}}^{(\beta )}_D\,
{}^4{\buildrel \circ \over {\Omega}}_{AB(\alpha )(\beta )}=
{}^4g_{AE}\, {}^4R^E{}_{BCD}=-{}^4R_{ABDC}=-{}^4R_{BACD}=
{}^4R_{CDAB}={1\over 2}(\partial_B\partial_D\, {}^4g_{AC}+\partial_A\partial_C
\, {}^4g_{BD}-\partial_A\partial_D\, {}^4g_{BC}-\partial_B\partial_C\, {}^4g
_{AD})+{}^4g_{EF}({}^4\Gamma^E_{AC}\, {}^4\Gamma^F_{BD}-{}^4\Gamma^E_{AD}\, 
{}^4\Gamma^F_{BC})$ has the following expression in the new basis

\begin{eqnarray}
{}^4R_{rsuv}&=&-\epsilon [{}^3R_{rsuv}+{}^3K_{ru}\, {}^3K_{sv}-
{}^3K_{rv}\, {}^3K_{su}]=\nonumber \\
&=&{}^3e_{(a)r}\, {}^3e_{(b)s}\, {}^4{\buildrel \circ \over {\Omega}}
_{uv(a)(b)}=\nonumber \\
&=&-\epsilon \Big[ {}^3e_{(a)r}\, {}^3e_{(b)s}\, {}^3\Omega_{uv(a)(b)}+
{}^3K_{ru}\, {}^3K_{sv}-{}^3K_{rv}\, {}^3K_{su}\Big] ,\nonumber \\
{}^4R_{\tau ruv}&=&{}^3e_{(a)u}\, {}^3e_{(b)v}\, {}^4{\buildrel \circ \over
{\Omega}}_{\tau r(a)(b)}
=N\, {}^3e_{(a)r}\, {}^4{\buildrel \circ \over {\Omega}}
_{uv(o)(a)}+N_{(a)}\, {}^3e_{(b)r}\, {}^4{\buildrel \circ \over {\Omega}}
_{uv(a)(b)}=\nonumber \\
&=&\epsilon \Big[ N ({}^3K_{ur|v}-{}^3K_{vr|u})-N_{(a)}({}^3e_{(b)r}\, 
{}^3\Omega_{uv(a)(b)}+({}^3K_{rv}\, {}^3K_{us}-{}^3K_{ru}\, {}^3K_{sv}){}^3e
^s_{(a)})\Big] =\nonumber \\
&=&-{{\epsilon}\over 2} \Big[ \partial_{\tau}(\partial_v\, {}^3g_{ru}-\partial
_u\, {}^3g_{rv})-\partial_r \Big( \partial_u({}^3e_{(a)v}N_{(a)})-\partial_v
({}^3e_{(a)u}N_{(a)})\Big) \Big] +\nonumber \\
&+&\epsilon \Big[ (N^2-N_{(a)}N_{(a)})({}^4\Gamma^{\tau}_{\tau v}\, {}^4\Gamma
^{\tau}_{ru}-{}^4\Gamma^{\tau}_{\tau u}\, {}^4\Gamma^{\tau}_{rv})-{}^3g_{mn}
({}^4\Gamma^m_{\tau v}\, {}^4\Gamma^n_{ru}-{}^4\Gamma^m_{\tau u}\, {}^4\Gamma^n
_{rv})-\nonumber \\
&-&{}^3e_{(a)m}N_{(a)}({}^4\Gamma^{\tau}_{\tau v}\, {}^4\Gamma^m_{ru}+
{}^4\Gamma^m_{\tau v}\, {}^4\Gamma^{\tau}_{ru}-{}^4\Gamma^{\tau}_{\tau u}\, 
{}^4\Gamma^m_{rv}-{}^4\Gamma^m_{\tau u}\, {}^4\Gamma^{\tau}_{rv}) \Big],
\nonumber \\
{}^4R_{\tau r\tau s}&=&N\, {}^3e_{(a)r}\, {}^4{\buildrel \circ \over
{\Omega}}_{\tau s(o)(a)}+N_{(a)}\, {}^3e_{(b)r}\, {}^4{\buildrel \circ \over
{\Omega}}_{\tau s(a)(b)}=-\epsilon \Big( N\, \Big[ \partial_{\tau}\, {}^3K_{rs} 
-\nonumber \\
&-&{}^3K_{su}(N_{(a)}\, {}^3e^u
_{(a)})_{|r}-{}^3K_{ru}(N_{(a)}\, {}^3e^u_{(a)})_{|s}-N_{(a)}\, {}^3e^u_{(a)}\, 
{}^3K_{ru|s}+N_{|r|s}\Big] +\nonumber \\
&+&N_{(a)}\, \Big[ {}^3e_{(b)r}\, \partial_{\tau}\, {}^3\omega_{s(a)(b)}+{1
\over 2}(\epsilon_{(a)(b)(c)}\epsilon_{(d)(e)(f)}-\epsilon_{(a)(b)(d)}\epsilon
_{(c)(e)(f)}){}^3e_{(b)r}\, {}^3e^w_{(e)}\cdot \nonumber \\
&\cdot& \Big[ \partial_{\tau}\, {}^3e_{(d)w}-(N_{(g)}\, {}^3e^u_{(g)} \partial
_u\, {}^3e_{(d)w}+{}^3e_{(d)u}\, \partial_w(N_{(g)}\, {}^3e^u_{(g)}))\Big]
{}^3\omega_{s(e)(f)}+\nonumber \\
&+&N_{(c)}\, {}^3e^w_{(c)}\, {}^3e_{(b)r} [{}^3\omega_w,{}^3\omega_s]_{(a)(b)}+
{}^3K_{sw}(\delta^u_r\, {}^3e^w_{(a)}-\delta^w_r\, {}^3e^u_{(a)}) \partial_uN+
\nonumber \\
&+&({}^3K_{rv}\, {}^3K_{su}-{}^3K_{uv}\, {}^3K_{rs}){}^3e^u_{(a)} N_{(c)}\,
{}^3e^v_{(c)}\Big] \Big) =\nonumber \\
&=&-{{\epsilon}\over 2}\Big[ -\partial^2_{\tau}\, {}^3g_{rs}+\partial_{\tau}
\Big( \partial_s({}^3e_{(a)r}N_{(a)})+\partial_r({}^3e_{(a)s}N_{(a)})\Big) -
\partial_r\partial_s(N^2-N_{(a)}N_{(a)})\Big] +\nonumber \\
&+&\epsilon \Big[ (N^2-N_{(a)}N_{(a)})({}^4\Gamma^{\tau}_{\tau s}\, {}^4\Gamma
^{\tau}_{\tau r}-{}^4\Gamma^{\tau}_{\tau\tau}\, {}^4\Gamma^{\tau}_{rs})-
{}^3g_{mn}({}^4\Gamma^m_{\tau s}\, {}^4\Gamma^n_{\tau r}-{}^4\Gamma^m
_{\tau\tau}\, {}^4\Gamma^n_{rs})-\nonumber \\
&-&{}^3e_{(a)m}N_{(a)}({}^4\Gamma^{\tau}_{\tau s}\, {}^4\Gamma^m_{\tau r}+
{}^4\Gamma^m_{\tau s}\, {}^4\Gamma^{\tau}_{\tau r}-{}^4\Gamma^{\tau}_{\tau\tau}
\, {}^4\Gamma^m_{rs}-{}^4\Gamma^m_{\tau\tau}\, {}^4\Gamma^{\tau}_{rs})\Big].
\label{d6}
\end{eqnarray}

While the expression of ${}^4R_{rsuv}$ in the holonomic basis coincides with 
the Gauss  equation (\ref{I6}) in the nonholonomic basis, the expressions of
${}^4R_{\tau ruv}$ and ${}^4R_{\tau r\tau s}$ are the analogue in the
holonomic basis of the Codazzi-Mainardi and Ricci equations respectively for
${}^4{\bar R}_{lruv}$ and ${}^4{\bar R}_{lrls}$ in the nonholonomic basis.
Moreover, we have [${\buildrel \circ \over =}$ refers to the use of vacuum
Einstein equations]

\begin{eqnarray}
{}^4R_{AB}&=&{}^4R_{BA}={}^4g^{CD}\, {}^4R_{CADB}=\nonumber \\
&=&{{\epsilon}\over {N^2}}\, {}^4R_{\tau A\tau B}-{{\epsilon N^r}\over {N^2}}
({}^4R_{\tau ArB}+{}^4R_{rA\tau B})-\epsilon ({}^3g^{rs}-{{N^rN^s}\over
{N^2}})\, {}^4R_{rAsB}\, {\buildrel \circ \over =}\, 0,\nonumber \\
{}^4R_{\tau\tau}&=&-\epsilon {}^3e^r_{(a)}\, {}^3e^s_{(b)}(\delta_{(a)(b)}-
{{N_{(a)}N_{(b)}}\over {N^2}})\, 
{}^4R_{r\tau s\tau}\, {\buildrel \circ \over =}\, 0,\nonumber \\
{}^4R_{\tau u}&=&{}^4R_{u\tau}={{\epsilon {}^3e^v_{(a)}N_{(a)}}\over {N^2}}\, 
{}^4R_{\tau u\tau v}-\epsilon {}^3e^r_{(a)}\, {}^3e^v_{(b)}(\delta_{(a)(b)}-
{{N_{(a)}N_{(b)}}\over {N^2}})\, 
{}^4R_{\tau ruv}\, {\buildrel \circ \over =}\, 0,\nonumber \\
{}^4R_{rs}&=&{}^4R_{sr}={{\epsilon}\over {N^2}}\, {}^4R_{\tau r\tau s}-
{{\epsilon {}^3e^u_{(a)}N_{(a)}}\over {N^2}} ({}^4R_{\tau rus}+
{}^4R_{\tau sur})-\nonumber \\
&-&\epsilon {}^3e^u_{(a)}\, {}^3e^v_{(b)}(\delta_{(a)(b)}
-{{N_{(a)}N_{(b)}}\over {N^2}})\, {}^4R_{urvs}
\, {\buildrel \circ \over =}\, 0,\nonumber \\
{}^4R&=&{}^4g^{AB}\, {}^4R_{AB}={{\epsilon}\over {N^2}}\, {}^4R_{\tau\tau}-
2{{\epsilon {}^3e^u_{(a)}N_{(a)}}\over {N^2}}\, {}^4R_{\tau u}-\epsilon 
{}^3e^r_{(a)}\, {}^3e^s_{(b)}(\delta_{(a)(b)}-{{N_{(a)}N_{(b)}}
\over {N^2}})\, {}^4R_{rs}=\nonumber \\
&=&-{2\over {N^2}} {}^3e^r_{(a)}\, {}^3e^s_{(a)}\, {}^4R_{\tau r\tau s}+
4{{{}^3e^u_{(c)}N_{(c)}}\over {N^2}} {}^3e^r_{(a)}\, {}^3e^v_{(b)}(\delta
_{(a)(b)}-{{N_{(a)}N_{(b)}}\over {N^2}}) {}^4R_{\tau ruv}+\nonumber \\
&+&{}^3e^r_{(a)}\, {}^3e_{(b)}(\delta_{(a)(b)}-{{N_{(a)}N_{(b)}}\over {N^2}})
{}^3e^u_{(c)}\, {}^3e_{(d)}^v(\delta_{(c)(d)}-{{N_{(c)}N_{(d)}}\over {N^2}})
{}^4R_{rusv}\, {\buildrel \circ \over =}\, 0,
\label{d7}
\end{eqnarray}

\begin{eqnarray}
{}^4C_{ABCD}&=&{}^4R_{ABCD}+{1\over 2}({}^4R_{AC}\, {}^4g_{BD}-{}^4R_{BC}\,
{}^4g_{AD}+{}^4R_{BD}\, {}^4g_{AC}-{}^4R_{AD}\, {}^4g_{BC})+\nonumber \\
&+&{1\over 6}({}^4g_{AC}\, {}^4g_{BD}-{}^4g_{AD}\, {}^4g_{BC})\, {}^4R\,
{\buildrel \circ \over =}\, {}^4R_{ABCD},\nonumber \\
{}^4C_{rsuv}&=&{}^4R_{rsuv}+{{\epsilon}\over 2} ({}^3g_{rv}\, {}^4R_{su}+{}^3g
_{su}\, {}^4R_{rv}-{}^3g_{ru}\, {}^4R_{sv}-{}^3g_{sv}\, {}^4R_{ru})+\nonumber \\
&+&{1\over 6} ({}^3g_{ru}\, {}^3g_{sv}-{}^3g_{rv}\, {}^3g_{su})\, {}^4R=
\nonumber \\
&=&{}^4R_{rsuv}-{1\over 2}{}^3e^m_{(a)}\, {}^3e^n_{(b)}(\delta_{(a)(b)}-
{{N_{(a)}N_{(b)}}\over {N^2}})\times \nonumber \\
&&\Big[ {}^3g_{rv}\, {}^4R_{msnu}+{}^3g_{su}\,
{}^4R_{mrnv}-{}^3g_{ru}\, {}^4R_{msnv}-{}^3g_{sv}\, {}^4R_{mrnu} \Big]
+\nonumber \\
&+&{1\over 6}({}^3g_{ru}\, {}^3g_{sv}-{}^3g_{rv}\, {}^3g_{su}){}^3e^m_{(a)}\,
{}^3e^n_{(b)}(\delta_{(a)(b)}-{{N_{(a)}N_{(b)}}\over {N^2}})\times \nonumber \\
&&{}^3e^w_{(c)}\,
{}^3e^t_{(d)}(\delta_{(c)(d)}-{{N_{(c)}N_{(d)}}\over {N^2}}) {}^4R_{mwnt}+
\nonumber \\
&+&{1\over {2N^2}}\Big( {}^3g_{rv}\, {}^4R_{\tau s\tau u}+{}^3g_{su}\, 
{}^4R_{\tau r\tau u}-{}^3g_{ru}\, {}^4R_{\tau s\tau v}-{}^3g_{sv}\, 
{}^4R_{\tau r\tau u}\Big)-\nonumber \\
&-&{1\over {3N^2}}({}^3g_{ru}\, {}^3g_{sv}-{}^3g_{rv}\, {}^3g_{su}){}^3e^m
_{(a)}\, {}^3e^n_{(a)}\, {}^4R_{\tau m\tau n}-\nonumber \\
&-&{1\over {2N^2}} {}^3e^m_{(a)}N_{(a)}\Big[ {}^3g_{rv} ({}^4R_{\tau smu}+
{}^4R_{\tau ums})+{}^3g_{su}({}^4R_{\tau rmv}+{}^4R_{\tau vmr})-\nonumber \\
&-&{}^3g_{ru}({}^4R_{\tau smv}+{}^4R_{\tau vms})-{}^3g_{sv}({}^4R_{\tau rmu}+
{}^4R_{\tau umr})\Big] +\nonumber \\
&+&{2\over {3N^2}} {}^3e^m_{(c)}N_{(c)} ({}^3g_{ru}\, {}^3g_{sv}-{}^3g_{rv}\, 
{}^3g_{su}) {}^3e^w_{(a)}\, {}^3e^n_{(b)}(\delta_{(a)(b)}-{{N_{(a)}N_{(b)}}
\over {N^2}}) {}^4R_{\tau wmn}
\, {\buildrel \circ \over =}\nonumber \\
&{\buildrel \circ \over =}&\, {}^4R_{rsuv},\nonumber \\
{}^4C_{\tau ruv}&=&{}^4R_{\tau ruv}+{{\epsilon}\over 2} [{}^3g_{ru}\, {}^4R
_{\tau v}-{}^3g_{rv}\, {}^4R_{\tau u}+N_{(a)} ({}^3e_{(a)v}\, {}^4R_{ru}-
{}^3e_{(a)u}\, {}^4R_{rv})]+\nonumber \\
&+&{1\over 6} N_{(a)} ({}^3e_{(a)u}\, {}^3g_{rv}-{}^3e_{(a)v}\, {}^3g_{ru})
{}^4R=\nonumber \\
&=&{}^4R_{\tau ruv}-{1\over 2}\Big( {}^3e^m_{(a)}\, {}^3e^n_{(b)}(\delta
_{(a)(b)}-{{N_{(a)}N_{(b)}}\over {N^2}})({}^3g_{ru}\, {}^4R_{\tau mvn}-
{}^3g_{rv}\, {}^4R_{\tau mun})-\nonumber \\
&-&{{N_{(a)}\, {}^3e^m_{(b)}N_{(b)}}\over {N^2}} \Big[ {}^3e_{(a)v}({}^4R
_{\tau rmu}+{}^4R_{\tau umr})-{}^3e_{(a)u}({}^4R_{\tau rmv}+{}^4R_{\tau vmr})
\Big] \Big) +\nonumber \\
&+&{2\over {3N^2}}N_{(a)}({}^3e_{(a)u}\, {}^3g_{rv}-{}^3e_{(a)v}\, {}^3g_{ru})
{}^3e^m_{(b)}N_{(b)}{}^3e^w_{(c)}\, {}^3e^n_{(d)}(\delta_{(c)(d)}-
{{N_{(c)}N_{(d)}}\over {N^2}}) {}^4r_{\tau wmn}+\nonumber \\
&+&{{N_{(a)}}\over {2N^2}}\Big[ {}^3e^m_{(a)}({}^3g_{ru}\, {}^4R_{\tau v\tau m}
-{}^3g_{rv}\, {}^4R_{\tau u\tau m})+{}^3e_{(a)v}\, {}^4R_{\tau r\tau u}-
{}^3e_{(a)u}\, {}^4R_{\tau r\tau v}\Big] -\nonumber \\
&-&{{N_{(a)}}\over {3N^2}}({}^3e_{(a)u}\, {}^3g_{rv}-{}^3e_{(a)v}\, {}^3g_{ru})
{}^3e^m_{(b)}\, {}^3e^n_{(b)}\, {}^4R_{\tau m\tau n}-\nonumber \\
&-&{1\over 2}N_{(a)}\, {}^3e^m_{(b)}\, {}^3e^n_{(c)}(\delta_{(b)(c)}-
{{N_{(b)}N_{(c)}}\over {N^2}}) ({}^3e_{(a)v}\, {}^4R_{mrnu}-{}^3e_{(a)u}\,
{}^4R_{mrnv})+\nonumber \\
&+&{1\over 6}N_{(a)}({}^3e_{(a)u}\, {}^3g_{rv}-{}^3e_{(a)v}\, {}^3g_{ru}) 
{}^3e^m_{(b)}\, {}^3e^n_{(c)}(\delta_{(b)(c)}-{{N_{(b)}N_{(c)}}\over {N^2}})
\times \nonumber \\
&&{}^3e^w_{(d)}\, {}^3e^t_{(e)}(\delta_{(d)(e)}-{{N_{(d)}N_{(e)}}\over {N^2}})
{}^4r_{mnwt}\, {\buildrel \circ \over =}\, {}^4R_{\tau ruv},\nonumber \\
{}^4C_{\tau r\tau s}&=&{}^4R_{\tau r\tau s}+{{\epsilon}\over 2} [N_{(a)} ({}^3e
_{(a)s}\, {}^4R_{\tau r}+{}^3e_{(a)r}\, {}^4R_{\tau s})- {}^3g_{rs}\,
{}^4R_{\tau\tau}+\nonumber \\
&+&(N^2-N_{(a)}N_{(a)})\, {}^4R_{rs}]-
{1\over 6} [{}^3g_{rs}\, (N^2-N_{(a)}N_{(a)})+N_{(a)}N_{(b)}
{}^3e_{(a)r}\, {}^3e_{(b)s}\, ]\, {}^4R=\nonumber \\
&=&(2-{{N_{(a)}N_{(a)}}\over {N^2}}) {}^4R_{\tau r\tau s}+{1\over 2}\Big(
{{N_{(a)}\, {}^3e^m_{(b)}N_{(b)}}\over {N^2}}({}^3e_{(a)s}\, {}^4R
_{\tau r\tau m}+{}^3e_{(a)r}\, {}^4R_{\tau s\tau m})+\nonumber \\
&+&{1\over 3}\Big[ {}^3g_{rs}(1-{{N_{(a)}N_{(a)}}\over {N^2}})+{{N_{(a)}N_{(b)}}
\over {N^2}} {}^3e_{(a)r}\, {}^3e_{(b)s}\Big] {}^3e^m_{(c)}\, {}^3e^n_{(c)}\,
{}^3R_{\tau m\tau n}\Big) -\nonumber \\
&-&{1\over 2}\Big[ N_{(a)}\, {}^3e^m_{(b)}\, {}^3e^n_{(c)}(\delta_{(b)(c)}-
{{N_{(b)}N_{(c)}}\over {N^2}})({}^3e_{(a)s}\, {}^4R_{\tau mrn}+{}^3e_{(a)r}\,
{}^4R_{\tau msn})+\nonumber \\
&+&(1-{{N_{(a)}N_{(a)}}\over {N^2}}) {}^3e^u_{(b)}N_{(b)} ({}^4R_{\tau rus}+
{}^4R_{\tau sur})\Big] -\nonumber \\
&-&{2\over 3}[{}^3g_{rs}(1-{{N_{(a)}N_{(a)}}\over {N^2}})+{{N_{(a)}N_{(b)}}
\over {N^2}} {}^3e_{(a)r}\, {}^3e_{(b)s}] {}^3e^u_{(c)}N_{(c)}\, \nonumber \\
&&{}^3e^m_{(d)}\,
{}^3e^n_{(e)}(\delta_{(d)(e)}-{{N_{(d)}N_{(e)}}\over {N^2}}){}^4R_{\tau mun}
+\nonumber \\
&+&(N^2-N_{(a)}N_{(a)}){}^3e^m_{(b)}\, {}^3e^n_{(c)}(\delta_{(b)(c)}-
{{N_{(b)}N_{(c)}}\over {N^2}}) {}^4R_{mrns}-\nonumber \\
&-&{1\over 6} [{}^3g_{rs}(N^2-N_{(a)}N_{(a)})+N_{(a)}N_{(b)}\, {}^3e_{(a)r}\,
{}^3e_{(b)s}] {}^3e^m_{(c)}\, {}^3e^n_{(d)}(\delta_{(c)(d)}-
{{N_{(c)}N_{(d)}}\over {N^2}})\nonumber \\
&&{}^3e^w_{(e)}\, {}^3e^t_{(f)}
(\delta_{(e)(f)}-{{N_{(e)}N_{(f)}}\over {N^2}}) {}^4R_{mwnt}
\, {\buildrel \circ \over =}\, {}^4R_{\tau r\tau s}.
\label{d8}
\end{eqnarray}

\noindent so that every quantity can be expressed in terms of ${}^4R_{rsuv}$,
${}^4R_{\tau ruv}$, ${}^4R_{\tau r\tau s}$.

For the electric ${}^4E_{AB}={}^4C_{A\tau B\tau}$ and magnetic 
${}^4H_{AB}={1\over 2}\epsilon_{B\tau EF}\, {}^4C_{A\tau}{}^{EF}$ 
components of the Weyl tensor (by assuming that the normals to $\Sigma_{\tau}$
are the privileged timelike 4-vectors) we have
${}^4E_{\tau\tau}={}^4E_{\tau r}={}^4H_{\tau\tau}={}^4H_{\tau r}=0$,
${}^4E_{rs}={}^4C_{r\tau s\tau}\, {\buildrel \circ \over =}\, {}^4R_{r\tau 
s\tau}$ and ${}^4H_{rs}={1\over 2}\epsilon_{s\tau}{}^{uv}\, {}^4C_{r\tau uv}\,
{\buildrel \circ \over =}\, {1\over 2}\epsilon_{s\tau}{}^{uv}\, {}^4R_{r\tau 
uv}$.

In the coordinates $\sigma^A=\{ \tau ,\vec \sigma \}$ the ``geodesic equation" 
is ${{d^2\sigma^A(s)}\over {ds^2}}+{}^4\Gamma^A_{BC} {{d\sigma
^B(s)}\over {ds}}{{d\sigma^C(s)}\over {ds}}=0$, while the ``geodesic deviation
equation" is $a^A={{d\sigma^B}\over {ds}}\, {}^4\nabla_B
({{d\sigma^C}\over {ds}}\, {}^4\nabla_C\, \triangle x^A)=-{}^4R^A{}_{BCD}\,
\triangle x^C\, {{d\sigma^B}\over {ds}}{{d\sigma^D}\over {ds}}$.
 The results of this Appendix allow the
identification of the dependence of 4-geodesics and of 4-geodesic deviations
on the gauge parameters of the theory [$\sigma^A(s)=(\tau (s); \vec \sigma 
(s))$]:

\begin{eqnarray}
&&{{d^2\tau (s)}\over {ds^2}}+{}^4\Gamma^{\tau}_{\tau\tau} ({{d\tau (s)}\over 
{ds}})^2+2\, {}^4\Gamma^{\tau}_{\tau r} {{d\tau (s)}\over {ds}}{{d\sigma^r(s)}
\over {ds}}+{}^4\Gamma^{\tau}_{rs} {{d\sigma^r(s)}\over {ds}}{{d\sigma^s(s)}
\over {ds}}=0,\nonumber \\
&&{{d^2\sigma^u(s)}\over {ds^2}}+{}^4\Gamma^u_{\tau\tau}({{d\tau (s)}\over 
{ds}})^2+2\, {}^4\Gamma^u_{\tau r} {{d\tau (s)}\over {ds}}{{d\sigma^r(s)}\over
{ds}}+{}^4\Gamma^u_{rs} {{d\sigma^r(s)}\over {ds}}{{d\sigma^s(s)}\over {ds}}
=0,\nonumber \\
&&{}\nonumber \\
a^{\tau}&=&-{{\epsilon}\over {N^2}}\Big( \Big[ {}^4R_{\tau m\tau n}{{d\sigma^m}
\over {ds}}{{d\sigma^n}\over {ds}}-{}^3e^r_{(a)}N_{(a)}(-{}^4R_{\tau 
r\tau n}{{d\tau}\over {ds}}+{}^4R_{rm\tau n}{{d\sigma^m}\over {ds}})
{{d\sigma^n}\over {ds}} \Big] \triangle x^{\tau}+\nonumber \\
&&+\Big[ (-{}^4R_{\tau m\tau s}{{d\tau}\over {ds}}+{}^4R_{\tau msn}{{d\sigma^n}
\over {ds}}){{d\sigma^m}\over {ds}}-{}^3e^r_{(a)}N_{(a)}\Big( {}^4R_{\tau r\tau
s}({{d\tau}\over {ds}})^2-\nonumber \\
&&-({}^4R_{\tau rsm}+{}^4R_{rm\tau s}){{d\tau}\over {ds}}{{d\sigma^m}\over 
{ds}}+{}^4R_{rmsn}{{d\sigma^m}\over {ds}}{{d\sigma^s}\over {ds}}\Big) \Big]
\triangle x^s\Big) ,\nonumber \\
a^u&=&\epsilon \Big( \Big[ {{{}^3e^u_{(a)}N_{(a)}}\over {N^2}} {}^4R_{\tau 
m\tau n}{{d\sigma^m}\over {ds}}{{d\sigma^n}\over {ds}}+{}^3e^u_{(a)}\, {}^3e^r
_{(b)}(\delta_{(a)(b)}-{{N_{(a)}N_{(b)}}\over {N^2}})\times \nonumber \\
&&(-{}^4R_{\tau r\tau n}{{d\tau}\over {ds}}+{}^4R_{rm\tau n}{{d\sigma^m}\over 
{ds}}){{d\sigma^n}\over {ds}}\Big] \triangle x^{\tau}+\nonumber \\
&&+\Big[ {{{}^3e^u_{(a)}N_{(a)}}\over {N^2}}(-{}^4R_{\tau m\tau s}{{d\tau}\over 
{ds}}+{}^4R_{\tau msn}{{d\sigma^n}\over {ds}}){{d\sigma^m}\over {ds}}+
\nonumber \\
&&+{}^3e^u_{(a)}\, {}^3e^r_{(b)}(\delta_{(a)(b)}-{{N_{(a)}N_{(b)}}\over {N^2}})
\Big( {}^4R_{\tau r\tau s}({{d\tau}\over {ds}})^2-\nonumber \\
&&-({}^4R_{\tau rsm}+{}^4R_{rm\tau s}){{d\tau}\over {ds}}{{d\sigma^m}\over {ds}}
+{}^4R_{rmsn}{{d\sigma^m}\over {ds}}{{d\sigma^n}\over {ds}}\Big) \Big]
\triangle x^s \Big) .
\label{d9}
\end{eqnarray}

More in general, to describe an arbitrary (not necessarily geodetic) congruence 
of timelike curves (congruence of observers; it is surface forming in absence of
vorticity) with tangent field $u^A= {}^4E^A_{(o)}[\varphi_{(a)}, N, N_{(a)}, 
{}^3e^r_{(a)}]$ (see Eq.(\ref{II14}); by varying the 3 functions $\varphi
_{(a)}(\tau ,\vec \sigma )$ we can describe any congruence) one uses [see for
instance Ref.\cite{uggla}, where there is a reformulation of Newman-Penrose 
formalism replacing the congruence of lightlike curves with one of timelike
ones; for the ``threading" viewpoint (3+1 decomposition with respect to an
arbitrary timelike congruence) see also Refs.\cite{bini,boer}]

\begin{eqnarray}
{}^4\nabla_A\, u_B &=& \epsilon u_A {\dot u}_B+\sigma_{AB} +{1\over 3} \Theta 
({}^4g_{AB}-\epsilon u_Au_B)-\omega_{AB},\nonumber \\
&&{\dot u}^A=u^B\, {}^4\nabla_B u^A,\quad acceleration,\nonumber \\
&&\Theta ={}^4\nabla_A u^A,\quad (volume)\, rate\, of\, expansion\, scalar,
\nonumber \\
&&\sigma_{AB}=\sigma_{BA}=-\epsilon {\dot u}_{(A} u_{B)}+{}^4\nabla_{(A} 
u_{B)}-{1\over 3}\Theta ({}^4g_{AB}-\epsilon u_Au_B),\nonumber \\
&&{}{}{}{} rate\, of\, shear\, tensor\, (with\, magnitude\, \sigma^2={1\over 2}
\sigma_{AB}\sigma^{AB}),\nonumber \\
&&\omega_{AB}=-\omega_{BA}=\epsilon_{ABCD}\omega^Cu^D=-u_{[A} {\dot u}_{B]}-
{}^4\nabla_{[A} u_{B]},\quad twist\, or\, vorticity\, tensor,\nonumber \\
&&{}{}{}{} \omega^A={1\over 2} \epsilon^{ABCD}\omega_{BC}u_D,\quad vorticity\,
vector.
\label{d10}
\end{eqnarray}

Associated quantities are: i) the representative length $l$ along the
worldlines of $u^A$, describing the volume expansion (contraction) behaviour 
of the congruence completely, by the equation ${1\over l}\, u^A\, {}^4\nabla_A 
l ={1\over 3} \Theta$; ii) the Hubble parameter H:  $H={1\over l}\, u^A\, 
{}^4\nabla_A l ={1\over 3} \Theta$; iii) the dimensionless (cosmological)
decelaration parameter: $q =-{{l\, u^A\, {}^4\nabla_A (u^B\, {}^4\nabla_B l)}
\over {(u^D\, {}^4\nabla_D l)^2}}=3 u^A\, {}^4\nabla_A {1\over {\Theta}} -1$.

If the congruence is geodesic, the geodesic deviation equation yields equations
for the rate of change of $\Theta$, $\sigma_{AB}$ and $\omega_{AB}$ along 
each geodesic in the congruence [see Ref.\cite{wald} for both timelike and
null congruences; the equation for $\Theta$ is the Raychauduri equation].

Let $\sigma^A(s)=\{ \tau (s); \vec \sigma (s)\}$ be a timelike geodesic $\Gamma$
with timelike tangent vector $u^A(s)={{d\sigma^A(s)}\over {ds}}$, $u^A(s)u_A(s)=
\epsilon$, $u^B(s)\, {}^4\nabla_B u^A(s)=0$ [the affine parameter s is the 
proper time]. Let us consider a tetrad field ${}^4E^A_{(\alpha )}(\tau ,\vec 
\sigma )[\varphi_{(a)},N,N_{(a)}, {}^3e^r_{(a)}]$, whose restriction to the 
geodesic $\Gamma$ has ${}^4E^A_{(o)}(\sigma (s))=u^A(s)$ [many tetrad fields
satisfy this requirement: they differ in the space axes ${}^4E^A_{(a)}(\sigma 
(s))$]: the tetrad ${}^4E^A_{(\alpha )}(\sigma (s))$ describes an accelerated 
observer with worldline $\Gamma$. By going to Riemann normal coordinates for 
$M^4$ [they are not uniquely determined: 
see Appendix A of Ref.\cite{russo2} for a review; in them we have at
the point $\sigma^A(s)$\cite {mtw}:\hfill\break
${}^4g_{AB}={}^3\eta_{AB}$, $\partial_C\, {}^4g_{AB}
=0$, ${}^4\Gamma^A_{BC}=0$, $\partial_C\partial_D\, {}^4g_{AB}=-{1\over 3}
({}^4R_{ACBD}+{}^4R_{ADBC})=-{2\over 3}\, {}^4J_{ABCD}$ (${}^4J$ is the Jacobi 
curvature tensor,carrying the same information of the Riemann tensor), 
$\partial_D\, {}^4\Gamma^A_{BC}=-{1\over 3}({}^4R^A{}_{BCD}+{}^4R^A{}_{CBD})$,
${}^4R_{ABCD}=\partial_B\partial_C\, {}^4g_{AD}-\partial_B\partial_D\, 
{}^4g_{AC}$]\hfill\break
such that the timelike
geodesic $\Gamma$ becomes a timelike straightline, we get the description of a 
``comoving inertial frame" for an observer in free fall at rest: by a suitable 
choice of the gauge parameters $\varphi_{(a)}$, $N$, $N_{(a)}$, ..., along
$\Gamma$ we can associate a fixed reference nonrotating tetrad 
(local Lorentz frame of the observer)
${}^4_{(in)}E^A_{(\alpha )}$ with this inertial observer so that 
${}^4_{(in)}E^A_{(o)}$ is his 4-velocity and the 4-acceleration vanishes
[the space axes ${}^4_{(in)}E^A_{(a)}$
are defined modulo a rigid rotation]. For $s=s_o$, in the point $\sigma_o^A=
\sigma^A(s_o)=\{ \tau_o=\tau (s_o); {\vec \sigma}_o=\vec \sigma (s_o)\}$, let
the tetrad ${}^4E^A_{(\alpha )}(\sigma (s))$ coincide with ${}^4_{(in)}E^A
_{(\alpha )}$: $\,\, {}^4E^A_{(\alpha )}(\sigma_o)={}^4_{(in)}E^A_{(\alpha )}$.
For $s > s_o$ the evolution of the tetrad ${}^4E^A_{(\alpha )}(\sigma (s))$ may 
be parametrized as a Lorentz transformation with respect to ${}^4_{(in)}E^A
_{(\alpha )}$: $\,\, {}^4E^A_{(\alpha )}(\sigma (s))={}^4_{(in)}E^A_{(\beta )}
\Lambda^{(\beta )}{}_{(\alpha )}(s)$. It is assumed that the measures made with
the clocks and rods of the accelerated observer are identical with those done
by a unaccelerated momentarily comoving inertial observer with his clocks and 
rods; in Minkowski spacetime this is called the ``locality hypothesis" in 
Ref.\cite{mashhoon} and it applies also in general relativity, because, due to 
the equivalence principle, an observer in a gravitational field is equivalent
to an accelerated observer in Minkowski spacetime.

Let $a^A(s)={{du^A(s)}\over {ds}}$,
$a^A(s) u_A(s)=0$, be the 4-acceleration of the accelerated observer. Among the
tetrads ${}^4E^A_{(\alpha )}(\sigma (s))$ with ${}^4E^A_{(o)}(\sigma (s))=
u^A(s)$,  the ``nonrotating" one ${}^4_{(FW)}E^A_{(\alpha )}(\sigma (s))$ is the
solution of the equations defining the ``Fermi-Walker transport" 
(gyroscope-type transport) of a vector 
along the worldline $\Gamma$ of the observer [see Ref.\cite{mtw}; in this case 
the infinitesimal Lorentz transformation $\Lambda^{(\beta )}{}_{(\alpha )}(s)=
\delta^{(\beta )}_{(\alpha )}+\omega^{(\beta )}{}_{(\alpha )}(s)$ generates only
the appropriate Lorentz transformation in the timelike 2-hyperplane spanned by 
$u^A(s)$ and $a^A(s)$; under Fermi-Walker transport ${}^4E^A_{(o)}$ remains
equal to $u^A$ and the triad ${}^4E^A_{(a)}$ is the correct relativistic 
generalization of Newtonian nonrotating frames]

\begin{eqnarray}
{{\delta}\over {\delta s}}&& 
{}^4_{(FW)}E^A_{(\alpha )}(\sigma (s)) =u^B(s)\, {}^4\nabla_B\, 
{}^4_{(FW)}E^A_{(\alpha )}(\sigma (s))=- \Omega^A_{(FW)}{}
_B(s) \, {}^4_{(FW)}E^B_{(\alpha )}(\sigma (s)),\nonumber \\
\Omega^{AB}_{(FW)}(s)&=&a^A(s) u^B(s) -a^B(s) u^A(s),\quad [\Omega^{AB}_{(FW)}
w_B=0\quad if\,\, w_Bu^B=w_Ba^B=0],
\label{d11}
\end{eqnarray}

\noindent where ${{\delta}\over {\delta s}}$ is the ``absolute derivative" of 
the vector field restricted to the timelike worldline $\Gamma$ (its vanishing 
defines ``parallel transport along $\Gamma$"). One speaks of ``Fermi transport" 
of a vector $F^A$ along $\Gamma$, if the vector is orthogonal to the 4-velocity
$u^A$ and it suffers Fermi-Walker transport, which reduces to ${{\delta}\over 
{\delta s}} F^A(\sigma (s))= u^A(s) a_B(s)F^B(\sigma (s))$ with ${{\delta}
\over {\delta s}} F^Au_A=0$. Therefore ${}^4_{(FW)}E^A_{(a)}$ is said a ``Fermi
triad" and ${}^4_{(FW)}E^A_{(\alpha)}$ a ``Fermi frame": this is the most
natural generalization of an inertial reference frame along the path of an
accelerated observer. In general, a Fermi frame cannot be extended to the 
whole spacetime manifold due to limitations imposed by curvature (tidal 
effects). The coordinated effort of many observers over an extended period of 
time can lead to a unique picture of natural phenomena (e.g. in astronomy) if
these observers occupy a finite region of spacetime over which an extended
nonrotating system can be defined \cite{mashhoon}; in practice, however, the
Newtonian framework is used for the sake of simplicity and relativistic effects
are treated as small perturbations in a post-Newtonian approximation scheme.

For any other tetrad field one has ${d\over {ds}}\, {}^4E^A_{(\alpha )}(\sigma 
(s))=-\Omega^A{}_B(s)\, {}^4E^B_{(\alpha )}(\sigma (s))$ with $\Omega^{AB}(s)=
\Omega^{AB}_{(FW)}+\Omega^{AB}_{(SR)}(s)$ with the spatial rotation part
$\Omega^{AB}_{(SR)}(s)=\epsilon^{ABCD}u_C(s)\omega_D(s)$, $\omega^Au_A=0$,
producing a rotation in the spacelike 2-hyperplane perpendicular to $u^A$ and
$\omega^A$ [$\Omega^{AB}_{(SR)}u_B=\Omega^{AB}_{(SR)}\omega_B=0$]. If at
$s=s_1$ one has $u^A(s_1)=(1;\vec 0)$, $\omega^A(s_1)=(0;\vec \omega )$, then
${d\over {ds}} [ {}^4E^r_{(a)}-{}^4_{(FW)}E^r_{(a)}](\sigma (s)){|}_{s=s_1}=
\epsilon^{rst} \omega^s \, {}^4E^t_{(a)}(\sigma (s_1))$.

Given the tetrad ${}^4E^A_{(\alpha )}(\sigma (s))$ along the worldline $\Gamma$,
the associated Frenet-Serret equations are\cite{vishve}

\begin{eqnarray}
{{\delta}\over {\delta s}}\, {}^4E^A_{(o)}(\sigma (s)) &=& \kappa (s)\, {}^4E^A
_{(1)}(\sigma (s)),\nonumber \\
{{\delta}\over {\delta s}}\, {}^4E^A_{(1)}(\sigma (s)) &=&\kappa (s)\, {}^4E^A
_{(o)}(\sigma (s)) +\tau_1(s)\, {}^4E^A_{(2)}(\sigma (s)),\nonumber \\
{{\delta}\over {\delta s}}\, {}^4E^A_{(2)}(\sigma (s)) &=&-\tau_1(s)\, {}^4E^A
_{(1)}(\sigma (s)) +\tau_2(s)\, {}^4E^A_{(3)}(\sigma (s)),\nonumber \\
{{\delta}\over {\delta s}}\, {}^4E^A_{(3)}(\sigma (s)) &=&-\tau_2(s)\, {}^4E^A
_{(2)}(\sigma (s)),
\label{d12}
\end{eqnarray}

\noindent where $\kappa (s)$, $\tau_1(s)$, $\tau_2(s)$ are the curvature and 
the first and second torsion of $\Gamma$ respectively [${}^4E^A_{(1)}$, ${}^4E
^A_{(2)}$, ${}^4E^A_{(3)}$ are said the normal and the first and second 
binormal respectively].

In Ref.\cite{mtw} [chapter 6 and section 13.6] there is the construction of the
``proper reference frame" of an accelerated observer, which uses ``Fermi
normal coordinates" $\tau_F$, ${\vec \sigma}_F$ [they are special Riemann
normal coordinates which are normal in all the points of the 4-geodesic
$\Gamma$]. This proper reference frame is both accelerated and rotating
relative to the local Lorentz frames along $\Gamma$ (as it can be shown with
accelerometer measurements and from the rotation of inertial-guidance
gyroscopes due to Coriolis and inertial forces). This proper reference frame
can be extended around the worldline $\Gamma$ till distances $l << {{c^2}\over
g},\quad {c\over \Omega}$ (the acceleration lengths for linear acceleration and
rotation respectively \cite{mashhoon}) due to inertial and tidal effects
[the hypothesis of locality requires that the intrinsic length and time scales 
of the phenomena under observation be negligibly small relative to the
corresponding acceleration scales associated with the observer].

The parameter $\tau_F$ is the proper time as measured by the accelerated 
observer's clock; the coordinates on the slice $\Sigma_{\tau_F}$ with normal
$l^A_F$ are the proper lengths (used as affine parameters) along 3-geodesics
emanating from $\Gamma$ (they are orthogonal to $l^A_F$ and determine locally
$\Sigma_{\tau_F}$). The line element is\cite{mtw,manasse}\hfill\break
\hfill\break
$ds^2=\epsilon \Big[ [1+{}^4R_{\tau r\tau s}\sigma_F^r\sigma_F^s] (d\tau_F)^2+
{4\over 3}\, {}^4R_{\tau mrn}\sigma_F^m\sigma_F^n\, d\tau_Fd\sigma_F^r -
(\delta_{rs}-{1\over 3}\, {}^4R_{rsmn}\sigma_F^m\sigma_F^n)d\sigma_F^rd\sigma_F
^n \Big] +O(| {\vec \sigma}_F |^3).$\hfill\break
\hfill\break
The observer carries with himself an orthonormal tetrad ${}^4E^A_{(\alpha )}$
with ${}^4E^A_{(o)}=l^A_F{|}_{\Gamma}=u^A$ (the 4-velocity of the observer),
which changes from point to point of $\Gamma$: ${{\delta}\over {\delta s}}\,
{}^4E^A_{(\alpha )}(\sigma (s))=-\Omega^A{}_B(s)\, {}^4E^B_{(\alpha )}(\sigma 
(s))$ [for $\omega^A=0$ the observer would Fermi-Walker transport his tetrad
(it would become a Fermi frame), while for $a^A=\omega^A=0$ he would be freely 
falling (geodesic motion with local Lorentz frames along all $\Gamma$) with
parallel transport of his tetrad: $u^B\, {}^4\nabla_B\, {}^4E^A_{(\alpha )}=0$].

An accelerated observer looking at a freely falling particle as it passes 
through the origin of his proper reference frame [$v^r={{dx^{(a)}}\over {dx
^{(o)}}}\, {}^3e^r_{(a)}$ is the 3-velocity of the particle; at the origin one
chooses ${}^3E^A_{(o)}=u^A=(1;\vec 0)$, ${}^4E^A_{(a)}=(0; {}^3e^r_{(a)})$],
sees the following 3-acceleration of the particle:\hfill\break
\hfill\break
${{d^2x^{(a)}}\over {dx^{(o)\, 2}}}\, {}^3e^r_{(a)}=-a^r-2(\vec \omega \times 
\vec v)^r+2(\vec a\cdot \vec v) v^r$,\hfill\break
\hfill\break
where $a^A(0;\vec a)$ is the observer's own 4-acceleration, $\vec \omega$ is the
angular velocity with which his spatial triad ${}^3e^r_{(a)}$ is rotating. The 
three terms are the inertial acceleration, the Coriolis acceleration and a 
relativistic correction to the inertial acceleration respectively.

In particular the 3+1 splitting (slicing) with the spacelike hypersurfaces 
$\Sigma_{\tau}$ has the associated $\Sigma_{\tau}$-adapted tetrads ${}^4
_{(\Sigma )}{\check {\tilde E}}^A_{(\alpha )}$ of Eq.(\ref{II7}) with
${}^4_{(\Sigma )}{\check {\tilde E}}^A_{(o)}=l^A={{\epsilon}\over N}(1; -N^r)$:
the unit normal vector field $l^A$ to $\Sigma_{\tau}$ can be interpreted as the
4-velocity field of observers instantaneously at rest in the slices $\Sigma
_{\tau}$, called ``Eulerian observers", because their motion follows the slices
with 4-acceleration ${}^3a^A$ tangent to $\Sigma_{\tau}$.
For this special surface forming ($\omega_{AB}=0$) nongeodesic congruence we 
have [we use the ${\hat b}^A_{\bar A}$]

\begin{eqnarray}
{}^4\nabla_A l_B&=& \epsilon \, {}^3a_Bl_A - {}^3K_{AB},\nonumber \\
&&{\dot l}_A={}^3a_A= {}^3a_r {\hat b}^r_A,\quad\quad {}^3a_r=\partial_r ln\, 
N,\nonumber \\
&&\Theta ={}^4\nabla_A l^A= \epsilon \, {}^3K,\nonumber \\
&&\sigma_{AB}= -[{}^3K_{rs}-{1\over 3} {}^3g_{rs}\, {}^3K] {\hat b}^r_A{\hat b}
^s_B.
\label{d13}
\end{eqnarray}

Let us remark that by a suitable choice of gauge it is possible to consider a
local foliation whose leaves $\Sigma_{\tilde \tau}$ are orthogonal to a
surface-forming timelike (or even spacelike) geodesic congruence [in general, 
this is possible only for a finite interval $\triangle \tau$, because coordinate
singularities appear for increasing $\tau$ due to the focusing property of
4-geodesics]. This case corresponds to a local system of ``Gaussian normal
coordinates"\cite{mtw} $\tilde \tau$, ${\vec {\tilde \sigma}}$ such that: 
\hfill\break
\hfill\break
i) the shift functions vanish: $N^r=0$ and ${}^4{\tilde g}
_{\tilde \tau r}=0$ [the 
surfaces $\Sigma_{\tilde \tau}$ are (locally) surfaces of simultaneity for the
observers moving along the geodesics of the congruence];\hfill\break
ii) the coordinate time $\tilde \tau$ measures proper time along the geodesics:
$d\tilde \tau =Nd\tau$, $ds^2= \epsilon \Big[ (d\tilde \tau )^2-{}^3{\tilde g}
_{rs}d{\tilde \sigma}^rd{\tilde \sigma}^s\Big]$.\hfill\break
\hfill\break
These coordinates are also called ``synchronous" coordinates; in cosmology, 
they are also said ``comoving", because the cosmological fluid (whose fluid 
lines are the geodesics of the congruence) is always at rest relative to $\Sigma
_{\tilde \tau}$.

\vfill\eject

\section{Hamiltonian Expression of 4-Tensors.}

By using Eq.(\ref{III9}) and (\ref{III18}) to eliminate the $\tau$-derivatives,
we get the Hamiltonian version of the quantities defined in Appendix B
[the symbol ``${\buildrel \circ \over =}$" identifies the components of the
4-tensors whose phase space expression requires the first half of the
Hamilton equations (\ref{III18})
for N, $N_{(a)}$, ${}^3e_{(a)r}$; remember that $\lambda_N\,
{\buildrel \circ \over =}\, \partial_{\tau} N$, $\lambda^{\vec N}_{(a)}\,
{\buildrel \circ \over =}\, \partial_{\tau} N_{(a)}$]. In this form we make
explicit the dependence of 4-tensors on the arbitrary lapse and shift
functions conjugate to the four first class constraints ${\tilde \pi}^N(\tau ,
\vec \sigma )\approx 0$, ${\tilde \pi}^{\vec N}_{(a)}(\tau ,\vec \sigma )
\approx 0$, but not yet the dependence on the further ten arbitrary functions 
conjugate to the remaining ten first class constraints, which have not yet
been used in the expression of the 4-tensors. Let us remark that the 4-tensors
of metric gravity do not depend on the three boost parameters $\varphi_{(a)}
(\tau ,\vec \sigma )$ [conjugate to ${\tilde \pi}^{\vec \varphi}_{(a)}(\tau 
,\vec \sigma )\approx 0$] and on the three angles conjugated to 
${}^3{\tilde M}_{(a)}(\tau ,\vec \sigma )\approx 0$.

We have [see Eqs.(\ref{II5a}) for the expressions of ${}^3\Gamma^u_{rs}$, 
${}^3\omega_{r(a)}$, ${}^3\Omega_{rs(a)}$, in terms of cotriads]

\begin{eqnarray}
{}^4\Gamma^{\tau}_{\tau\tau}\,&=&\, {1\over {N}}
\Big[ \lambda_N +N_{(a)}\, {}^3e^r_{(a)}\partial_rN-{{\epsilon}\over {4k\, 
{}^3e}} {}^3G_{o(a)(b)(c)(d)}N_{(a)}N_{(b)} 
\, {}^3e_{(c)u}\, {}^3{\tilde \pi}^u_{(d)} \Big] ,\nonumber \\
{}^4\Gamma^{\tau}_{r\tau}&=&{}^4\Gamma^{\tau}_{\tau r}={1\over {N}}
\Big[ \partial_rN-{{\epsilon}\over {4k\, {}^3e}} {}^3G_{o(a)(b)(c)(d)}\, {}^3e
_{(a)r}N_{(b)}\, {}^3e_{(c)u}\, {}^3{\tilde \pi}^u_{(d)} \Big],\nonumber \\
{}^4\Gamma^{\tau}_{rs}&=&{}^4\Gamma^{\tau}_{sr}=-{1\over {N}}{{\epsilon}\over 
{4k\, {}^3e}}{}^3G_{o(a)(b)(c)(d)}\, {}^3e_{(a)r}\, {}^3e_{(b)s}\, {}^3e_{(c)u}
\, {}^3{\tilde \pi}^u_{(d)},\nonumber \\
{}^4\Gamma^u_{\tau\tau}\, &{\buildrel \circ \over =}&\, {}^3e^u_{(a)}
\Big[ \lambda^{\vec N}_{(a)}-{{N_{(a)}}\over N} \lambda_N\Big]+\nonumber \\
&+&N (\delta_{(a)(b)}-{{N_{(a)}N_{(b)}}\over {N^2}}){}^3e^u_{(a)}\, {}^3e^v
_{(b)} \partial_vN+ {}^3e^v_{(b)}N_{(b)} ({}^3e^u_{(a)}\, N_{(a)})_{|v}-
\nonumber \\
&-&{{\epsilon N}\over {2k\, {}^3e}}\, {}^3G_{o(a)(b)(c)(d)}N_{(a)} (\delta
_{(b)(e)}-{{N_{(b)}N_{(e)}}\over {2N^2}}) {}^3e^u_{(e)}\, {}^3e_{(c)v}\,
{}^3{\tilde \pi}^v_{(d)}-\nonumber \\
&-&N_{(a)}\, {}^3e^u_{(b)} \Big[
{{\epsilon N}\over {4k\, {}^3e}} {}^3G_{o(a)(b)(c)
(d)}\, {}^3e_{(c)r}\, {}^3{\tilde \pi}^r_{(d)}+\nonumber \\
&+&{}^3e^v_{(a)} (N_{(c)}\, {}^3e_{(c)}^r \partial_r\, {}^3e_{(b)v}+
{}^3e_{(b)r}\, \partial_v(N_{(c)}\, {}^3e^r_{(c)}))+\nonumber \\
&+&\epsilon_{(a)(b)(c)} {\hat \mu}_{(c)}\Big],\nonumber \\
{}^4\Gamma^u_{r\tau}&=&{}^4\Gamma^u_{\tau r}={}^3e^u_{(a)}\Big[ N_{(a)|r}-
{{N_{(a)}}\over N} \partial_rN\Big] -\nonumber \\
&-&{{\epsilon N}\over {4k\, {}^3e}}(\delta_{(a)(b)}-{{N_{(a)}N_{(b)}}\over
{N^2}}){}^3e^u_{(a)}\, {}^3G_{o(b)(c)(d)(e)}\, {}^3e_{(c)r}\, {}^3e_{(d)s}\,
{}^3{\tilde \pi}^s_{(e)},\nonumber \\
{}^4\Gamma^u_{rs}&=&{}^4\Gamma^u_{sr}={}^3\Gamma^u_{rs}+{{N_{(e)}}\over {N}}\,
{{\epsilon}\over {4k\, {}^3e}}\, {}^3e^u_{(e)}
{}^3G_{o(a)(b)(c)(d)}\, {}^3e_{(a)r}\, {}^3e_{(b)s}
\, {}^3e_{(c)v}\, {}^3{\tilde \pi}^v_{(d)},
\label{e1}
\end{eqnarray}

\begin{eqnarray}
{}^4{\buildrel \circ \over {\omega}}_{\tau (o)(a)}&=&-{}^4{\buildrel \circ 
\over {\omega}}_{\tau (a)(o)}=\nonumber \\
&=&-\epsilon {}^3e^r_{(a)}\partial_rN-{1\over {4k\, {}^3e}}{}^3G
_{o(a)(b)(c)(d)}\, N_{(b)}\, {}^3e_{(c)u}\, {}^3{\tilde \pi}^u_{(d)},
\nonumber \\
{}^4{\buildrel \circ \over {\omega}}_{\tau (a)(b)}&=&-{}^4{\buildrel \circ 
\over {\omega}}_{\tau (b)(a)}\, {\buildrel \circ \over =}\, -\epsilon \,
[{}^3\omega_{r(a)(b)}\, {}^3e^r_{(c)}N_{(c)}+\epsilon_{(a)(b)(c)}{\hat \mu}
_{(c)}],\nonumber \\
{}^4{\buildrel \circ \over {\omega}}_{r(o)(a)}&=&-{}^4{\buildrel \circ \over 
{\omega}}_{r(a)(o)}=-{1\over {4k\, {}^3e}}{}^3G_{o(a)(b)(c)(d)}{}^3e_{(b)r}\,
{}^3e_{(c)u}\, {}^3{\tilde \pi}^u_{(d)},\nonumber \\
{}^4{\buildrel \circ \over {\omega}}_{r(a)(b)}&=&-{}^4{\buildrel \circ \over 
{\omega}}_{r(b)(a)}=-\epsilon {}^3\omega_{r(a)(b)}=
{1\over 2}\Big[ {}^3e^s_{(a)}(\partial_r\, {}^3e_{(b)s}-
\partial_s\, {}^3e_{(b)r})+\nonumber \\
&+&{}^3e^s_{(b)}(\partial_s\, {}^3e_{(a)r}-\partial_r\, {}^3e_{(a)s})+{}^3e^u
_{(a)}\, {}^3e^v_{(b)}\, {}^3e_{(c)r}(\partial_v\, {}^3e_{(c)u}-\partial_u\,
{}^3e_{(c)v}) \Big] .
\label{e2}
\end{eqnarray}

\begin{eqnarray}
{}^4{\buildrel \circ \over {\Omega}}_{rs(a)(b)}&=&{}^3e^u_{(a)}\, {}^3e^v
_{(b)}\, {}^4R_{uvrs}=-\epsilon \Big[ {}^3\Omega_{rs(a)(b)}+\nonumber \\
&+&{1\over {(4k\, {}^3e)^2}}\, {}^3G_{o(a)(c)(d)(e)}\, {}^3G_{o(b)(f)(g)(h)}
\nonumber \\
&\cdot& ({}^3e_{(c)r}\, {}^3e_{(f)s}-{}^3e_{(c)s}\, {}^3e_{(f)r}) {}^3e_{(d)u}
\, {}^3{\tilde \pi}^u_{(e)}\, {}^3e_{(g)v}\, {}^3{\tilde \pi}^v_{(h)}\Big]
,\nonumber \\
{}^4{\buildrel \circ \over {\Omega}}_{rs(o)(a)}&=&{1\over N}{}^3e^v_{(a)}
({}^4R_{\tau vrs}-N^u\, {}^4R_{uvrs})=\nonumber \\
&=&{1\over {4k}} \, {}^3e^u_{(a)}\Big[ ({1\over {{}^3e}} {}^3G_{o(b)(c)(d)(e)}\,
{}^3e_{(b)r}\, {}^3e_{(c)u}\, {}^3e_{(d)v}\, {}^3{\tilde \pi}^v_{(e)})_{|s}-
\nonumber \\
&-&({1\over {{}^3e}} {}^3G_{o(b)(c)(d)(e)}\, {}^3e_{(b)s}\, {}^3e_{(c)u}\,
{}^3e_{(d)v}\, {}^3{\tilde \pi}^v_{(e)})_{|r}\Big] ,\nonumber \\
{}^4{\buildrel \circ \over {\Omega}}_{\tau r(a)(b)}&=&{}^3e^u_{(a)}\, {}^3e^v
_{(b)}\, {}^4R_{uv\tau r}\, {\buildrel \circ \over =}\, \nonumber \\
&{\buildrel \circ \over =}&\, -\epsilon \Big( \partial_{\tau}\, {}^3\omega
_{r(a)(b)}+{1\over 2}(\epsilon_{(a)(b)(c)}\epsilon_{(d)(e)(f)}-\epsilon
_{(a)(b)(d)}\epsilon_{(c)(e)(f)})\cdot \nonumber \\
&&{}^3e^s_{(c)} \Big[ {{\epsilon N}\over {4k\, {}^3e}} {}^3G_{o(d)(l)(m)(n)}\,
{}^3e_{(l)s}\, {}^3e_{(m)v}\, {}^3{\tilde \pi}^v_{(n)}+\nonumber \\
&+&N_{(l)}\, {}^3e^u_{(l)}\partial_u\, {}^3e_{(d)s}+{}^3e_{(d)u}\partial_s
(N_{(l)}\, {}^3e^u_{(l)})+\epsilon_{(d)(m)(n)}{\hat \mu}_{(m)}\, {}^3e_{(n)s}-
\nonumber \\
&-&N_{(g)}\, {}^3e^u_{(g)}\partial_u\, {}^3e_{(d)s}-{}^3e_{(d)u}\partial_s
(N_{(g)}\, {}^3e^u_{(g)})\Big] {}^3\omega_{r(e)(f)}+\nonumber \\
&+&N_{(c)}\, {}^3e^s_{(c)}\, [{}^3\omega_s,{}^3\omega_r]_{(a)(b)}+\nonumber \\
&+&{{\epsilon}\over {4k\, {}^3e}} {}^3G_{o(c)(d)(e)(f)}\, {}^3e_{(c)r}\, 
{}^3e_{(e)u}\, {}^3{\tilde \pi}^u_{(f)}(\delta_{(a)(d)}\, {}^3e^u_{(b)}-
\delta_{(b)(d)}\, {}^3e^u_{(a)}) \partial_uN+\nonumber \\
&+&{1\over {(4k\, {}^3e)^2}}(\delta_{(a)(l)}\delta_{(b)(d)}-\delta_{(a)(d)}
\delta_{(b)(l)}){}^3G_{o(d)(e)(f)(g)}\, {}^3G_{o(h)(l)(m)(n)}\cdot
\nonumber \\
&\cdot&  {}^3e_{(h)r}
N_{(e)}\, {}^3e_{(f)w}\, {}^3{\tilde \pi}^w_{(g)}\, {}^3e_{(m)v}\, {}^3{\tilde
\pi}^v_{(n)} \Big) , \nonumber \\
{}^4{\buildrel \circ \over {\Omega}}_{\tau r(o)(a)}&=&{1\over N}\, {}^3e^u
_{(a)}({}^4R_{\tau u\tau r}-N^s\, {}^4R_{su\tau r})\, {\buildrel \circ \over
=}\nonumber \\
&{\buildrel \circ \over =}&\, -\epsilon \, {}^3e^s_{(a)} \Big[ \partial_{\tau}
\, {}^3K_{rs} + N_{|s|r}-\nonumber \\
&-&{{\epsilon}\over {4k\, {}^3e}}\, {}^3G_{o(c)(d)(e)(f)}\, {}^3e_{(d)u}\, 
{}^3e_{(e)w}\, {}^3{\tilde \pi}^w_{(f)} \Big( {}^3e_{(c)r}(N_{(b)}\, {}^3e^u
_{(b)})_{|s} + {}^3e_{(c)s}(N_{(b)}\, {}^3e^u_{(b)})_{|r}\Big) -\nonumber \\
&-&{{\epsilon}\over {4k}} N_{(b)}\, {}^3e^u_{(b)} \Big( {}^3G_{o(c)(d)(e)(f)}\, 
{}^3e_{(c)s}\, {}^3e_{(d)u}\, {}^3e_{(e)w}\, {}^3{\tilde \pi}^w_{(f)}\Big)_{|r}
\, \Big] .
\label{e3}
\end{eqnarray}

In the last two equations  the quantities $\partial_{\tau}\, {}^3\omega
_{r(a)(b)}$ and $\partial_{\tau}\, {}^3K_{rs}$ are a shorthand for their 
expression which is given in Eqs.(\ref{III18a}). Let us remark that, since
$\partial_{\tau}\, {}^3K_{rs}$ depends on $\partial_{\tau}\, {}^3{\tilde \pi}^r
_{(a)}$, the quantities ${}^4{\buildrel \circ \over {\Omega}}_{\tau r(o)(a)}$ 
[and, therefore, ${}^4R_{\tau r\tau s}$] 
are dynamical, because they require the use of the second half of
the Hamilton equations (\ref{III18}) [i.e. of the Einstein equations ${}^4{\bar
G}_{rs}\, {\buildrel \circ \over =}\, 0$] for their
explicit phase space determination. Then, we get

\begin{eqnarray}
{}^4R_{rsuv}&=&{}^3e_{(a)r}\, {}^3e_{(b)s}\, {}^4{\buildrel \circ \over {\Omega
}}_{uv(a)(b)}=-\epsilon \Big[ {}^3R_{rsuv}+{1\over {(4k\, {}^3e)^2}}
{}^3G_{o(a)(b)(c)(d)}\, {}^3G_{o(e)(f)(g)(h)}\nonumber \\
&\cdot& {}^3e_{(a)r}\, {}^3e_{(e)s} ({}^3e_{(b)u}\, {}^3e_{(f)v}-{}^3e_{(b)v}\,
{}^3e_{(f)u})\, {}^3e_{(c)t}\, {}^3{\tilde \pi}^t_{(d)}\, {}^3e_{(g)w}\,
{}^3{\tilde \pi}^w_{(h)}\Big] ,\nonumber \\
{}^4R_{\tau ruv}&=&{}^3e_{(a)u}\, {}^3e_{(b)v}\, {}^4{\buildrel \circ \over
{\Omega}}_{\tau r(a)(b)}
=N\, {}^3e_{(a)r}\, {}^4{\buildrel \circ \over {\Omega}}
_{uv(o)(a)}+N_{(a)}\, {}^3e_{(b)r}\, {}^4{\buildrel \circ \over {\Omega}}
_{uv(a)(b)},\nonumber \\
{}^4R_{\tau r\tau s}\, &=&\, N\, {}^3e_{(a)r}\, 
{}^4{\buildrel \circ \over {\Omega}}_{\tau s(o)(a)}+N_{(a)}\, {}^3e_{(b)r}\,
{}^4{\buildrel \circ \over {\Omega}}_{\tau s(a)(b)}.
\label{e4}
\end{eqnarray}

By using Eqs.(\ref{d7}) and (\ref{d8}),  we can get the phase
space expression of ${}^4R_{AB}\, {\buildrel \circ \over =}\, 0$, ${}^4R
\, {\buildrel \circ \over =}\, $, ${}^4C_{ABCD}\, {\buildrel \circ \over =}\, 
{}^4R_{ABCD}$. Let us remember
that the acceleration of the integral curves with tangent vector $l^{\mu}(\tau
,\vec \sigma )$ [the normal to $\Sigma_{\tau}$ in $z^{\mu}(\tau ,\vec \sigma 
)$] is ${}^3a_r(\tau ,\vec \sigma ) = \partial_r\, ln\, N(\tau ,\vec \sigma )$.

\vfill\eject

\end{document}